\documentclass[11pt,a4paper]{article}
\pdfoutput=1

\usepackage{jheppub}
\usepackage{subfigure}
\usepackage{epstopdf}
\usepackage[T1]{fontenc}
\usepackage{extarrows}

\title{Anomalous transport from holography: Part I}
\author[a]{Yanyan Bu,}
\author[a,b]{Michael Lublinsky,}
\author[a]{and Amir Sharon}
\affiliation[a]{Department of Physics, Ben-Gurion University of the Negev,
Beer-Sheva 84105, Israel}
\affiliation[b]{Physics Department, University of Connecticut, 2152 Hillside
Road, Storrs, CT 06269-3046, USA}

\emailAdd{yybu@post.bgu.ac.il}
\emailAdd{lublinm@bgu.ac.il}
\emailAdd{sharon.amir24@gmail.com}

\abstract{We revisit the transport properties induced by the chiral anomaly in a charged plasma holographically dual to anomalous $U(1)_V\times U(1)_A$ Maxwell theory in  Schwarzschild-$AdS_5$. Off-shell constitutive relations for vector and axial currents are derived using various approximations generalising most of known in the literature anomaly-induced phenomena and revealing some new ones. In a weak external field approximation, the constitutive relations have all-order derivatives resummed into six momenta-dependent transport coefficient functions: the diffusion, the electric/magnetic conductivity, and three anomaly induced functions. The latter generalise the chiral magnetic and chiral separation effects. Nonlinear transport is  studied assuming presence of constant background external fields. The chiral magnetic effect, including all order nonlinearity in magnetic field, is proven to be exact when the magnetic field is the only external field that is turned on. Non-linear corrections to the constitutive relations due to electric and axial external fields are computed.}


\arxivnumber{1608.08595}
\begin{document}
\maketitle

\flushbottom

\section{Introduction and summary} \label{section1}

Hydrodynamics is an effective long-distance description of most QFTs at nonzero temperature. Within the hydrodynamic approximation, the entire dynamics of a microscopic theory is reduced to that of  macroscopic currents, such as of  charge current operators computed in a locally near equilibrium thermal state. An essential element of any hydrodynamics is a constitutive relation which relates the macroscopic currents to fluid-dynamic variables, such as charge densities,  and to external forces. The most simple example of constitutive relation is the diffusion approximation for the electric current $\vec J$
\begin{equation}\label{cr1}
\vec{J}=-\mathcal{D}_0\vec{\nabla}\rho
\end{equation}
where $\rho$ is the corresponding  charge density. Derivative expansion in the fluid-dynamic variables  accounts for deviations from thermal equilibrium.  At each order, the derivative expansion is fixed by thermodynamics and symmetries, up to a finite number of transport coefficients. The latter are not calculable from hydrodynamics itself, but have to be determined from underlying  microscopic theory or experimentally.

Chiral anomalies emerge and play an important role in relativistic QFTs with massless fermions. The chiral anomaly is reflected in three-point functions of currents associated with global symmetries.
%and this is the triangle anomaly.
When the global $U(1)$ currents are coupled to external electromagnetic fields, the triangle anomaly renders the axial current into non-conserved,
\begin{equation} \label{continuity}
\partial_\mu J^\mu=0,~~~~~~~~~~~\partial_\mu J^\mu_5=4\kappa \left(3\vec{E}\cdot \vec{B}+ \vec{E}^a\cdot \vec{B}^a\right),
\end{equation}
where $J^\mu$/$J^\mu_5$ are vector/axial currents, and $\kappa$ is the anomaly coefficient. In $SU(N_c)$ gauge theory with a massless Dirac fermion in fundamental
representation, $\kappa=e N_c/(24\pi^2)$ and $e$ is electric charge. Here $\vec{E},\vec{B}$ ($\vec{E}^a,\vec{B}^a$) are vector (axial) external electromagnetic fields\footnote{A possibility of experimentally creating axial electromagnetic fields in a laboratory was recently discussed in~\cite{1607.03491}.}.

In presence of external fields, the triangle anomaly modifies the usual constitutive relations for the vector/axial currents. One such example is the chiral magnetic effect (CME)~\cite{hep-ph/0406125,0706.1026,0711.0950,0808.3382, 0911.3715}\footnote{See also \cite{Vilenkin:1980fu,hep-ph/9710234,cond-mat/9803346} for earlier related works.}:  appearance of electric current directed  along  applied magnetic field and  is due to    nonzero topological charge. In QCD coupled to electromagnetism, CME is realised via chirality imbalance between left- and right-handed quarks, usually parametrised by an axial chemical potential. In perturbative QCD, the chiral magnetic conductivity was computed in~\cite{0907.5007,1002.2495,1406.1150,1406.3584,1504.05866}. Lattice simulations of CME can be found in~\cite{0907.0494,0911.1348,0912.2961,1011.3795,1105.0385,1303.6266}. In strongly coupled regime, holographic AdS/CFT correspondence~\cite{hep-th/9711200,hep-th/9802109,hep-th/9802150} was used to compute chiral magnetic conductivity in~\cite{0908.4189,0909.4782,0910.3722,1003.2293, 1005.1888,1005.2587,1102.4334,1103.3773,1112.4227,1207.5309,1305.3949,1603.08770}.

The chiral separation effect (CSE)~\cite{hep-ph/0405216,hep-ph/0505072} is another interesting phenomenon induced by triangle anomalies. It is reflected in separation of chiral charges along external magnetic field at finite density of vector charges. Chiral charges can  be also separated along  external electric field, when both vector and axial charge densities are nonzero, the so-called chiral electric separation effect (CESE)~\cite{1303.7192,1409.6395}.

Without external fields, triangle anomalies affect transport properties through hydrodynamic flows.
%When a fluid flows vortically,
Particularly, there exists an anomaly induced chiral vortical effect~\cite{0809.2488,0809.2596,0906.5044}, which relates the current to fluid's vorticity. In the fluid's local rest frame, the chiral vortical effect is  $\vec{J}=\frac{1}{2}\xi \vec{\nabla} \times \vec{u}$, where $\vec{u}$ is the fluid velocity. The transport coefficient $\xi$ was first calculated in~\cite{0809.2488,0809.2596} using the fluid/gravity correspondence~\cite{0712.2456,0905.4352,1107.5780}. Later,  in~\cite{0906.5044} it was shown that the chiral vortical term is required by existence of a positive-definite entropy current associated with the hydrodynamic system and furthermore that
$\xi$ is uniquely determined by the anomaly coefficient $\kappa$.

In heavy ion collisions, experimentally observable effects induced by the anomalies  were discussed in~\cite{1002.0804,1010.0038,1103.1307,1311.2574,1608.00982}. We refer the reader to~\cite{1312.3348,1509.04073,1511.04050} for comprehensive reviews on the subject of anomalous transports.

In~\cite{1511.08789} we derived the most general off-shell constitutive relation for a globally conserved $U(1)$ current driven by non-dynamical external electromagnetic fields. The derivation involved a resummation of all-order gradient terms in $U(1)$ current. The gradient resummation was implemented via the technique of~\cite{1406.7222,1409.3095,1502.08044,1504.01370}. The latter was devised to resum, in linear approximation, all-order velocity derivatives in the energy-momentum tensor of a holographic conformal fluid. For a holographically defined theory involving a probe Maxwell field in the Schwarzschild-$AdS_5$ geometry, the constitutive relation for the boundary  current was found to be parameterised by three momenta-dependent transport coefficient functions: diffusion and two conductivities. The key element in the derivation was   "off-shellness" of our method.  That is the transport
coefficients were uniquely determined via solution of dynamical components of the
Maxwell equations in the bulk.  The constraint component was shown to be equivalent to continuity equation of thus derived current.
%were split into dynamical and constraint components. While dynamical components were solved to precisely determine all the three transport coefficient functions, constraint component is equivalent to continuity equation of thus derived current.

In the present work we extend the study of~\cite{1511.08789} and account for the effects  induced by the triangle anomaly: when the triangle anomaly is present for both left/right-handed chiralities,
we derive off-shell constitutive relations for vector/axial currents.
%As seen from~(\ref{continuity}), the vector current is strictly conserved but the axial one is anomalous.
As mentioned above, anomalies contribute to the stress-energy tensor~\cite{0809.2488,0809.2596}. However, as in~\cite{1511.08789} we chose to
work in the probe limit in which the currents and  stress-energy tensor decouple. In the dual gravity, the probe limit ignores the backreaction of the gauge dynamics on the bulk geometry. The holographic model in study consists of two Maxwell fields in the Schwarzschild-$AdS_5$ black brane geometry. The triangle anomaly is holographically modeled via the gauge Chern-Simons action for both Maxwell fields (with opposite signs).
This holographic setup can be realised via a top-down brane construction of $D4/D8/\overline{D8}$~\cite{hep-th/0412141}.

Below we will consider the charge densities (chemical potentials) as constant with small inhomogeneous fluctuations on top:
%That is, the vector (axial) charge density $\rho$ ($\rho_{_5}$) and chemical potential $\mu$ ($\mu_{_5}$) are split into
\begin{equation}
\begin{split}
&\rho(x_\alpha)=\bar{\rho}+\epsilon \delta \rho(x_\alpha), ~~~~~~~~~~~~\rho_{_5}(x_\alpha) = \bar{\rho}_{_5} +\epsilon \delta \rho_{_5}(x_\alpha),\\
&\mu(x_\alpha)=\bar{\mu}+\epsilon \delta \mu(x_\alpha),~~~~~~~~~~~ \mu_{_5}(x_\alpha) = \bar{\mu}_{_5} +\epsilon \delta \mu_{_5}(x_\alpha),
\end{split}
\end{equation}
where $\bar{\rho},\bar{\rho}_{_5},\bar{\mu},\bar{\mu}_{_5}$ are constant backgrounds and $\delta \rho,\delta \rho_{_5},\delta\mu, \delta \mu_{_5}$ are the inhomogeneous fluctuations. The parameter $\epsilon$ is formally introduced as a  small parameter. It will be used to set up a perturbative procedure.
We will be particularly interested in linearisation in inhomogeneous fluctuations and most of our results will be accurate up to first order in $\epsilon$.
%Notice that $\mu(\mu_{_5})$ and $\rho (\rho_{_5})$ are not independent.
The charge densities  $\rho,\rho_{_5}$ are the hydrodynamic variables. They can be related to corresponding chemical potentials via (\ref{def potentials}).
%The dependence of transport coefficients on $\rho,\rho_{_5}$ can be obtained with the help of (\ref{def potentials}).
For constant parts,  $\bar{\mu}=\bar{\rho}/2$ and $\bar{\mu}_{_5}=\bar{\rho}_{_5}/2$.

%In accord with different treatments for external fields,
Our study is divided into two largely independent parts. In the first part, the external fields are assumed to be weak and
%have inhomogeneous fluctuations only. The fields are assumed to
scale linearly with $\epsilon$
% linear order in fluctuations,
\begin{equation}\label{study1}
E_i(x_\alpha)\to \epsilon E_i(x_\alpha),~~~B_i(x_\alpha)\to \epsilon B_i(x_\alpha),~~~ E_i^a(x_\alpha)\to \epsilon E_i^a(x_\alpha),~~~B_i(x_\alpha)\to \epsilon B_i^a(x_\alpha).
\end{equation}
To first order in $\epsilon$, we are able to derive the most general all order off-shell constitutive relations for the vector/axial currents.

In the second part of our work,
%a more general setup is chosen:
the external fields are assumed to have constant background values  $\vec{\bf E},\vec{\bf B},\vec{\bf E}^a, \vec{\bf B}^a$
plus small  inhomogeneous fluctuations $\delta \vec{E},\delta \vec{B},\delta \vec{E}^a,\delta \vec{B}^a$,
\begin{equation}\label{study2}
\begin{split}
\vec{E}(x_\alpha)&=\vec{\bf E}+\epsilon\delta \vec{E}(x_\alpha),~~~~~~~~~~~~ \vec{B}(x_\alpha)=\vec{\bf B}+\epsilon \delta \vec{B}(x_\alpha),~~\\
\vec{E}^a(x_\alpha)&=\vec{\bf E}^a+\epsilon \delta \vec{E}^a(x_\alpha), ~~~~~~~~~
\vec{B}^a(x_\alpha)=\vec{\bf B}^a+\epsilon\delta \vec{E}^a(x_\alpha).
\end{split}
\end{equation}
%where $\vec{\bf E},\vec{\bf B},\vec{\bf E}^a, \vec{\bf B}^a$ are constant backgrounds and $\delta \vec{E},\delta \vec{B},\delta \vec{E}^a,\delta \vec{B}^a$ are fluctuations.
We will see that the constant backgrounds  induce interesting nonlinear anomaly-induced structures in both currents. %Obviously, these two independent studies are identical once constant backgrounds $\vec{\bf E},\vec{\bf B},\vec{\bf E}^a, \vec{\bf B}^a$ are set to zero.

Our study goes in several directions beyond the results reported in the literature.

$\bullet$ For the linearised setup, we present a rigorous  derivation of off-shell constitutive relations for the vector/axial currents, coupled to non-dynamical external vector/axial electromagnetic fields. Apart from the well-known chiral magnetic/separation effects, we obtain two additional anomaly-related transport coefficients,  magnetic conductivity and its axial analogue. Furthermore, all the transport coefficients are generalised to momenta-dependent functions as a result of exact all order gradient resummation.
%, which were not paid much attention in the literature.

$\bullet$
Beyond the linear regime, we calculate some nonlinear effects induced by constant background external fields. While  the phenomena that
we discover are largely not new, most of them have not been reported for the present holographic model.  In the absence of external electric and axial fields,  the chiral magnetic/separation effects are proven to be exact for arbitrary strong constant magnetic field.

$\bullet$
In a follow-on publication \cite{BLS},  we will extend the study of non-linear CME to spatially-varying magnetic field and will evaluate derivative corrections to it in the constitutive relation.
%As a future work, further investigation along this line will reveal the dependence of momenta-dependent anomalous transports on the magnetic field.
Furthermore, we will consider a case when the axial chemical potential is dynamically generated through $\vec E\cdot\vec B$ term for the case of constant magnetic field and a weak time-dependent electric fields. Such setup is experimentally realisable in condensed matter systems\footnote{We thank Dmitri Kharzeev for proposing us this study.}.   Dependence of AC conductivity on magnetic field is in the focus of this study.

In the first, "linear", part of our study, all-order derivatives are resummed into the following constitutive relations for the vector/axial currents\footnote{An axial analogue of $\sigma_e$ was found to vanish in the present holographic model.}
\begin{equation}\label{jmu}
J^t=\rho,~~~~~~~~~
\vec{J}=-\mathcal{D}\vec{\nabla}\rho +\sigma_e \vec{E}+\sigma_m \vec{\nabla}\times \vec{B}+\sigma_{\chi} \vec{B}+\sigma_a \vec{\nabla}\times \vec{B}^a+\sigma_\kappa \vec{B}^a,
\end{equation}
where $\mathcal{D}$, $\sigma_{e/m}$, $\sigma_{\chi/\kappa}$ and $\sigma_a$ are scalar functionals of spacetime derivative operators
$$
\mathcal{D}[\partial_t,\vec{\partial}^2],~~~~~\sigma_{e/m}[\partial_t, \vec{\partial}^2], ~~~~~\sigma_{\chi/\kappa}[\partial_t,\vec{\partial}^2],
~~~~~\sigma_a[\partial_t, \vec{\partial}^2].
$$
Transforming to the Fourier space via the replacement $(\partial_t,\vec{\partial})\to (-i\omega,i\vec{q})$, these scalar functionals become functions of the frequency $\omega$ and momentum squared $q^2$. We refer to these functions as transport coefficient functions (TCFs). Here $\omega$ and $\vec{q}$ are dimensionless, while the dimensionfull momenta are $\pi T \omega$ and $\pi T \vec{q}$ with $T$ being the temperature. The axial current is
\begin{equation}\label{jmu5}
J^t_5=\rho_{_5},~~~~~~~
\vec{J}_5=-\mathcal{D}\vec{\nabla}\rho_{_5}+\sigma_e \vec{E}^a+ \sigma_m \vec{\nabla} \times \vec{B}^a+\sigma_\chi \vec{B}^a+\sigma_a \vec{\nabla} \times \vec{B}+\sigma_\kappa \vec{B}.
\end{equation}

TCFs contain information on infinitely many derivatives (and transport coefficients) in conventional hydrodynamic expansion.  The latter is recovered
via small momenta expansion. While most of the results on transport coefficients reported in the literature are  obtained order-by-order in the expansion, our results  are exact to all orders. Furthermore, the TCFs account for collective effects of non-hydrodynamic modes, which never emerge
in the strict low frequency/momentum expansion. The diffusion $\mathcal{D}$, electric conductivity $\sigma_e$ and magnetic conductivity $\sigma_m$ were studied in~\cite{1511.08789}. The additional three TCFs are induced by the anomaly: $\sigma_{\chi}$ is the momenta-dependent chiral magnetic conductivity~\cite{0907.5007}; $\sigma_\kappa$ generalises the chiral separation effect~\cite{hep-ph/0405216,hep-ph/0505072}; $\sigma_a$ is an axial analogue of the magnetic conductivity $\sigma_m$.
%As in~\cite{0907.5007,1312.1204}, our $\sigma_\chi$ ($\sigma_\kappa$) has dependence on vector (axial) chemical potential.

(\ref{jmu},\ref{jmu5}) can be equivalently presented using the chemical potentials $\mu,\mu_{_5}$ instead of the charge densities
\begin{equation} \label{jmu chem}
J^t=\alpha_1\mu+\alpha_2\partial_kE_k,~~~~~
\vec{J}=-\mathcal{D}^\prime\vec{\nabla}\mu + \sigma_e^\prime \vec{E}+ \sigma_m^\prime \vec{\nabla}\times \vec{B}+ \sigma_\chi \vec{B}+\sigma_a \vec{\nabla}\times \vec{B}^a + \sigma_\kappa \vec{B}^a,
\end{equation}
\begin{equation} \label{jmu5 chem}
J^t_5=\alpha_1\mu_{_5}+\alpha_2\partial_kE_k^a,~~~~
\vec{J}_5=-\mathcal{D}^\prime\vec{\nabla}\mu_{_5} + \sigma_e^\prime \vec{E}^a+ \sigma_m^\prime \vec{\nabla}\times \vec{B}^a+ \sigma_\chi \vec{B}^a+\sigma_a \vec{\nabla}\times \vec{B} + \sigma_\kappa \vec{B}.
\end{equation}
When the triangle anomaly is switched off ($\kappa=0$),  (\ref{jmu chem}) for $J^\mu$ coincides with the constitutive relation of~\cite{1511.03646}  derived  from an effective action.

In the present holographic model, we succeeded to uniquely determine all the TCFs in~(\ref{jmu},\ref{jmu5}).  As for (\ref{jmu chem},\ref{jmu5 chem}),
the coefficients $\alpha_1$, $\alpha_2$, $\mathcal{D}^\prime$, $\sigma_{e/m}^\prime$ are in fact frame-dependent. We postpone further discussion
until section \ref{section4}  where these TCFs are presented in a certain frame (see (\ref{alpha12/D prime},\ref{sigma prime})).

In the hydrodynamic limit $\omega,q\ll 1$, the TCFs are series expandable:
\begin{equation} \label{D expansion}
\mathcal{D}=\frac{1}{2}+\underline{\frac{1}{8}\pi i\omega+\frac{1}{48}\left[-\pi^2 \omega^2+ q^2 \left(6\log 2-3\pi\right)\right]}+\cdots,
\end{equation}
\begin{equation}
\sigma_e=1+\frac{\log 2}{2}i\omega+\frac{1}{24}\left[\pi^2\omega^2\underline{-q^2 \left(3\pi+ 6\log 2 \right)} \right]+\cdots,
\end{equation}
\begin{equation}
\sigma_m=72\kappa^2\left(\bar{\mu}^2+\bar{\mu}_{_5}^2\right)\left(2\log 2-1\right) +i\omega\left[\underline{\frac{1}{16}(2\pi-\pi^2+4\log 2)}+ \mathcal{O} \left(\bar{\mu}^2,\bar{\mu}_{_5}^2\right)\right]+\cdots,
\end{equation}
\begin{equation}
\begin{split}
\sigma_\chi&=12\kappa \bar{\mu}_{_5} \left\{1+ i\omega \log 2 -\frac{1}{4} \omega^2 \log^22 -\frac{q^2}{24} \left[\pi^2-1728\kappa^2\left(\bar{\mu}^2_{_5}+ 3\bar{\mu}^2 \right)\left(\log 2-1\right)^2\right]\right\}+\cdots,
\end{split}
\end{equation}
\begin{equation}
\sigma_a=144\kappa^2\bar{\mu}\bar{\mu}_{_5}\left(2\log 2-1\right)+\cdots,
\end{equation}
\begin{equation} \label{sigma_kappa expansion}
\begin{split}
\sigma_\kappa&=12\kappa \bar{\mu} \left\{1+ i\omega\log 2 -\frac{1}{4} \omega^2 \log^22 - \frac{q^2}{24}\left[\pi^2-1728\kappa^2\left(\bar{\mu}^2+ 3\bar{\mu}_{_5}^2\right) \left(\log 2-1\right)^2\right]\right\}+\cdots.
\end{split}
\end{equation}
We were unable to obtain an analytical result for the anomalous correction to $i\omega$-term in $\sigma_m$. In section \ref{sss422} we will reveal that anomalous correction to $i\omega$-term in $\sigma_m$ is linear in $(\bar{\mu}^2+\bar{\mu}_{_5}^2)$.
%Yet, based on the numerical study it could be guessed to be equal the zeroth order term.
Turning the anomaly off, the magnetic conductivity $\sigma_m$  coincides with that of~\cite{1511.08789}. Interestingly,  $\sigma_m$ is being
corrected by the anomaly\footnote{This fact was previously noticed in~\cite{1304.5529} where the authors went beyond the probe limit and
included  backreaction of the bulk gauge fields onto the geometry. However, taking the probe limit in~\cite{1304.5529}
does not seem to coincide with our results.  The reasons behind this discrepancy remain unclear to us.}.
Appearance of anomalous corrections in $\sigma_m$ could be explained as the effect of two triangular anomaly-generating Feynman diagrams
inserted in the current-current correlator\footnote{We thank Ho-Ung Yee for bringing this argument to our attention.} \cite{Gross:1972pv}.

In~\cite{1511.08789}, we made a full comparison of $\mathcal{D},\sigma_e$ and anomaly-free part of $\sigma_m$ with known results in the literature,
particularly with the ones that could be extracted from the current-current correlators.
The underlined terms in $\mathcal{D},\sigma_{e/m}$ cannot be fixed from the  correlators, while the constant pieces of $\sigma_m, \sigma^a$  can be. In~\cite{1310.5160}, the constant piece $\sigma_m^0$ was evaluated for a pure QED plasma with one Dirac fermion at one loop level: the result was found to be positive and was interpreted as an anti-screening of electric currents in the plasma medium. In contrast, in \cite{1003.0699} $\sigma_m^0$ was argued to be zero based on Boltzmann equations. In strongly coupled regime, to our best knowledge, $\sigma_m, \sigma^a$ have not been  reported in the literature.

In the same holographic model as considered here, the constant terms in $\sigma_\chi$ and $\sigma_\kappa$ were originally presented in~\cite{1005.2587}. We find full agreement with those results. The constitutive relations~(\ref{jmu},\ref{jmu5}) can be used to derive Kubo formula~(\ref{kubo formulae}) for $\sigma_{\chi/\kappa}$,
which comes out to be consistent with~\cite{0907.5007}.
Beyond the constant terms, the analytical results in $\sigma_{\chi/\kappa}$ are new as far as we can tell.

Away from the hydrodynamic limit, the TCFs are known numerically only: $\mathcal{D}$, $\sigma_e$ and the anomaly-free part of $\sigma_m$ were already reported in~\cite{1511.08789}; numerical results for $\sigma_m,\sigma_a,\sigma_\chi,\sigma_\kappa$ will be presented below, in section~\ref{sss422}. Comparison of $\sigma_\chi$  with the results of~\cite{0908.4189,1102.4577,1312.1204} will be discussed as well.

In the second part of our study,  new nonlinear structures  emerge in the constitutive relations for both currents.
The complete set of results will be displayed in section~\ref{section5}. Meanwhile we will only flash  the results with the axial background  fields turned off.
To zeroth order in fluctuations, the vector/axial currents are
\begin{equation} \label{jmu EB exact}
J^t_{(0)}=\bar{\rho},~~~~~~~
J^i_{(0)}={\bf E}_i+12\kappa \mu_{_5} {\bf B}_i -12\kappa \epsilon^{ijk} \mathbb{A}_j^{(0)}(1) {\bf E}_k,
\end{equation}
\begin{equation} \label{jmu5 EB exact}
J^t_{5(0)}=\bar{\rho}_{_5},~~~~~~~~
J^i_{5(0)}=12\kappa \mu {\bf B}_i -12\kappa\epsilon^{ijk}\mathbb{V}_j^{(0)}(1){\bf E}_k,
\end{equation}
where the subscript $_{(0)}$ denotes zeroth order in fluctuations ($\mathcal{O}(\epsilon^0$)).
%$\mu$/$\mu_{_5}$ are vector/axial chemical potentials.
$\mathbb{V}_i^{(0)}(1)$ and $\mathbb{A}_i^{(0)}(1)$ are functions of $\vec{\bf E}$, $\vec{\bf B}$, $\mu$, $\mu_{_5}$, and are perturbatively computed in section~\ref{subsection51}. When ${\bf E}=0$, (\ref{jmu EB exact}) confirms exactness of the CME~\cite{hep-ph/0406125,0706.1026,0711.0950,0808.3382,0911.3715} for arbitrary constant magnetic field, in agreement with \cite{1012.6026,1010.1550}.
% demonstrated this exact relation based on field theoretic analysis.
Both $\mu,\mu_{_5}$  depend on ${\bf E}, {\bf B}$ and these dependences account for nonlinearity of the CME with respect to external fields.
The electric field ${\bf E}$ introduces corrections to the original form of the CME and is a source of new structures.
% to be explained below.

In the weak field limit, $\mathbb{V}_j^{(0)}(1)$, $\mathbb{A}_j^{(0)}(1)$ are expandable in the amplitudes of ${\bf E},{\bf B}$.
We quote the results up to third order:
\begin{equation} \label{jmu EB}
\vec{J}_{(0)}=\vec{\bf E}+12\kappa \mu_{_5} \vec{\bf B}-72 \log 2 \,\kappa^2 \mu \vec{\bf B} \times \vec{\bf E} + 18\pi^2 \kappa^3 \mu_{_5} \left(\vec{\bf B}\times \vec{\bf E}\right)\times \vec{\bf E}+\cdots,
\end{equation}
\begin{equation} \label{jmu5 EB}
\vec{J}_{5(0)}=12\kappa \mu \vec{\bf B}-72 \log 2 \,\kappa^2 \mu_{_5} \vec{\bf B} \times \vec{\bf E}+18\pi^2 \kappa^3 \mu \left(\vec{\bf B}\times \vec{\bf E}\right)\times \vec{\bf E}+\cdots,
\end{equation}
where $\mu,\mu_{_5}$ are also expandable in ${\bf E}$ and ${\bf B}$,
\begin{equation} \label{chem/den2}
\begin{split}
\mu=\mu[E,B]&=\frac{1}{2}\bar{\rho}+18\left(1-2\log 2\right)\kappa^2\bar{\rho}{\bf B}^2+\cdots,\\
\mu_{_5}=\mu_5[E,B]&=\frac{1}{2}\bar{\rho}_{_5}+18\left(1-2\log 2\right)\kappa^2\bar{\rho}_{_5}{\bf B}^2+\frac{3}{2}\left(\pi -2\log 2\right)\kappa \vec{\bf B}\cdot \vec{\bf E}+\cdots,
\end{split}
\end{equation}
where $\cdots$ are terms of higher powers in $\bf E$ and $\bf B$. The second order term $\vec{\bf E} \times \vec{\bf B}$ is the Hall current induced by the anomaly, referred to as chiral Hall effect in~\cite{1407.3168}. In classical electromagnetism, the Hall current is generated by the Lorentz force and the imposition of steady state condition. However, the usual Hall current is generated non-linearly and cannot be generated in a holographic model with the Maxwell action only, which does not induce any nonlinearity. Beyond the probe limit, the Hall current does emerge~\cite{1105.6360,1304.5529}.

The last term in~(\ref{jmu EB}) is induced by the anomaly and was derived in~\cite{1603.03442} within the chiral kinetic theory. Quite naturally, the transport coefficient associated with the last term in~(\ref{jmu EB}) calculated in \cite{1603.03442}  is different from our result. Alternatively, using the identity $(\vec{\bf B}\times \vec{\bf E})\times \vec{\bf E}= \vec{\bf E}(\vec{\bf E}\cdot \vec{\bf B})-{\bf E}^2 \vec{\bf B}$, the last term in~(\ref{jmu EB}) can be split into two pieces: one represents ${\bf E}^2$-correction to the chiral magnetic conductivity; the other contributes to the chiral electric effect~\cite{1106.3576}. Analogous to~(\ref{jmu EB}), the last term in~(\ref{jmu5 EB}) is of interest too. On the one hand, it makes ${\bf E}^2$-correction to the chiral separation effect; on the other hand, it contributes to the chiral separation effect induced by the electric field. Our study implies that, via higher order corrections, the chiral electric separation effect exists even when there is no axial chemical potential.

At order $\mathcal{O}(\epsilon^0)$, the continuity equation for $J^\mu_5$ in~(\ref{continuity}) is in tension with (\ref{jmu5 EB}) if $\vec{\bf E}\cdot \vec{\bf B}\neq 0$. That is,  the axial charge density will linearly grow with time leading to instability.  It will manifest itself in a breakdown of the constitutive relations (\ref{jmu EB}) and (\ref {jmu5 EB}), which would have to be amended by derivative terms.

%However, when density fluctuations depend on both time and space, the continuity equation (\ref{continuity}) is put into sourced wave equations for $\delta \rho,\delta \rho_{_5}$. Setting $\vec{\bf E}=0$, these wave equations predict the chiral magnetic wave of \cite{1012.6026}. So, when $\vec{\bf E}\cdot \vec{\bf B}\neq 0$ inhomogeneous fluctuations for charge densities are necessary for consistency.

%A very interesting and presumably experimentally realisable setup is when $\bar\rho=\bar\rho_5=0$, that is when the axial chemical
%potential is generated solely  through  the $\vec {\bf E}\cdot \vec {\bf B}$ term.

%Including contributions to boundary currents from inhomogeneous fluctuations of external fields and charge densities, we first reproduce results of the first part and then generate more (nonlinear) structures.
Beyond the constant background field approximation, numerous new structures emerge, which involve  inhomogeneous fluctuations of
the external fields and charge densities. Particularly, we notice
the emergence of $\delta \rho_{_5} \vec{\bf B}$ ($\delta \rho \vec{\bf B}$) in $\vec{J}$ ($\vec{J}_5$), see~(\ref{structure of jmu},\ref{structure of jmu5}). The interplay between these two terms predicts the chiral magnetic wave~\cite{1012.6026}. We list the new structures
 in section~\ref{subsection52}, but leave computation of corresponding transport coefficients to future work.

%We postpone more results to section~\ref{subsection52}.

%In section \ref{subsection53} we will give a formal proof of exactness for chiral magnetic effect when the magnetic field depends on space only and $\vec{E}=0$.

This paper is structured as follows. In section~\ref{section2} we present the holographic model. In section~\ref{section3} we outline the strategy of deriving the boundary currents through solving the anomalous Maxwell equations in the bulk.
%given unspecified external fields and charge densities.
Section~\ref{section4} presents the first part of our study. In section~\ref{subsection41}, we derive the constitutive relations~(\ref{jmu},\ref{jmu5}) by  solving the dynamical components of the bulk anomalous Maxwell equations near the conformal boundary.
The boundary external fields and charge densities
appear as source terms in these equations. The main technique is based on decomposition of  the bulk gauge fields in terms of basis constructed from  the external fields and charge densities. Dynamics of the Maxwell equations is translated into ordinary differential equations (ODEs) for the decomposition coefficients, which are nothing else but the components of the inverse Green function matrix. In section~\ref{subsection42} we determine the TCFs by  solving these ODEs. In section~\ref{section5} we turn to the second part of this work, corresponding to the scheme~(\ref{study2}). Finally in section~\ref{section6} we again outline the main results of this work and make some discussions. Some technical details are deposited into several appendices. In appendix~\ref{appendix1}, we write down all the  ODEs for the decomposition coefficients  and derive constraint relations among them. In appendix~\ref{appendix2} we summarise analytic perturbative solutions for these coefficients.
%, generating conventional hydrodynamic expansion of transport coefficient functions.
In appendix~\ref{appendix5} we prove that the TCFs in~(\ref{jmu},\ref{jmu5}) are frame-independent.

\section{The holographic model: from $U(1)_L\times U(1)_R$ to $U(1)_V\times U(1)_A$} \label{section2}

The holographic model is the $U(1)_L\times U(1)_R$ theory in the Schwarzschild-$AdS_5$ black brane spacetime.
The triangle anomaly of the boundary field theory is introduced via the Chern-Simons terms (of opposite signs for left/right fields) in the bulk action
\begin{equation}
S_1=\int d^5x \sqrt{-g}\mathcal{L}_1+S_{\textrm{c.t.}},
\end{equation}
where the Lagrangian density $\mathcal{L}_1$ is
\begin{equation}\label{LLR}
\begin{split}
\mathcal{L}_1=&-\frac{1}{4}(F_{L})_{MN}(F_{L})^{MN}-\frac{1}{4}(F_R)_{MN} (F_R)^{MN} +\frac{\kappa_1\, \epsilon^{MNPQR}}{4\sqrt{-g}}\\
&\times\left[(A_L)_M (F_L)_{NP} (F_L)_{QR} - (A_R)_M (F_R)_{NP}(F_R)_{QR}\right].
\end{split}
\end{equation}
In the ingoing Eddington-Finkelstein coordinates, the spacetime metric is
\begin{equation}
ds^2=g_{_{MN}}dx^Mdx^N=2dtdr-r^2f(r)dt^2+r^2\delta_{ij}dx^idx^j,
\end{equation}
where $f(r)=1-1/r^4$, so that the Hawking temperature (identified as temperature of the boundary theory) is normalised to $\pi T=1$.
On the constant $r$ hypersurface $\Sigma$, the induced metric $\gamma_{\mu\nu}$ is
\begin{equation}
ds^2|_{\Sigma}=\gamma_{\mu\nu}dx^\mu dx^\nu=-r^2f(r)dt^2+r^2\delta_{ij}dx^idx^j.
\end{equation}
$\epsilon^{MNPQR}$ is the Levi-Civita symbol with the convention $\epsilon^{rtxyz}=+1$, and the Levi-Civita tensor is $\epsilon^{MNPQR}/\sqrt{-g}$. The counter-term action $S_{\textrm{c.t.}}$ is~\cite{hep-th/0002125,0910.3722,1004.3541}
\begin{equation}\label{ct LR}
S_{\textrm{c.t.}}=\frac{1}{4}\log r \int d^4x \sqrt{-\gamma}\left\{(F_L)_{\mu\nu} (F_L)^{\mu\nu} +(F_R)_{\mu\nu} (F_R)^{\mu\nu}\right\}.
\end{equation}

The bulk theory can be reformulated as $U(1)_V\times U(1)_A$ via the combination
\begin{equation}\label{trans}
A_L=\frac{eV+e^\prime A}{\sqrt{2}},~~~A_R=\frac{eV-e^\prime A}{\sqrt{2}},
\end{equation}
where the gauge coupling $e$ ($e^\prime$) is associated with the vector (axial) field $V_M$ ($A_M$). In terms of $V$ and $A$ fields, the Lagrangian density $\mathcal{L}_1$ becomes
\begin{equation}\label{L1VA}
\begin{split}
\mathcal{L}_1=&-\frac{1}{4}e^2 (F^V)_{MN} (F^V)^{MN}-\frac{1}{4} e^{\prime 2} (F^a)_{MN} (F^a)^{MN}+ \frac{\kappa_1e^{\prime}\, \epsilon^{MNPQR}}{4\sqrt{2}\sqrt{-g}}\\
&\times \left\{ 2e^2  V_M (F^a)_{NP} (F^V)_{QR}+ e^2 A_M (F^V)_{NP} (F^V)_{QR}+ e^{\prime 2} A_M (F^a)_{NP} (F^a)_{QR}  \right\},
\end{split}
\end{equation}
which can equivalently be written as
\begin{equation}\label{L2VA}
\begin{split}
\mathcal{L}_1=&-2\kappa e^2e^\prime\, \underline{\nabla_M \widetilde{V}^M} -\frac{1}{4} e^2 (F^V)_{MN} (F^V)^{MN}-\frac{1}{4}e^{\prime 2} (F^a)_{MN}(F^a)^{MN}\\
&+\frac{\kappa\,\epsilon^{MNPQR}}{2\sqrt{-g}}\left[3e^2e^\prime A_M (F^V)_{NP} (F^V)_{QR} + e^{\prime 3} A_M (F^a)_{NP}(F^a)_{QR}\right],
\end{split}
\end{equation}
where $F^{V,a}$ are field strengths of $V,A$, respectively. $\kappa$ and $\widetilde{V}^M$ are defined as
\begin{equation}
\begin{split}
\kappa&=\frac{\kappa_1}{2\sqrt{2}},\\
\widetilde{V}^M&=\frac{\epsilon^{MNPQR}}{\sqrt{-g}} V_NA_P (F^V)_{QR}.
\end{split}
\end{equation}
While the underlined total derivative term in~(\ref{L2VA}) does not affect the equations of motion, it results in non-conservation of the vector current (dual to $V_M$). Following \cite{0908.4189} we are to cancel this total derivative by adding the Bardeen counter-term so that the vector current becomes conserved, as in real electromagnetic theory. In terms of $V$ and $A$ fields, the counter-term action (\ref{ct LR}) is
\begin{equation}\label{ct VA}
S_{\textrm{c.t.}}=\frac{1}{4}\log r \int d^4x \sqrt{-\gamma}\left[e^2 (F^V)_{\mu\nu} (F^V)^{\mu\nu} +e^{\prime 2} (F^a)_{\mu\nu}(F^a)^{\mu\nu}\right].
\end{equation}

From now on we will work with a new action $S$
\begin{equation}
S=\int d^5x \sqrt{-g}\mathcal{L}+S_{\textrm{c.t.}},
\end{equation}
where
\begin{equation}\label{LPVA}
\begin{split}
\mathcal{L}=&-\frac{1}{4} e^2 (F^V)_{MN} (F^V)^{MN}-\frac{1}{4}e^{\prime 2} (F^a)_{MN} (F^a)^{MN}\\
&+\frac{\kappa\,\epsilon^{MNPQR}}{2\sqrt{-g}}\left[3e^2e^\prime A_M (F^V)_{NP} (F^V)_{QR} + e^{\prime 3} A_M (F^a)_{NP}(F^a)_{QR}\right].
\end{split}
\end{equation}
Equations of motion for $V$ and $A$ fields are derived via standard variational procedure. Under the variation
\begin{equation}\label{var VA}
V\to V+\delta V,~~~A\to A+\delta A,
\end{equation}
from~(\ref{LPVA}) we have
\begin{equation}\label{delLP VA}
\begin{split}
\delta \mathcal{L}&=\delta V_M\left\{e^2\nabla_N(F^V)^{NM}+\frac{3\kappa e^2e^\prime \epsilon^{MNPQR}}{\sqrt{-g}}(F^a)_{NP} (F^V)_{QR}\right\}+\delta A_M\\
&\times\left\{e^{\prime 2}\nabla_N(F^a)^{NM} +\frac{3\kappa e^\prime \epsilon^{MNPQR}} {2\sqrt{-g}}\left[e^2 (F^V)_{NP} (F^V)_{QR}+ e^{\prime 2} (F^a)_{NP} (F^a)_{QR} \right] \right\}\\
&-e^2\nabla_M\left[\delta V_N (F^V)^{MN}+\frac{6\kappa e^\prime\epsilon^{MNPQR}} {\sqrt{-g}}A_N\delta V_P (F^V)_{QR}\right]\\
&-e^{\prime 2}\nabla_M \left[\delta A_N (F^a)^{MN}+\frac{2\kappa e^\prime \epsilon^{MNPQR}} {\sqrt{-g}}A_N\delta A_P (F^a)_{QR}\right],
\end{split}
\end{equation}
and the variation of $S_{\textrm{c.t.}}$ (cf.~(\ref{ct VA}))
\begin{equation}
\delta S_{\textrm{c.t.}}=-\int d^4x\sqrt{-\gamma} \left\{e^2\delta V_\mu \widetilde{\nabla}_\nu (F^V)^{\nu\mu}+e^{\prime 2}\delta A_\mu \widetilde{\nabla}_\nu (F^a)^{\nu\mu} \right\}\log r,
\end{equation}
where $\widetilde{\nabla}_\mu$ is covariant derivative compatible with the induced metric $\gamma_{\mu\nu}$.
Then, equations of motion for $V$ and $A$ fields are
\begin{equation}\label{eom VAmu}
\textrm{dynamical~~equations}:~~~\textrm{EV}^\mu=\textrm{EA}^\mu=0,
\end{equation}
\begin{equation}\label{eom VAr}
\textrm{constraint~~equations}:~~~\textrm{EV}^r=\textrm{EA}^r=0,
\end{equation}
where
\begin{equation}\label{EV}
\textrm{EV}^M\equiv e^2\nabla_N(F^V)^{NM}+\frac{3\kappa e^2e^\prime \epsilon^{MNPQR}} {\sqrt{-g}} (F^a)_{NP} (F^V)_{QR},
\end{equation}
\begin{equation}\label{EA}
\textrm{EA}^M\equiv e^{\prime 2}\nabla_N(F^a)^{NM} +\frac{3\kappa e^\prime\epsilon^{MNPQR}} {2\sqrt{-g}} \left[e^2 (F^V)_{NP} (F^V)_{QR}+ e^{\prime 2} (F^a)_{NP} (F^a)_{QR}\right].
\end{equation}
Imposing the dynamical equations (\ref{eom VAmu}), the action variation $\delta S$ reduces to
\begin{equation} \label{action variation}
\begin{split}
\delta S&=\int d^4x \sqrt{-\gamma}\, n_{_M}\left\{-e^2 \delta V_N (F^V)^{MN} -\frac{6\kappa e^2 e^\prime \epsilon^{MNPQR}}{\sqrt{-g}} A_N \delta V_P (F^V)_{QR} \right.\\
&~~~~~~~~~~~~~~~~~~~~~~~~~~\left.-e^{\prime 2} \delta A_N (F^a)^{MN}- \frac{2\kappa e^{\prime 3} \epsilon^{MNPQR}}{\sqrt{-g}} A_N \delta A_P (F^a)_{QR}\right\}\\
&-\int d^4x\sqrt{-\gamma} \left\{e^2\delta V_\mu \widetilde{\nabla}_\nu (F^V)^{\nu\mu}+e^{\prime 2}\delta A_\mu \widetilde{\nabla}_\nu (F^a)^{\nu\mu} \right\}\log r\\
&+\int d^5x \sqrt{-g}\left(\delta V_r\, \textrm{EV}^r+\delta A_r\, \textrm{EA}^r\right),
\end{split}
\end{equation}
where $n_{_M}$ is the outpointing unit normal vector of the slice $\Sigma$. The last line of~(\ref{action variation})  vanishes either using the constraint equations~(\ref{eom VAr}) or through a radial gauge choice $\delta V_r=\delta A_r=0$. The boundary currents are defined as
\begin{equation} \label{current definition}
J^\mu\equiv \lim_{r\to\infty}\frac{\delta S}{\delta V_\mu},~~~~~~~~~~~~~
J^\mu_5\equiv \lim_{r\to\infty}\frac{\delta S}{\delta A_\mu}.
\end{equation}
In terms of the bulk fields, the boundary currents are
\begin{align}\label{j bct}
J^\mu&=\lim_{r\to\infty}\sqrt{-\gamma}\,e^2 \left\{(F^V)^{\mu M}n_{_M}+ \frac{6\kappa e^\prime \epsilon^{M\mu NQR}}{\sqrt{-g}}n_{_M} A_N (F^V)_{QR}- \widetilde{\nabla}_\nu (F^V)^{\nu\mu}\log r \right\},\nonumber\\
J_5^\mu&=\lim_{r\to\infty}\sqrt{-\gamma}\,e^{\prime 2} \left\{(F^a)^{\mu M}n_{_M}+ \frac{2\kappa e^\prime \epsilon^{M\mu NQR}}{\sqrt{-g}}n_{_M} A_N (F^a)_{QR}- \widetilde{\nabla}_\nu (F^a)^{\nu\mu}\log r \right\}.
\end{align}

It is important to stress that the currents in~(\ref{current definition}) are defined independently of the  constraint equations~(\ref{eom VAr}). Throughout this work, the radial gauge $V_r=A_r=0$ will be assumed. Thus, in order to completely determine the boundary currents~(\ref{j bct}) it is sufficient
to solve the dynamical equations~(\ref{eom VAmu}) for the bulk gauge fields $V_\mu,A_\mu$ only, leaving the constraints aside.
The constraint equations~(\ref{eom VAr}) give rise to the continuity equations~(\ref{continuity}). In this way, the currents' constitutive relations to be derived below are off-shell. In subsequent presentation, the couplings $e$ and $e^\prime$ will be absorbed into redefinition of $V$ and $A$ fields,
while the notations for $V$ and $A$ will remain unchanged for convenience.

It is useful to reexpress the currents~(\ref{j bct}) in terms of coefficients of
near boundary asymptotic expansion of the bulk gauge fields. Near $r=\infty$,
\begin{equation}\label{asmp cov1}
V_\mu=\mathcal{V}_\mu + \frac{V_\mu^{(1)}}{r}+ \frac{V_\mu^{(2)}}{r^2}- \frac{2V_\mu^{\textrm{L}}}{r^2} \log r+\mathcal{O}\left(\frac{\log r}{r^3}\right),
\end{equation}
\begin{equation}\label{asmp cov2}
A_\mu=\mathcal{A}_\mu + \frac{A_\mu^{(1)}}{r}+ \frac{A_\mu^{(2)}}{r^2}- \frac{2A_\mu^{\textrm{L}}}{r^2} \log r+\mathcal{O}\left(\frac{\log r}{r^3}\right),
\end{equation}
where
\begin{equation}\label{asmp cov3}
V_\mu^{(1)}=\mathcal{F}_{t\mu}^V,~~~4V_\mu^{\textrm{L}}=\partial^\nu \mathcal{F}_{\mu\nu}^V,
\end{equation}
\begin{equation}\label{asmp cov4}
A_\mu^{(1)}=\mathcal{F}_{t\mu}^a,~~~4A_\mu^{\textrm{L}}=\partial^\nu \mathcal{F}_{\mu\nu}^a.
\end{equation}
The holographic dictionary implies that $\mathcal{V}_\mu,\mathcal{A}_\mu$ are gauge potentials of the
external fields $\vec {E}$, $\vec{B}$, $\vec{E}^a$ and $\vec{B}^a$,
\begin{equation}
\begin{split}
&E_i=\mathcal{F}_{it}^V=\partial_i\mathcal{V}_t-\partial_t \mathcal{V}_i,~~~~ B_i=\frac{1}{2}\epsilon_{ijk}\mathcal{F}_{jk}^V=\epsilon_{ijk}\partial_{j}\mathcal{V}_k,\\
&E_i^a=\mathcal{F}_{it}^a=\partial_i\mathcal{A}_t-\partial_t \mathcal{A}_i,~~~ B_i^a=\frac{1}{2}\epsilon_{ijk}\mathcal{F}_{jk}^a=\epsilon_{ijk}\partial_{j}\mathcal{A}_k.
\end{split}
\end{equation}
As mentioned above, only the dynamical equations~(\ref{eom VAmu}) were utilized in order to get (\ref{asmp cov1}-\ref{asmp cov4}).
The near-boundary data $V_\mu^{(2)}$ and $A_\mu^{(2)}$ have to be determined via integrating of the dynamical equations (\ref{eom VAmu}) from the horizon to the boundary. The currents (\ref{j bct}) become
\begin{equation}\label{bdry currents}
\begin{split}
J^{\mu}	&=\eta^{\mu\nu}(2V_{\nu}^{(2)}+2V^{\textrm{L}}_{\nu}+\eta^{\sigma t} \partial_{\sigma} \mathcal{F}_{t\nu}^V) +6\kappa \epsilon^{\mu\nu\rho\lambda} \mathcal{A}_{\nu}\mathcal{F}_{\rho\lambda}^V,\\
J_{5}^{\mu}	&=\eta^{\mu\nu}(2A_{\nu}^{(2)}+2A_{\nu}^{\textrm{L}}+\eta^{\sigma t} \partial_{\sigma}\mathcal{F}_{t\nu}^{a})+2\kappa\epsilon^{\mu\nu\rho\lambda} \mathcal{A}_{\nu}\mathcal{F}_{\rho\lambda}^{a}.
\end{split}
\end{equation}

Note  explicit dependence of the currents $J^\mu$ and $J^\mu_5$ on the axial gauge potential $\mathcal{A}_\mu$. The last term in $J^\mu$ of~(\ref{bdry currents}) is crucial in guaranteeing conservation of $J^\mu$, that is the gauge invariance  under the vector gauge transformation $\mathcal{V}_\mu \to \mathcal{V}_\mu +\partial_\mu \phi$. Clearly, explicit dependence of physical quantities on the axial potential $\mathcal{A}_\mu$ is because that the transformation $\mathcal{A}_\mu\to \mathcal{A}_\mu+\partial_\mu\varphi$ is not a symmetry,

In  presence of anomaly one distinguishes between consistent current and covariant current \cite{Bardeen:1984pm}. Consistent current is defined as a functional derivative of effective action with respect to external gauge field. Covariant current is obtained by subtracting a suitably chosen Chern-Simons current from the consistent one, so that the current becomes invariant under both vector and axial gauge transformations.
%Note the consistent current is the one coupling to external gauge potential.
The currents defined in~(\ref{current definition},\ref{bdry currents}) are consistent. The associated covariant currents are
\begin{equation}
J^\mu_{\textrm{cov}}=J^\mu-6\kappa \epsilon^{\mu\nu\rho\lambda} \mathcal{A}_{\nu} \mathcal{F}_{\rho\lambda}^V,~~~~~~~~~~~~
J^\mu_{5\textrm{cov}}=J_5^\mu-2\kappa\epsilon^{\mu\nu\rho\lambda} \mathcal{A}_{\nu}\mathcal{F}_{\rho\lambda}^{a}.
\end{equation}
Obviously, when the axial field $\mathcal{A}_\mu=0$, both consistent and covariant currents coincide.

\section{Anomalous Maxwell equations in the bulk}\label{section3}

To derive constitutive relations for the currents $J^\mu$ and $J_5^\mu$, we consider  finite vector/axial charge densities exposed to external vector and axial electromagnetic fields. Holographically, the charge densities and external fields are encoded in asymptotic behaviors of the bulk gauge fields. In the bulk, we will solve the dynamical equations (\ref{eom VAmu}) assuming some charge densities and external fields, but without specifying them explicitly. In this section we outline the strategy for deriving of currents' constitutive relations.

Following~\cite{1511.08789}, we start with the most general static and homogeneous profiles for the bulk gauge fields
which solve the dynamical equations~(\ref{eom VAmu}),
\begin{equation}\label{homogeneous solution}
V_\mu=\mathcal{V}_\mu-\frac{\rho}{2r^2}\delta_{\mu t}, ~~~~A_\mu=\mathcal{A}_\mu-\frac{\rho_{_5}}{2r^2}\delta_{\mu t},
\end{equation}
where $\mathcal{V}_\mu,\mathcal{A}_\mu,\rho,\rho_{_5}$ are all constants for the moment. Regularity requirement at $r=1$ fixes one integration constant for each $V_i$ and $A_i$. Through~(\ref{bdry currents}), the boundary currents are
\begin{equation}
J^t=\rho,~~~J^i=0;~~~~~~~J_5^t=\rho_{_5},~~~J_5^i=0.
\end{equation}
Hence, $\rho$ and $\rho_{_5}$ are identified as the vector/axial charge densities.

Next, following the idea of fluid/gravity correspondence~\cite{0712.2456}, we promote $\mathcal{V}_\mu,\mathcal{A}_\mu,\rho,\rho_{_5}$ into arbitrary functions of the boundary coordinates
\begin{equation}
\begin{split}
\mathcal{V}_\mu\to \mathcal{V}_\mu(x_\alpha),~~\rho \to \rho(x_\alpha);~~~~~~~~~~
\mathcal{A}_\mu \to \mathcal{A}_\mu(x_\alpha),~~\rho_{_5}\to \rho_{_5}(x_\alpha).
\end{split}
\end{equation}
(\ref{homogeneous solution}) ceases to be a solution of the dynamical equations (\ref{eom VAmu}).
To have them  satisfied, suitable corrections  in $V_\mu$ and $A_\mu$ have to be introduced:
\begin{equation} \label{corrections}
\begin{split}
V_\mu(r,x_\alpha)&=\mathcal{V}_\mu(x_\alpha)-\frac{\rho(x_\alpha)}{2r^2}\delta_{\mu t}+ \mathbb{V}_\mu(r,x_\alpha),\\
A_\mu(r,x_\alpha)&=\mathcal{A}_\mu(x_\alpha)-\frac{\rho_{_5}(x_\alpha)}{2r^2}\delta_{\mu t} + \mathbb{A}_\mu(r,x_\alpha),
\end{split}
\end{equation}
where $\mathbb{V}_\mu,\mathbb{A}_\mu$ will be determined from solving (\ref{eom VAmu}). Appropriate boundary conditions have to be specified.
First, $\mathbb{V}_\mu$ and $\mathbb{A}_\mu$ have to be regular over the whole integration interval of $r$, from one to infinity.
Second, at the conformal boundary $r=\infty$, we require
\begin{equation}\label{AdS constraint}
\mathbb{V}_\mu\to 0,~~~~~~\mathbb{A}_\mu \to 0~~~~~~~\textrm{as}~~~~~~r\to \infty,
\end{equation}
which amounts to fixing external gauge potentials to be $\mathcal{V}_\mu$ and $\mathcal{A}_\mu$. Additional integration constants will be fixed by a frame choice. In this work we adopt the Landau frame convention for covariant currents,
\begin{equation}\label{Landau frame}
J^t_{\textrm{cov}}=\rho(x_\alpha),~~~~~~~~~~~J^t_{5\textrm{cov}}=\rho_{_5}(x_\alpha).
\end{equation}
The Landau frame choice can be identified as  a residual gauge fixing for the bulk fields. Most of our results, however, would be independent of this choice.  Appendix~\ref{appendix5} is entirely devoted to this discussion.

The vector/axial chemical potentials are defined as
\begin{equation} \label{def potentials}
\begin{split}
&\mu=V_t(r=\infty)-V_t(r=1)=\frac{1}{2}\rho-\mathbb{V}_t(r=1),\\
&\mu_{_5}=A_t(r=\infty)-A_t(r=1)=\frac{1}{2}\rho_{_5}-\mathbb{A}_t(r=1).
\end{split}
\end{equation}
For the homogeneous case, the definition~(\ref{def potentials}) results in $\mu=\rho/2, ~ \mu_{_5}=\rho_{_5}/2$. Beyond the homogeneous case, $\mu,\mu_{_5}$ are nonlinear functions of densities and external fields.

For generic configurations of external fields and charge densities, (\ref{eom VAmu},\ref{eom VAr}) become rather involved. In terms of $\mathbb{V}_\mu$ and $\mathbb{A}_\mu$, the dynamical equations~(\ref{eom VAmu}) are
\begin{equation}\label{eom Vt}
\begin{split}
0&=r^3\partial_r^2 \mathbb{V}_t+3r^2 \partial_r \mathbb{V}_t+r\partial_r \partial_k \mathbb{V}_k+ 12\kappa \epsilon^{ijk} \partial_r \mathbb{A}_i\left( \partial_j \mathcal{V}_k + \partial_j \mathbb{V}_k\right)\\
&+12\kappa \epsilon^{ijk}\partial_r\mathbb{V}_i \left(\partial_j \mathcal{A}_k +\partial_j \mathbb{A}_k\right),
\end{split}
\end{equation}
\begin{equation}\label{eom Vi}
\begin{split}
0&=(r^5-r)\partial_r^2 \mathbb{V}_i+(3r^4+1)\partial_r \mathbb{V}_i+2r^3\partial_r \partial_t \mathbb{V}_i-r^3 \partial_r\partial_i \mathbb{V}_t+ r^2\left(\partial_t \mathbb{V}_i- \partial_i \mathbb{V}_t\right)\\
&+r(\partial^2 \mathbb{V}_i - \partial_i\partial_k\mathbb{V}_k)-\frac{1}{2}\partial_i \rho +r^2\left(\partial_t\mathcal{V}_i-\partial_i\mathcal{V}_t\right)+ r\left(\partial^2 \mathcal{V}_i- \partial_i \partial_k \mathcal{V}_k\right)\\
&+12\kappa r^2\epsilon^{ijk}\left(\frac{1}{r^3}\rho_{_5}\partial_j \mathcal{V}_k + \frac{1}{r^3}\rho_{_5}\partial_j\mathbb{V}_k +\partial_r \mathbb{A}_t \partial_j \mathcal{V}_k+\partial_r \mathbb{A}_t \partial_j \mathbb{V}_k\right)\\
&-12\kappa r^2\epsilon^{ijk} \partial_r\mathbb{A}_j\left[\left(\partial_t \mathcal{V}_k- \partial_k \mathcal{V}_t\right)+\left(\partial_t \mathbb{V}_k- \partial_k \mathbb{V}_t\right)+\frac{1}{2r^2}\partial_k \rho\right]\\
&-12\kappa r^2\epsilon^{ijk} \partial_r\mathbb{V}_j \left[\left(\partial_t \mathcal{A}_k- \partial_k \mathcal{A}_t\right)+\left(\partial_t \mathbb{A}_k- \partial_k \mathbb{A}_t\right)+\frac{1}{2r^2}\partial_k \rho_{_5}\right]\\
&+12\kappa r^2 \epsilon^{ijk}\left(\frac{1}{r^3}\rho \partial_j \mathcal{A}_k +\frac{1}{r^3}\rho \partial_j \mathbb{A}_k+\partial_r \mathbb{V}_t \partial_j \mathcal{A}_k +\partial_r \mathbb{V}_t \partial_j \mathbb{A}_k\right),
\end{split}
\end{equation}
\begin{equation}\label{eom At}
\begin{split}
0&=r^3\partial_r^2 \mathbb{A}_t+ 3r^2 \partial_r \mathbb{A}_t+r\partial_r \partial_k \mathbb{A}_k+ 12\kappa \epsilon^{ijk} \partial_r \mathbb{V}_i\left( \partial_j \mathcal{V}_k+ \partial_j \mathbb{V}_k\right)\\
&+12\kappa \epsilon^{ijk}\partial_r\mathbb{A}_i \left(\partial_j \mathcal{A}_k +\partial_j \mathbb{A}_k\right),
\end{split}
\end{equation}
\begin{equation}\label{eom Ai}
\begin{split}
0&=(r^5-r)\partial_r^2 \mathbb{A}_i+(3r^4+1)\partial_r \mathbb{A}_i+2r^3\partial_r \partial_t \mathbb{A}_i-r^3 \partial_r\partial_i \mathbb{A}_t+ r^2\left(\partial_t \mathbb{A}_i- \partial_i \mathbb{A}_t\right)\\
&+r(\partial^2 \mathbb{A}_i - \partial_i\partial_k\mathbb{A}_k)-\frac{1}{2}\partial_i \rho_{_5} +r^2\left(\partial_t\mathcal{A}_i-\partial_i\mathcal{A}_t\right)+ r\left(\partial^2 \mathcal{A}_i- \partial_i \partial_k \mathcal{A}_k\right)\\
&+12\kappa r^2\epsilon^{ijk}\left(\frac{1}{r^3}\rho\partial_j \mathcal{V}_k +\frac{1}{r^3}\rho\partial_j\mathbb{V}_k +\partial_r \mathbb{V}_t \partial_j \mathcal{V}_k+\partial_r \mathbb{V}_t \partial_j \mathbb{V}_k\right)\\
&-12\kappa r^2\epsilon^{ijk} \partial_r\mathbb{V}_j\left[\left(\partial_t \mathcal{V}_k- \partial_k \mathcal{V}_t\right)+\left(\partial_t \mathbb{V}_k- \partial_k \mathbb{V}_t\right)+\frac{1}{2r^2}\partial_k \rho\right]\\
&-12\kappa r^2\epsilon^{ijk} \partial_r\mathbb{A}_j \left[\left(\partial_t \mathcal{A}_k- \partial_k \mathcal{A}_t\right)+\left(\partial_t \mathbb{A}_k- \partial_k \mathbb{A}_t\right)+\frac{1}{2r^2}\partial_k \rho_{_5}\right]\\
&+12\kappa r^2 \epsilon^{ijk}\left(\frac{1}{r^3}\rho_{_5} \partial_j \mathcal{A}_k +\frac{1}{r^3}\rho_{_5} \partial_j \mathbb{A}_k+\partial_r \mathbb{A}_t \partial_j \mathcal{A}_k +\partial_r \mathbb{A}_t \partial_j \mathbb{A}_k\right).
\end{split}
\end{equation}

Triangle anomaly is a source of nonlinearity in all these equations. In the context of fluid/gravity correspondence~\cite{0712.2456}, external fields $\mathcal{V}_\mu,\mathcal{A}_\mu$ and charge densities $\rho,\rho_{_5}$ are assumed to vary  slowly from point to point. Consequently,
the corrections $\mathbb{V}_\mu$ and $\mathbb{A}_\mu$ can be constructed through order by order expansion in  derivatives of the external fields and charge densities. Contrary to the approach adopted below,  the method of \cite{0712.2456} is implemented using ``on-shell'' relations. That is, the
bulk solutions are constructed with the help of the constraint equations.
%have to be imposed in order to construct solutions for bulk fields.

To extract the TCFs  to all order in derivative expansion, we do linearisation in external fields and charge densities.
%Particularly, solving dynamical equations~(\ref{eom Vt}-\ref{eom Ai}) suffice to determine $\mathbb{V}_\mu, \mathbb{A}_\mu$ and thus fully fixes all the transports. Leaving constraint equations~(\ref{eom VAr}) unsolved, external fields and charge densities are kept as arbitrary.
As announced in section~\ref{section1}, we will solve~(\ref{eom Vt}-\ref{eom Ai}) under two different linearization schemes~(\ref{study1}) and (\ref{study2}).
%, which separates further presentations into two independent parts.

\section{Study I: linear transport} \label{section4}
In this section we study linear TCFs corresponding to the linearisation scheme (\ref{study1}),
\begin{equation}\label{linear shceme1}
\rho(x_\alpha)=\bar{\rho}+\epsilon \delta\rho(x_\alpha),~~~~~\mathcal{V}_\mu \to \epsilon \mathcal{V}_\mu;~~~~~
\rho_{_5}(x_\alpha)=\bar{\rho}_{_5}+ \epsilon \delta\rho_{_5}(x_\alpha), ~~~~~\mathcal{A}_\mu \to \epsilon \mathcal{A}_\mu,
\end{equation}
where $\bar{\rho},\bar{\rho}_{_5}$ are constants.
%Recall that $\epsilon$ is to denote the linearization:
All calculations below are accurate  to linear order in $\epsilon$. Obviously, $\mathbb{V}_\mu$ and $\mathbb{A}_\mu$ scale as $\epsilon$ too,
\begin{equation}
\mathbb{V}_\mu\to\epsilon \mathbb{V}_\mu,~~~~~~~~\mathbb{A}_\mu\to\epsilon \mathbb{A}_\mu.
\end{equation}
The presentation is split into two subsections: one is devoted to derivation of  the constitutive relations~(\ref{jmu},\ref{jmu5}) while the other
one focuses on determination of  transport coefficients.

\subsection{Derivation of constitutive relations from the dynamical equations} \label{subsection41}

Under the scheme~(\ref{linear shceme1}), the dynamical equations~(\ref{eom Vt}-\ref{eom Ai}) are
\begin{equation}\label{eom Vt scheme1}
0=r^2\partial_r^2 \mathbb{V}_t+3r \partial_r \mathbb{V}_t+\partial_r \partial_k \mathbb{V}_k,
\end{equation}
\begin{equation}\label{eom Vi scheme1}
\begin{split}
0&=(r^5-r)\partial_r^2 \mathbb{V}_i+(3r^4+1)\partial_r \mathbb{V}_i+2r^3\partial_r \partial_t \mathbb{V}_i-r^3 \partial_r\partial_i \mathbb{V}_t+ r^2\left(\partial_t \mathbb{V}_i- \partial_i \mathbb{V}_t\right)\\
&+r(\partial^2 \mathbb{V}_i - \partial_i\partial_k\mathbb{V}_k)-\frac{1}{2}\partial_i \delta\rho +r^2\left(\partial_t\mathcal{V}_i-\partial_i\mathcal{V}_t\right)+ r\left(\partial^2 \mathcal{V}_i- \partial_i \partial_k \mathcal{V}_k\right)\\
&+\frac{12\kappa}{r}\epsilon^{ijk} \left[\bar{\rho}_{_5}\left(\partial_j \mathcal{V}_k +\partial_j\mathbb{V}_k \right)+\bar{\rho} \left(\partial_j \mathcal{A}_k+ \partial_j\mathbb{A}_k\right)\right],
\end{split}
\end{equation}
\begin{equation}\label{eom At scheme1}
0=r^2\partial_r^2 \mathbb{A}_t+ 3r \partial_r \mathbb{A}_t+\partial_r \partial_k \mathbb{A}_k,
\end{equation}
\begin{equation}\label{eom Ai scheme1}
\begin{split}
0&=(r^5-r)\partial_r^2 \mathbb{A}_i+(3r^4+1)\partial_r \mathbb{A}_i+2r^3\partial_r \partial_t \mathbb{A}_i-r^3 \partial_r\partial_i \mathbb{A}_t+ r^2\left(\partial_t \mathbb{A}_i- \partial_i \mathbb{A}_t\right)\\
&+r(\partial^2 \mathbb{A}_i - \partial_i\partial_k\mathbb{A}_k)-\frac{1}{2}\partial_i \delta \rho_{_5} +r^2\left(\partial_t\mathcal{A}_i-\partial_i\mathcal{A}_t\right)+ r\left(\partial^2 \mathcal{A}_i- \partial_i \partial_k \mathcal{A}_k\right)\\
&+\frac{12\kappa}{r}\epsilon^{ijk} \left[\bar{\rho}\left(\partial_j \mathcal{V}_k +\partial_j\mathbb{V}_k \right)+\bar{\rho}_{_5} \left(\partial_j \mathcal{A}_k+ \partial_j\mathbb{A}_k\right)\right].
\end{split}
\end{equation}
At linear level, $\mathbb{V}_\mu$ and $\mathbb{A}_\mu$ are still coupled together through the anomaly-induced terms.

To order $\mathcal{O}(\epsilon)$, the constraint equations~(\ref{eom VAr}) are
\begin{equation}
\begin{split}
0&=r^3\partial_r\partial_t \mathbb{V}_t+r\left(\partial^2 \mathbb{V}_t- \partial_t \partial_k \mathbb{V}_k \right)- r^3f(r) \partial_r \partial_k \mathbb{V}_k +\partial_t \delta\rho-\frac{1}{2r}\partial^2 \delta\rho\\
&+r\left(\partial^2 \mathcal{V}_t- \partial_t\partial_k\mathcal{V}_k \right),
\end{split}
\end{equation}
\begin{equation}
\begin{split}
0&=r^3\partial_r\partial_t \mathbb{A}_t+r\left(\partial^2 \mathbb{A}_t -\partial_t \partial_k\mathbb{A}_k \right)- r^3f(r) \partial_r \partial_k \mathbb{A}_k +\partial_t \delta\rho_{_5}-\frac{1}{2r}\partial^2 \delta \rho_{_5}\\
&+r\left(\partial^2 \mathcal{A}_t - \partial_t\partial_k\mathcal{A}_k \right),
\end{split}
\end{equation}
which do not feel the effect of triangle anomaly at this order in $\epsilon$.

The corrections $\mathbb{V}_\mu$ and $\mathbb{A}_\mu$ are decomposed as
\begin{equation}\label{decomp V}
\begin{split}
\mathbb{V}_t&=S_1 \mathcal{V}_t+S_2\partial_k\mathcal{V}_k+S_3\delta\rho+S_4 \mathcal{A}_t +S_5\partial_k\mathcal{A}_k+S_6 \delta\rho_{_5},\\
\mathbb{V}_i&=V_1\mathcal{V}_i+V_2\partial_i\mathcal{V}_t+V_3 \partial_i\partial_k \mathcal{V}_k+V_4\partial_i \delta\rho+V_5 \epsilon^{ijk}\partial_j \mathcal{V}_k\\
&+V_6\mathcal{A}_i+V_7\partial_i \mathcal{A}_t+V_8\partial_i \partial_k \mathcal{A}_k +V_9 \partial_i\delta\rho_{_5}+V_{10} \epsilon^{ijk}\partial_j \mathcal{A}_k,
\end{split}
\end{equation}
\begin{equation}\label{decomp A}
\begin{split}
\mathbb{A}_t&=\bar{S}_1 \mathcal{V}_t+ \bar{S}_2\partial_k\mathcal{V}_k+\bar{S}_3\delta \rho+ \bar{S}_4 \mathcal{A}_t +\bar{S}_5\partial_k\mathcal{A}_k+\bar{S}_6 \delta \rho_{_5},\\
\mathbb{A}_i&=\bar{V}_1\mathcal{V}_i+\bar{V}_2\partial_i\mathcal{V}_t+\bar{V}_3 \partial_i\partial_k \mathcal{V}_k+\bar{V}_4\partial_i \delta\rho+\bar{V}_5 \epsilon^{ijk} \partial_j \mathcal{V}_k\\
&+\bar{V}_6\mathcal{A}_i+\bar{V}_7\partial_i \mathcal{A}_t+\bar{V}_8\partial_i \partial_k \mathcal{A}_k +\bar{V}_9 \partial_i\delta\rho_{_5}+\bar{V}_{10} \epsilon^{ijk}\partial_j \mathcal{A}_k,
\end{split}
\end{equation}
where $S_i,V_i,\bar{S}_i,\bar{V}_i$ are elements of the inverse Green function matrix. They are
scalar functionals of the boundary derivative operators and functions of radial coordinate $r$. In  momentum space, the derivative operators turn into scalar functions of frequency $\omega$ and momentum squared $q^2$:
$$
S_i\left(r,\partial_t,\vec{\partial}^2\right)\to S_i(r,\omega,q^2),~~~~~~~~ \bar{S}_i\left(r,\partial_t,\vec{\partial}^2\right)\to \bar{S}_i(r,\omega,q^2)$$
$$
V_i\left(r,\partial_t,\vec{\partial}^2\right)\to S_i(r,\omega,q^2),~~~~~~~~ \bar{V}_i\left(r,\partial_t,\vec{\partial}^2\right)\to \bar{V}_i(r,\omega,q^2),
$$
which satisfy partially decoupled ODEs~(\ref{eom 1}-\ref{eom 22}). Dynamics of the bulk theory is now reflected by these ODEs.
Accordingly, the boundary conditions for the decomposition coefficients in~(\ref{decomp V},\ref{decomp A}) are
\begin{equation} \label{asymp cond}
S_i\to 0,~~~~~~~~~\bar{S}_i\to 0,~~~~~~~~~~V_i\to 0,~~~~~~~~~~\bar{V}_i\to 0~~~~~~~~~~~~\textrm{as}~r\to 0.
\end{equation}
\begin{equation} \label{regular cond}
S_i,~\bar{S}_i,~V_i,~\bar{V}_i~~\textrm{are regular over the whole interval of } ~ r \in[1,\infty].
\end{equation}
Additional  integration constants will be fixed by the frame convention~(\ref{Landau frame}).

Pre-asymptotic expansions (\ref{asmp cov1}-\ref{asmp cov4}) translate into pre-asymptotic behaviour of the
decomposition coefficients in~(\ref{decomp V},\ref{decomp A}). Near $r=\infty$,
\begin{equation}
\begin{split}
&S_i\to \frac{s_{_i}^1}{r}+\frac{s_{_i}}{r^2}+ \frac{s_{_i}^\textrm{L}\log r}{r^2}+\cdots,~~~~~~~~~~~
V_i\to\frac{v_{_i}^1}{r}+\frac{v_{_i}}{r^2}+ \frac{v_{_i}^\textrm{L}\log r}{r^2}+\cdots,\\
&\bar{S}_i\to \frac{\bar{s}_{_i}^1}{r}+ \frac{\bar{s}_{_i}}{r^2} + \frac{\bar{s}_{_i}^\textrm{L}\log r}{r^2}+\cdots,~~~~~~~~~~~
\bar{V}_i\to \frac{\bar{v}_{_i}^1}{r}+ \frac{\bar{v}_{_i}}{r^2}+ \frac{\bar{v}_{_i}^\textrm{L}\log r}{r^2}+\cdots,
\end{split}
\end{equation}
where $s_{_i}^{1,\textrm{L}}$, $v_{_i}^{1,\textrm{L}}$, $\bar{s}_{_i}^{1,\textrm{L}}$, $\bar{v}_{_i}^{1,\textrm{L}}$ are uniquely fixed in near-boundary analysis. $s_{_i}$, $v_{_i}$, $\bar{s}_{_i}$ and $\bar{v}_{_i}$ will be determined once the ODEs (\ref{eom 1}-\ref{eom 22}) are solved. The boundary currents~(\ref{bdry currents}) are
\begin{align} \label{jmu/jmu5}
J^t=&\rho-\left(2s_{_1}+\frac{1}{2}q^2\right)\mathcal{V}_t-\left(2s_{_2}-\frac{1}{2}i \omega\right)\partial_k \mathcal{V}_k-2s_{_3} \delta\rho - 2s_{_4} \mathcal{A}_t- 2s_{_5}\partial_k \mathcal{A}_k-2s_{_6}\delta \rho_{_5},\nonumber\\
J^i=&\left[2v_{_1}+\frac{1}{2}\left(\omega^2+q^2\right)\right]\mathcal{V}_i+\left(2v_{_2}- \frac{1}{2}i\omega\right)\partial_i\mathcal{V}_t\,+\left(2v_{_3}+\frac{1}{2}\right) \partial_i \partial_k\mathcal{V}_k+2v_{_4}\partial_i\delta \rho \nonumber\\
&+2v_{_5}\epsilon^{ijk}\partial_j \mathcal{V}_k+2v_{_6}\mathcal{A}_i +2v_{_7} \partial_i\mathcal{A}_t+2v_{_8}\partial_i\partial_k\mathcal{A}_k+2v_{_9} \partial_i\delta\rho_{_5}+2v_{_{10}}\epsilon^{ijk}\partial_j\mathcal{A}_k;\\
J^t_5=&\rho_{_5}-2\bar{s}_{_1}\mathcal{V}_t-2\bar{s}_{_2}\partial_k \mathcal{V}_k- 2\bar{s}_{_3} \delta\rho- \left(2\bar{s}_{_4}+\frac{1}{2}q^2\right)\mathcal{A}_t - \left(2\bar{s}_{_5}-\frac{1}{2}i\omega\right) \partial_k\mathcal{A}_k-2\bar{s}_{_6}\delta \rho_{_5},\nonumber\\
J^i_5=&2\bar{v}_{_1}\mathcal{V}_i+2\bar{v}_{_2}\partial_i\mathcal{V}_t+2\bar{v}_{_3} \partial_i \partial_k\mathcal{V}_k+2\bar{v}_{_4}\partial_i\delta \rho+2\bar{v}_{_5} \epsilon^{ijk}\partial_j\mathcal{V}_k+\left[2\bar{v}_{_6}+\frac{1}{2}\left(\omega^2+q^2 \right)\right]\mathcal{A}_i\nonumber\\
&+\left(2\bar{v}_{_7}-\frac{1}{2}i\omega\right)\partial_i\mathcal{A}_t\,+\left(2 \bar{v}_{_8} +\frac{1}{2}\right)\partial_i \partial_k\mathcal{A}_k +2\bar{v}_{_9} \partial_i\delta\rho_{_5}+2\bar{v}_{_{10}}\epsilon^{ijk}\partial_j\mathcal{A}_k.\nonumber
\end{align}
The frame convention~(\ref{Landau frame}) leads to the following relations
\begin{equation}\label{Lcons1}
s_{_4}=s_{_6}=\bar{s}_{_1}=\bar{s}_{_3}=0.
\end{equation}
\begin{equation}\label{Lcons2}
s_{_1}=\bar{s}_{_4}=-\frac{1}{4}q^2,~~~~~~~~~~~~~s_{_3}=\bar{s}_{_6}=0.
\end{equation}
\begin{equation}\label{Lcons3}
s_{_2}=\bar{s}_{_5}=\frac{1}{4}i\omega,~~~~~~~~~~~~~~\bar{s}_{_2}=s_{_5}=0.
\end{equation}

Combined with the ODEs~(\ref{eom 1}-\ref{eom 22}), (\ref{Lcons1}-\ref{Lcons3}) imply constraints among the decomposition coefficients in~(\ref{decomp V},\ref{decomp A}), see (\ref{constraint1}, \ref{constraint2}, \ref{constraint3}, \ref{constraint4}, \ref{symmt1}, \ref{symmt2}). Via~(\ref{LD cons}, \ref{symmt1}, \ref{symmt2}), the boundary currents~(\ref{jmu/jmu5}) are eventually cast into the constitutive relations~(\ref{jmu},\ref{jmu5}) with the TCFs expressed in terms of $v_{_i},\bar{v}_{_i}$,
\begin{equation}\label{transport def}
\begin{split}
&\mathcal{D}=-2v_{_4}=-2\bar{v}_{_9},~~~~~\sigma_e=2v_{_2}-\frac{1}{2}i\omega=2\bar{v}_{_7}- \frac{1}{2}i\omega,~~~~~\sigma_m=2v_{_3}+\frac{1}{2}=2\bar{v}_{_8}+\frac{1}{2},\\
&\sigma_{\chi}=2v_{_5}=2\bar{v}_{_{10}},~~~~~\sigma^a=2v_{_8}=2 \bar{v}_{_3},~~~~~ \sigma_{\kappa} =2v_{_{10}}=2\bar{v}_{_5}.
\end{split}
\end{equation}
There are crossing rules for the anomaly-related TCFs: $\sigma_{m/a}$ are invariant under the interchange of $\bar{\mu}$ and $\bar{\mu}_{_5}$; $\sigma_{\chi}$ and $\sigma_\kappa$ are related via $\sigma_\chi\left[\bar{\mu}\leftrightarrow \bar{\mu}_{_5} \right]= \sigma_\kappa$.

Thanks to the linearisation, both $J^\mu$ and $J^\mu_5$ are conserved. That is, $J^\mu$ and $J^\mu_5$ are invariant under the gauge  transformations $\mathcal{V}_\mu \to \mathcal{V}_\mu  +\partial_\mu \phi$ and $\mathcal{A}_\mu \to \mathcal{A}_\mu + \partial_\mu \varphi$. In appendix~\ref{appendix5} we prove that all the TCFs in~(\ref{jmu},\ref{jmu5}), in fact, are uniquely fixed without imposing the frame convention~(\ref{Landau frame}).

Following the definition~(\ref{def potentials}), the chemical potentials $\mu,\mu_{_5}$ are
\begin{equation} \label{chem/den first}
\mu=\frac{1}{2}\rho-S_3\delta\rho+S_2 \partial_t^{-1}\partial_kE_k,~~~~~
\mu_{_5}=\frac{1}{2}\rho_{_5}-S_3\delta\rho_{_5}+S_2 \partial_t^{-1}\partial_kE_k^a.
\end{equation}
These relations can be used  to replace the charge densities $\rho,\rho_{_5}$ in~(\ref{jmu},\ref{jmu5}) in favour of $\mu,\mu_{_5}$.
The results are presented in~(\ref{jmu chem},\ref{jmu5 chem}) with the coefficients given by
\begin{equation} \label{alpha12/D prime}
\alpha_1=\frac{2}{1-2S_3(r=1)},~~~~~~\alpha_2=-\frac{2S_2(r=1)\partial_t^{-1}}{1-2S_3(r=1)},
~~~~~~\mathcal{D}^\prime=\frac{2\mathcal{D}}{1-2S_3(r=1)},
\end{equation}
\begin{equation}\label{sigma prime}
\sigma_e^\prime=\sigma_e+\frac{2\mathcal{D} S_2(r=1)}{1-2S_3(r=1)} \partial_t^{-1} \partial^2,~~~~~~
\sigma_m^\prime= \sigma_m-\frac{2\mathcal{D}S_2(r=1)}{1-2S_3(r=1)},
\end{equation}
where $S_2,S_3$ are to be determined in section~\ref{subsection42}.

Putting the currents on-shell, the constitutive relations~(\ref{jmu},\ref{jmu5}) bear a standard form of linear response theory, from which current-current correlators read
\begin{equation}
\langle J^t J^t\rangle=\langle J^t_5 J^t_5 \rangle=-\frac{\sigma_e q^2} {i\omega-q^2\mathcal{D}},
\end{equation}
\begin{equation}
\langle J^t J^t_5\rangle =0,
\end{equation}
\begin{equation}
\langle J^t J^i\rangle=\langle J^t_5 J^i_5 \rangle=-\frac{\sigma_e \omega q_i} {i\omega-q^2\mathcal{D}},
\end{equation}
\begin{equation}
\langle J^t J^i_5 \rangle=\langle J^t_5 J^i \rangle=0,
\end{equation}
\begin{equation} \label{correlator1}
\langle J^i J^j \rangle=\langle J^i_5 J^j_5 \rangle= \left(i\omega\sigma_e+ q^2 \sigma_m \right) \left(\delta_{ij} -\frac{q_i q_j} {q^2}\right) -\frac{\omega^2\sigma_e} {i\omega-q^2\mathcal{D}} \cdot\frac{q_iq_j}{q^2}+ \sigma_\chi\epsilon_{ijk}iq_k,
\end{equation}
\begin{equation}\label{correlator2}
\langle J^i J^j_5 \rangle= q^2\sigma_a \left(\delta_{ij}-\frac{q_i q_j}{q^2}\right)+ \sigma_\kappa \epsilon_{ijk}iq_k.
\end{equation}
While the  dispersion relation $i\omega-q^2\mathcal{D}(\omega,q^2)=0$ is not affected by the anomaly, residues of the correlators~(\ref{correlator1},\ref{correlator2}) get modified.   Kubo formulas are usually used to relate  transport coefficients to thermal correlators.
However, evaluated on-shell, the correlators partially lose information about dynamics of off-shell one-point currents. As a consequence,
they are insufficient to determine all order transport coefficients. For example, beyond their constant values, $\mathcal{D},\sigma_{e/m}$ cannot be fully extracted from the current-current correlators \cite{1511.08789}.

Yet, there are exact relations between the correlators and $\sigma_\chi,\sigma_\kappa$,
\begin{equation}\label{kubo formulae}
\sigma_\chi(\omega,\vec{q})=-\frac{i}{2q_n} \sum_{i,j}\epsilon_{nij}\langle J^i J^j \rangle,~~~~~~~~~~~~~
\sigma_\kappa(\omega,\vec{q})=-\frac{i}{2q_n} \sum_{i,j}\epsilon_{nij}\langle J^i J^j_5 \rangle
\end{equation}
which are valid for arbitrary $\omega$ and $\vec{q}$. The relation~(\ref{kubo formulae}) for $\sigma_\chi$ was first derived in~\cite{0907.5007} by promoting constant magnetic field in the original CME into an inhomogeneous perturbation.  Our constitutive relation translates into
rigorous derivation of  (\ref{kubo formulae}).
%While $\sigma_a$ is an axial analogue of $\sigma_m$, it is
Contrary to $\sigma_m$,  $\sigma_a$ can be determined from the correlators, particularly from the mixed correlator $\langle J^i J^j_5 \rangle$.
This became possible thanks to the absence of the $\vec E^a$ ($\vec{E}$) term in the constitutive relations for $J^\mu$ ($J^\mu_5$). We suspect this is accidental and specific to the model in study.

\subsection{Results: solving the bulk equations} \label{subsection42}

To determine all the TCFs  in~(\ref{jmu},\ref{jmu5}), we merely need to solve the following ODEs: (\ref{eom 1},\ref{eom 2}), (\ref{eom 11},\ref{eom 12}) and (\ref{eom 5},\ref{eom 6},\ref{eom 9},\ref{eom 10}) (see appendix~\ref{appendix1} for detailed analysis). Part of the  ODEs, (\ref{eom 1},\ref{eom 2}), (\ref{eom 11},\ref{eom 12}) were already solved in~\cite{1511.08789}. In principle, we  only need to solve (\ref{eom 5},\ref{eom 6},\ref{eom 9},\ref{eom 10}) and then $S_2,V_3,\bar{V}_3$ would be extracted via the relations~(\ref{constraint3}). In practice, however, we solve (\ref{eom 3}, \ref{eom 5}-\ref{eom 10}). In this way, we avoid numerically problematic special points $\omega,q=0$ when making use of the relations~(\ref{constraint3}).

We first solve the ODEs analytically  in the hydrodynamic limit  and then numerically for arbitrary $\omega$ and $q$.

\subsubsection{Hydrodynamic expansion: analytical results}\label{sss421}

In the hydrodynamic limit  $\omega, q \ll 1$, the ODEs (\ref{eom 3}, \ref{eom 5}-\ref{eom 10}) can be solved perturbatively. Let introduce a formal expansion parameter $\lambda$
\begin{equation}
\omega \to \lambda \omega,~~~~~~~~~~~\vec{q} \to \lambda \vec{q}.
\end{equation}
Note that $\bar{S}_2=0$ from (\ref{constraint3}). The functions to be solved for are $\left\{S_2,V_1,\bar{V}_1,V_3,\bar{V}_3,V_5,\bar{V}_5\right\}$, which are expanded in powers of $\lambda$,
\begin{equation}
\begin{split}
&S_2=\sum_{n=0}^{\infty}\lambda^n S_2^{(n)},~~~~~V_1=\sum_{n=0}^{\infty}\lambda^n V_1^{(n)},~~~~~\bar{V}_1=\sum_{n=0}^{\infty}\lambda^n \bar{V}_1^{(n)},~~~~~
V_3=\sum_{n=0}^{\infty}\lambda^n V_3^{(n)}\\
&\bar{V}_3=\sum_{n=0}^{\infty}\lambda^n \bar{V}_3^{(n)},~~~~~
V_5=\sum_{n=0}^{\infty}\lambda^n V_5^{(n)},~~~~~\bar{V}_5=\sum_{n=0}^{\infty}\lambda^n \bar{V}_5^{(n)}.
\end{split}
\end{equation}
At each order in $\lambda$, the solutions are expressed as double integrals over $r$, see appendix~\ref{appendix2}. Below we collect the series expansions of $v_{_3}, \bar{v}_{_3}, v_{_5}, \bar{v}_{_5}$,
%\begin{equation}
%v_{_1}=\frac{1}{2}i\omega+\frac{1}{4} \left\{\omega^2\left(1+\log 2\right)+ q^2 \left[1-144\kappa^2\left(\bar{\mu}^2+\bar{\mu}_{_5}^2\right) \left(2\log 2-1\right) \right] \right\}+\cdots,
%\end{equation}
%
%\begin{equation}
%\bar{v}_{_1}=72\kappa^2\bar{\mu}\bar{\mu}_{_5}q^2\left(2\log2-1\right)+\cdots,
%\end{equation}
%
\begin{equation}
v_{_3}=-\frac{1}{4}\left[1-144\kappa^2\left(\bar{\mu}^2+\bar{\mu}_{_5}^2\right) \left(2\log2-1\right)\right]+ \frac{i\omega}{32}\left[(2\pi-\pi^2+4\log 2)+ \mathcal{O}\left(\bar{\mu}^2,\bar{\mu}_{_5}^2\right)\right]+\cdots,
\end{equation}
\begin{equation}
\bar{v}_{_3}=72\kappa^2\bar{\mu}\bar{\mu}_{_5}\left(2\log 2-1\right)+\cdots,
\end{equation}
\begin{equation}
\begin{split}
v_{_5}=&6\kappa\bar{\mu}_{_5}+6\kappa \bar{\mu}_{_5}i\omega\log 2-\frac{1}{4}\kappa \bar{\mu}_{_5}\left\{6\omega^2\log^22+q^2\left[\pi^2-1728\kappa^2 \left(\bar{\mu}_{_5}^2 + 3\bar{\mu}^2 \right) \right.\right.\\
&\left.\left.\times\left(\log2-1\right)^2\right]\right\}+\cdots,
\end{split}
\end{equation}
\begin{equation}
\bar{v}_{_5}=v_{_5}[\bar{\mu}\leftrightarrow \bar{\mu}_{_5}].
\end{equation}
The series expansions of $v_{_2},v_{_4}$ were worked out in~\cite{1511.08789},
\begin{equation}
v_{_2}=\frac{1}{2}+\frac{1}{4}i\omega \left(1+\log 2\right)+\frac{1}{48}\left(\pi^2 \omega^2 -3\pi q^2-6q^2 \log 2\right)+\cdots,
\end{equation}
\begin{equation}
v_{_4}=-\frac{1}{4}-\frac{\pi}{16}i\omega +\frac{1}{96}\left[\pi^2\omega^2-q^2\left(6\log 2-3\pi\right)\right]+\cdots.
\end{equation}
Once substituted into~(\ref{transport def}), these perturbative results generate the hydrodynamic expansion of all the TCFs   as quoted in (\ref{D expansion}-\ref{sigma_kappa expansion}).

\subsubsection{Beyond the hydro limit: numerical results} \label{sss422}

To proceed with the  all-order derivative resummation, we have to go beyond the conventional hydrodynamic limit and solve the ODEs (\ref{eom 3}, \ref{eom 5}-\ref{eom 10}) for generic values of momenta. We were able to do it numerically only. We deal with a boundary value problem for a system of second order linear ODEs. We performed our numerical calculations within two different numerical methods, a shooting technique and a  spectral method.
Within numerical accuracy, both approaches give the same results.

Within the shooting technique our numerical procedure is much like that of~\cite{1511.08789}. One starts with a trial initial value for the functions to be solved for at the horizon $r = 1$ and integrates the ODEs up to the conformal boundary $r=\infty$. The solutions generated in this step have to fulfil the boundary conditions at $r=\infty$. If not, the trial initial data have to be adjusted and the procedure is repeated until the requirements at the boundary are satisfied with a satisfactory  numerical accuracy. The fine-tuning process of finding the correct initial data is reduced to a root-finding routine, which can be implemented by the Newton's method.

The spectral method converts the continuous boundary value problem of linear ODEs into that of discrete linear algebra. We distribute a number of points on the integration domain. These points are collectively referred to as collation grid. The functions to be solved for are then represented by their values on the grid. For given  values of the functions on the grid, their derivatives at the grid are approximated by differentiating interpolation functions (normally based on polynomial or trigonometric interpolation). Thus, a differential operation is mapped into a  matrix. Eventually, this collation procedure allows to discretise the original continuous problem and turns it into a system of algebraic equations involving values of the functions on the grid. The boundary conditions are mapped into algebraic relations among the values of the functions at the outermost grid points. They will be imposed by replacing suitable equations in above-mentioned algebraic equations. For more details on spectral method, we recommend the references~\cite{spectral1,spectral2,spectral3,0706.2286}. As for other non-periodic problems, we choose a Chebyshev grid and use polynomial interpolation to calculate differentiation matrices.

Since $\mathcal{D}$ and $\sigma_e$ are not affected by the anomaly, they are the same as those presented in~\cite{1511.08789}, and we would not display them here.  As $\sigma_\kappa$ can be obtained from $\sigma_\chi$ via the crossing rule as pointed out below~(\ref{transport def}),
we will focus on numerical plots for $\sigma_m,\sigma_a,\sigma_\chi$ only.

\begin{figure}[htbp]
\centering
\includegraphics[scale=0.5]{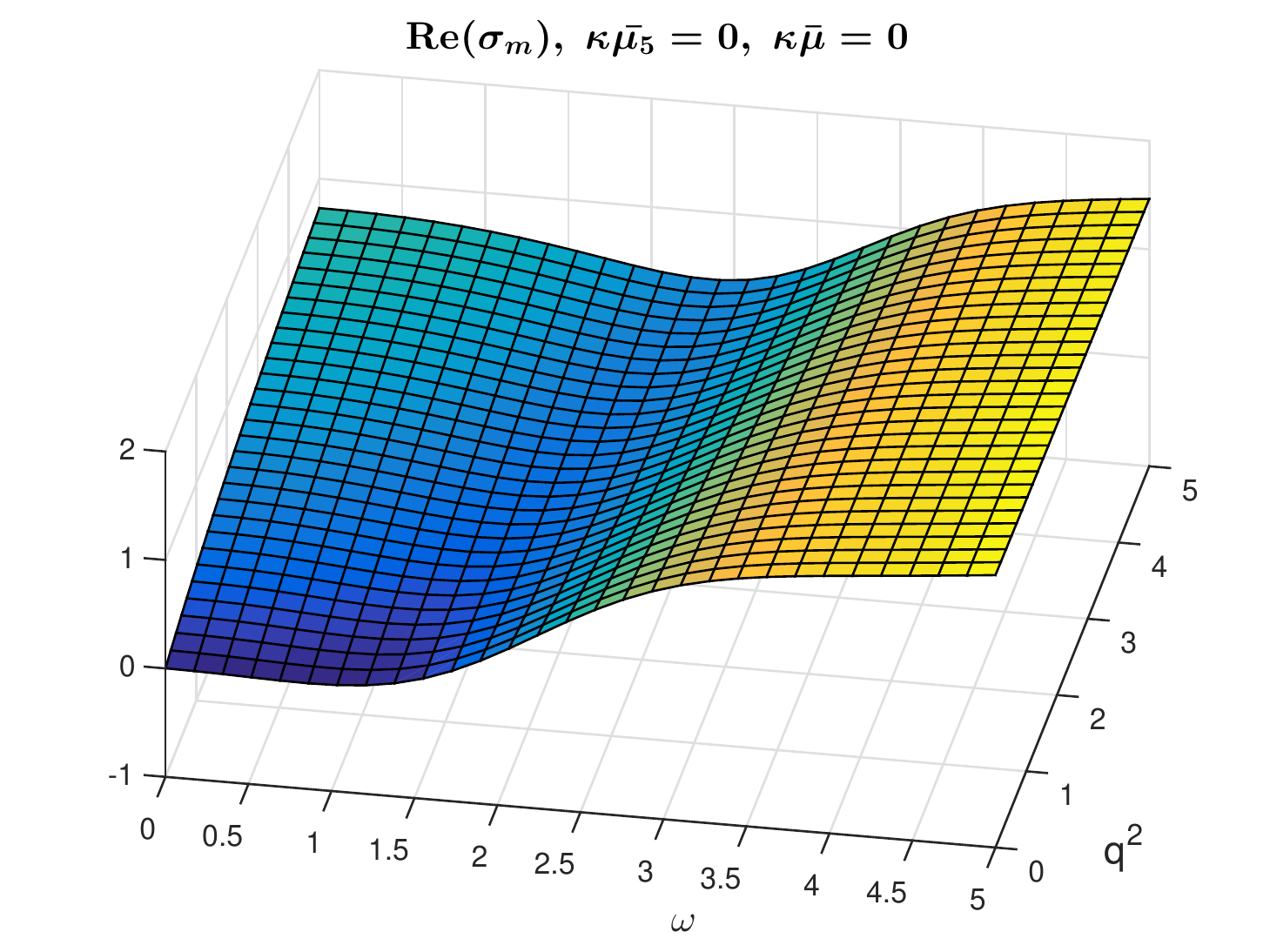}
\includegraphics[scale=0.5]{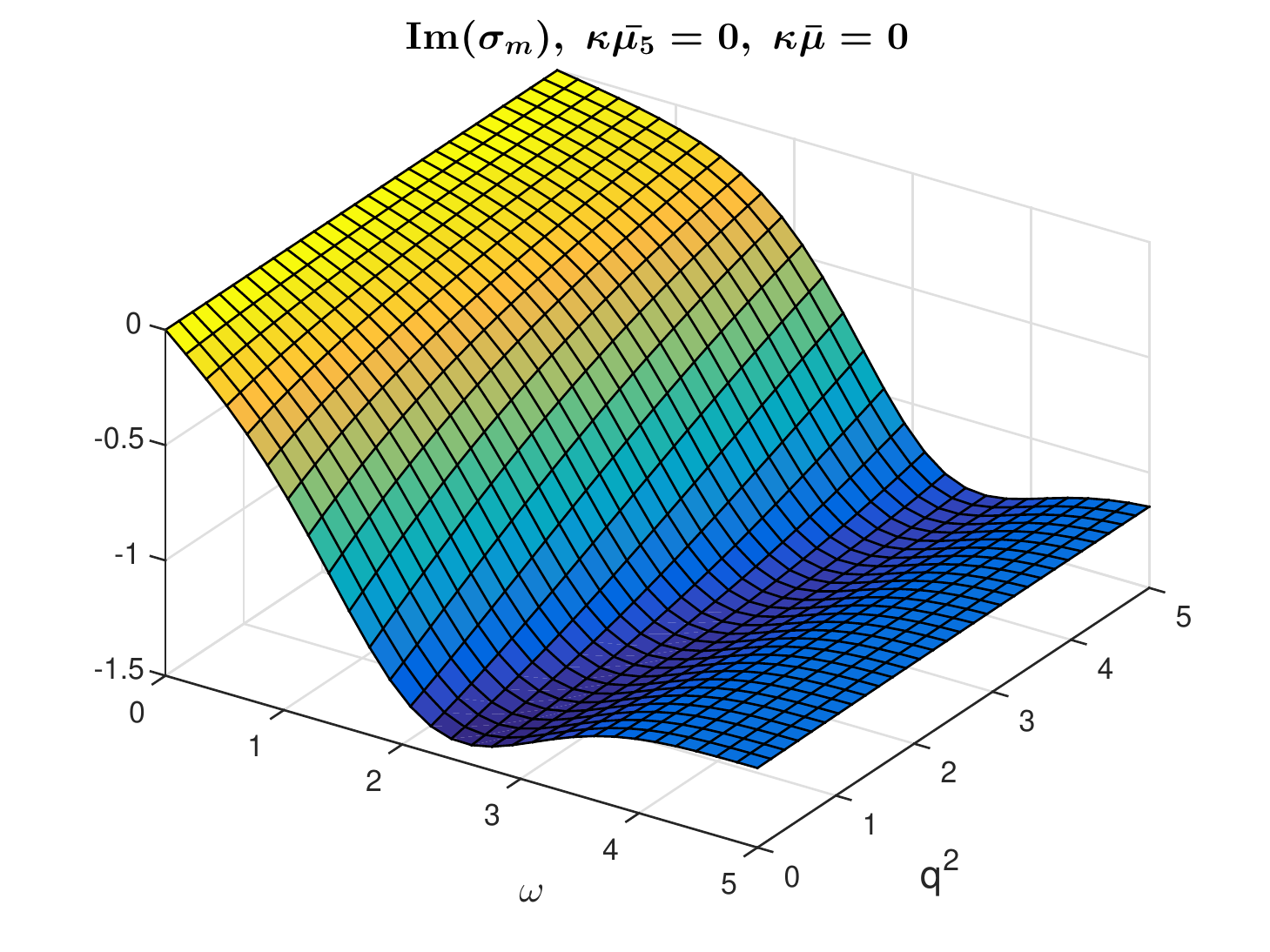}
\caption{Magnetic conductivity $\sigma_m$ as a function of $\omega$ and $q^2$ when $\kappa\bar{\mu}=\kappa\bar{\mu}_{_5}=0$.}
\label{magnetic1}
\end{figure}

\begin{figure}[htbp]
\centering
\includegraphics[scale=0.5]{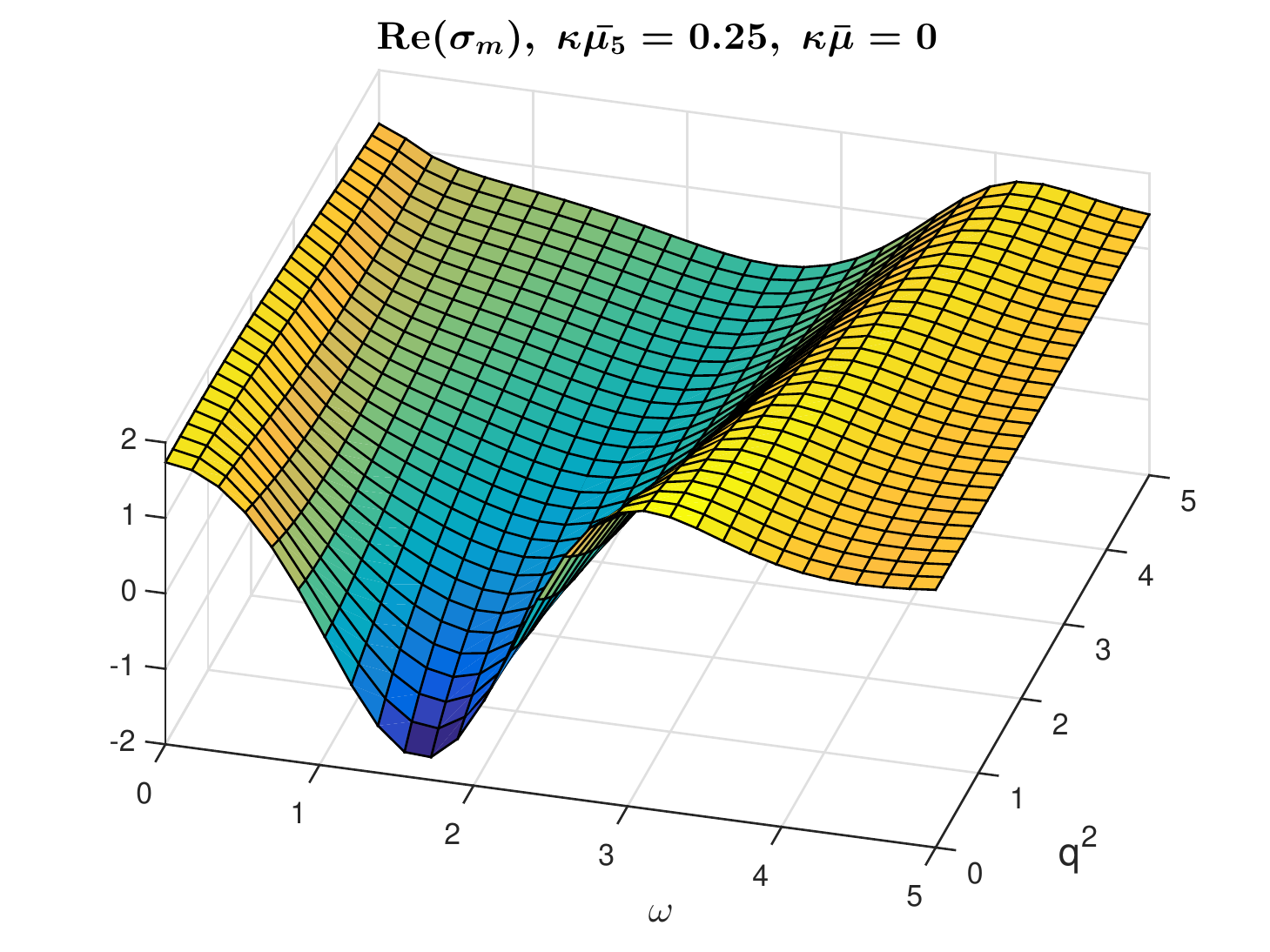}
\includegraphics[scale=0.5]{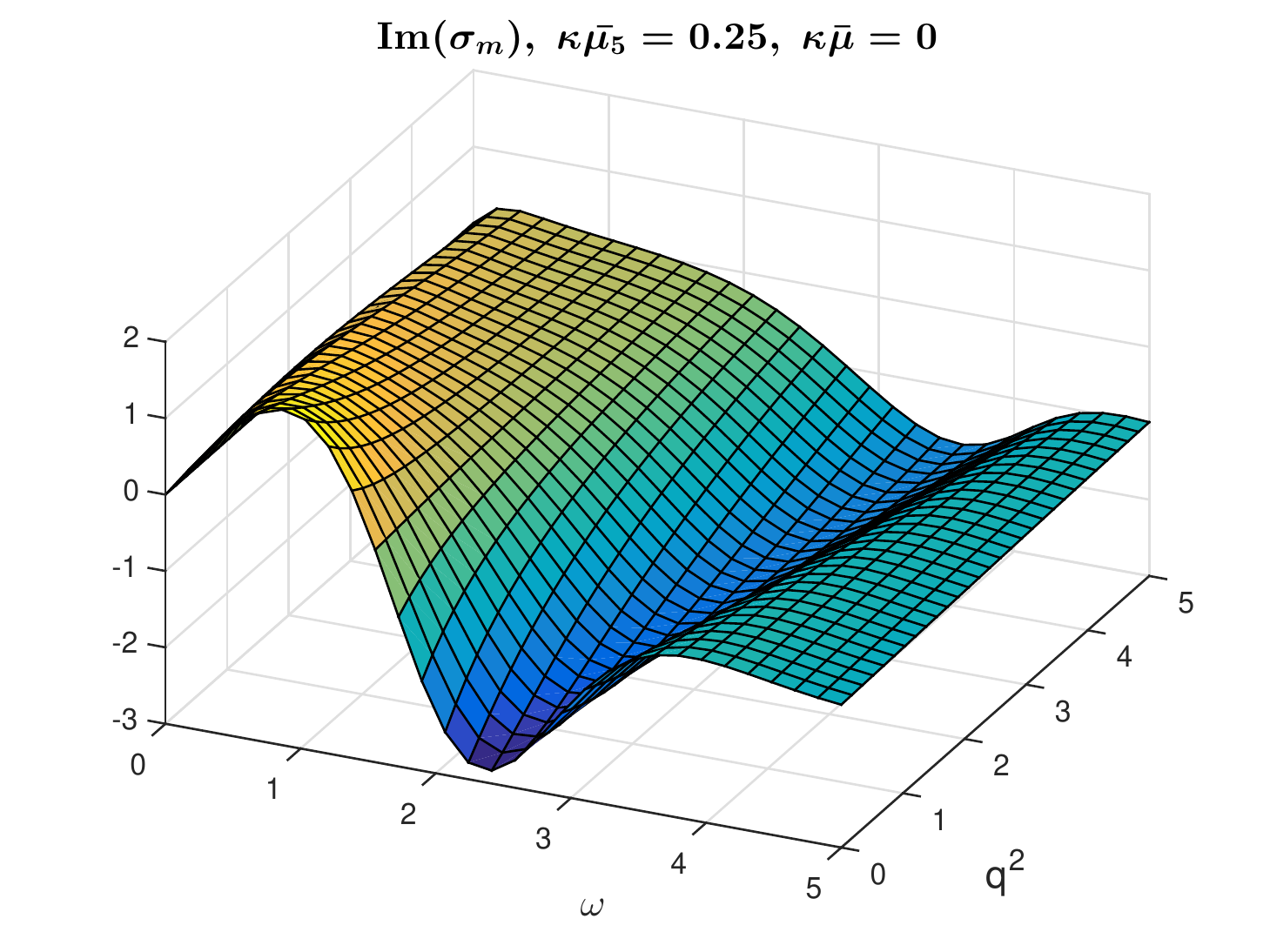}
\caption{Magnetic conductivity $\sigma_m$ as a function of $\omega$ and $q^2$ when $\kappa\bar{\mu}=0, \kappa\bar{\mu}_{_5}=1/4$.}
\label{magnetic2}
\end{figure}

\begin{figure}[htbp]
\centering
\includegraphics[scale=0.5]{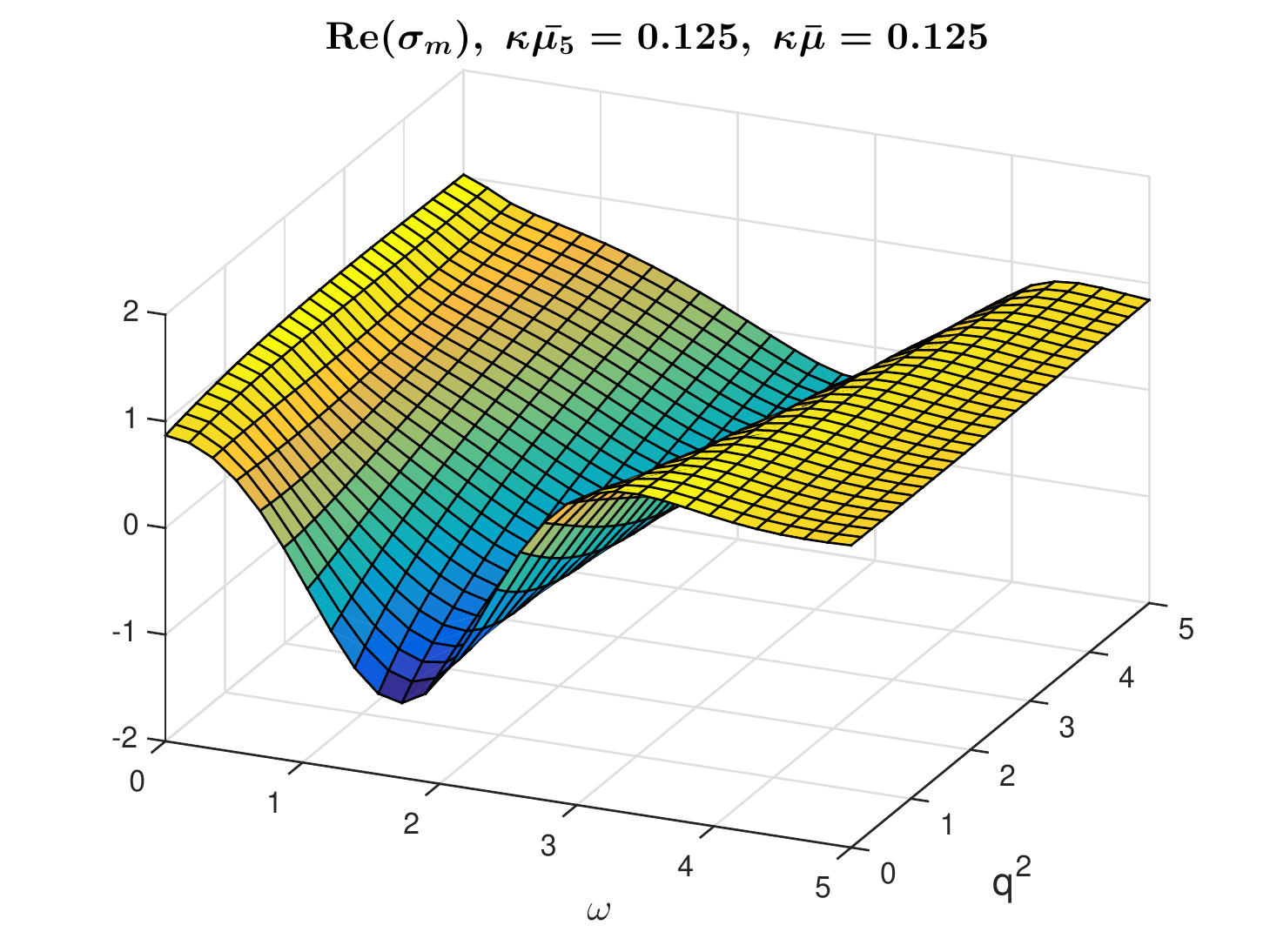}
\includegraphics[scale=0.5]{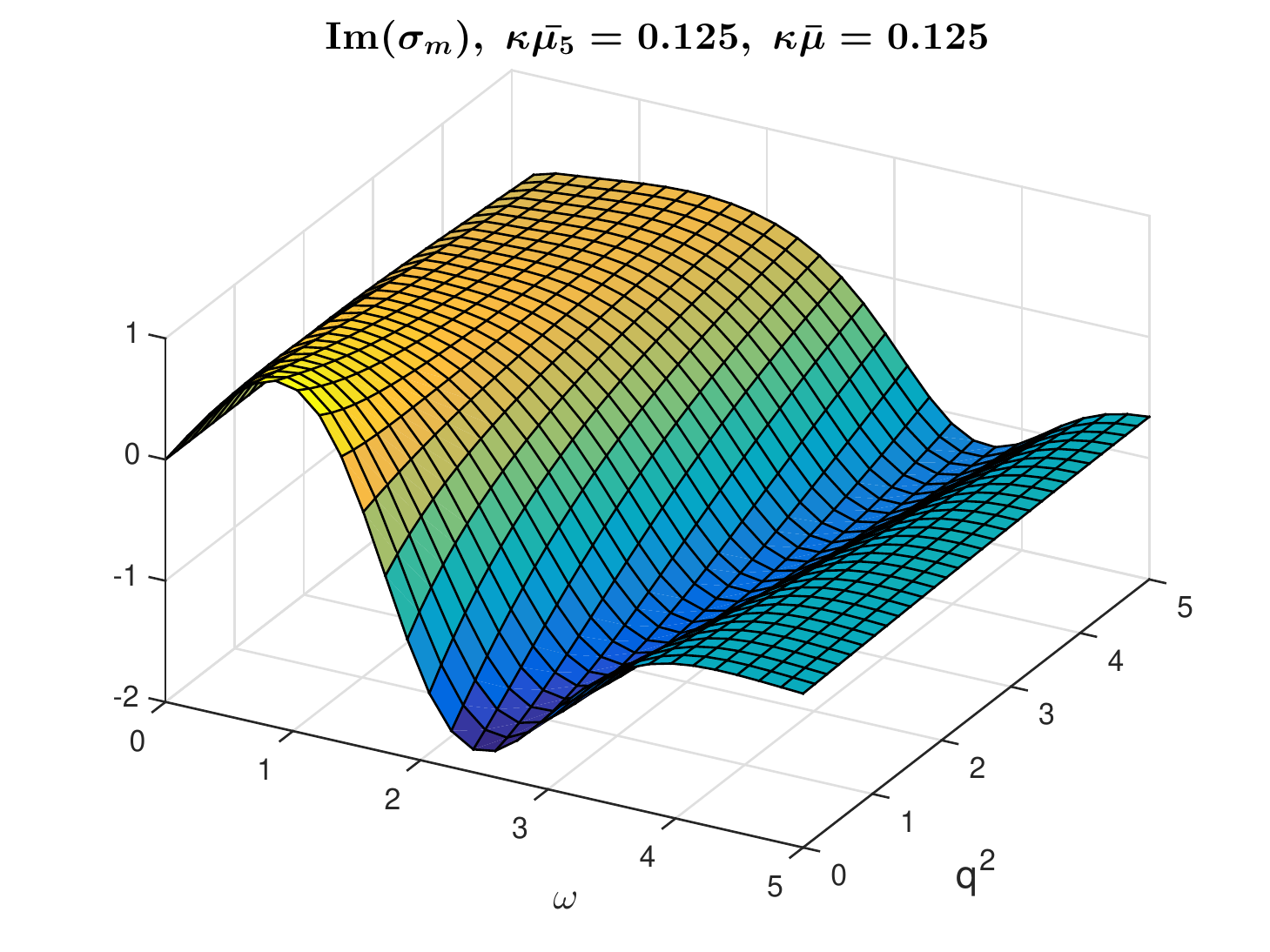}
\caption{Magnetic conductivity $\sigma_m$ as function of $\omega$ and $q^2$ when $\kappa\bar{\mu}=\kappa\bar{\mu}_{_5}=1/8$.}
\label{magnetic3}
\end{figure}

\begin{figure}[htbp]
\centering
\includegraphics[scale=0.5]{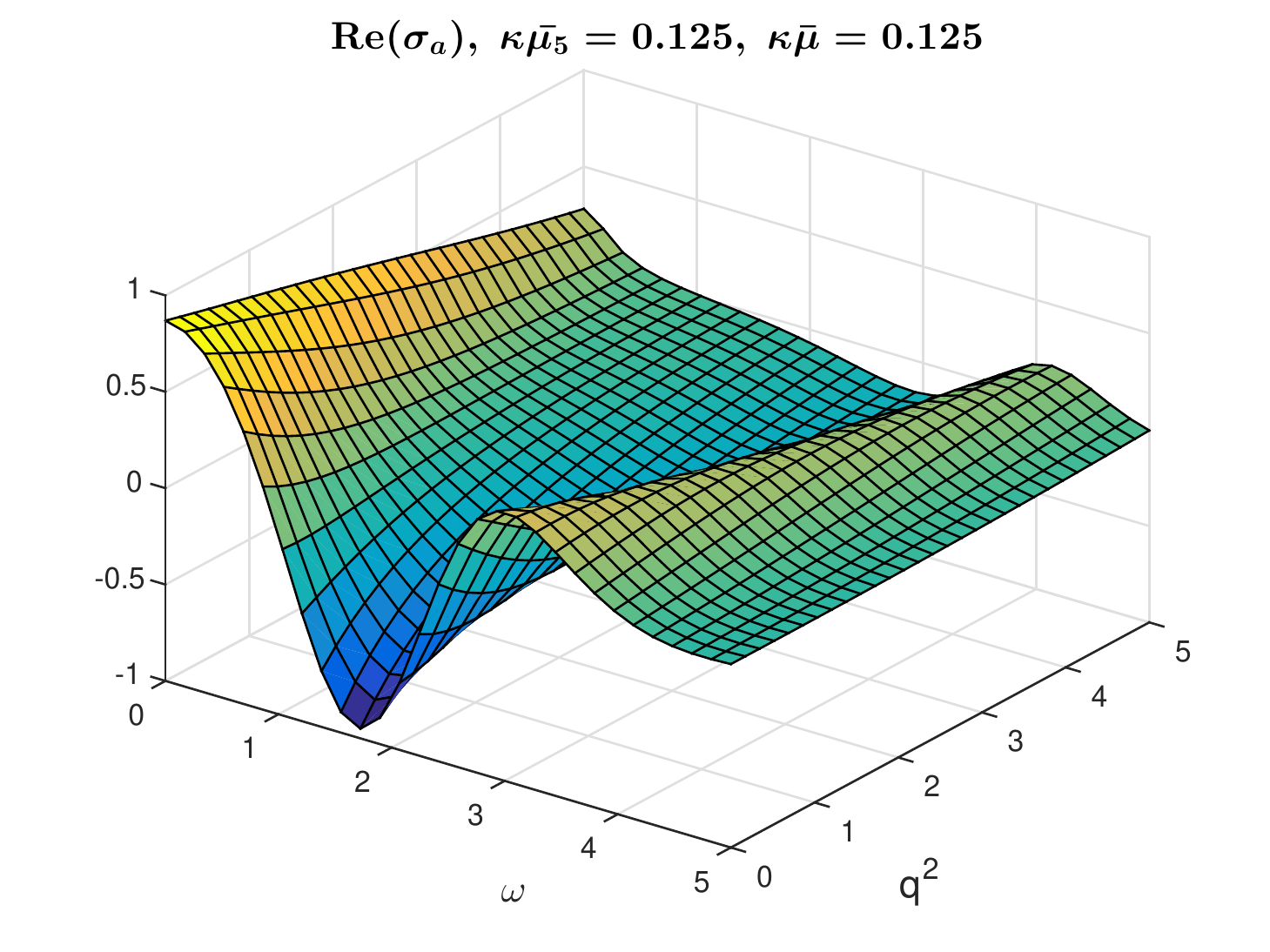}
\includegraphics[scale=0.5]{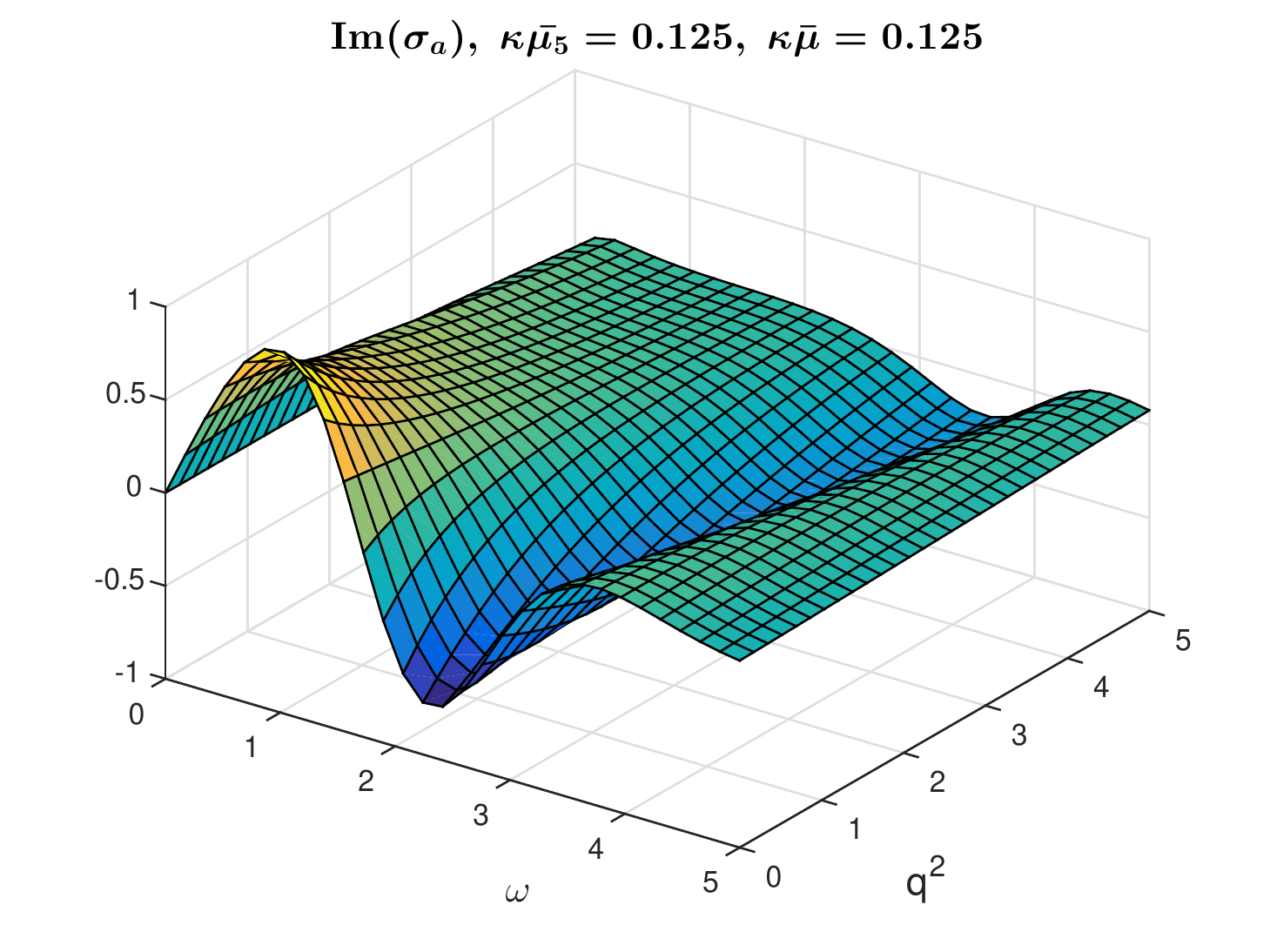}
\caption{$\sigma_a$ as function of $\omega$ and $q^2$ when $\kappa\bar{\mu}=\kappa\bar{\mu}_{_5}=1/8$.}
\label{axial-magnetic}
\end{figure}

Figure~\ref{magnetic1} is a reproduction of a 3D plot for the magnetic conductivity $\sigma_m$ from~\cite{1511.08789}: compared to \cite{1511.08789}, we extend the plot domain to larger momenta so that asymptotic regime is more clearly seen. Figures~\ref{magnetic2},\ref{magnetic3} show anomaly-modified $\sigma_m$ for sample values of $\kappa \bar{\mu}$ and $\kappa \bar{\mu}_{_5}$.
%Due to anomalous correction, both $\textrm{Re}(\sigma_m)$ and $\textrm{Im}(\sigma_m)$ become more negative around their minima, which are roughly located at $\left\{\omega\simeq 1.7,q=0\right\}$ for $\textrm{Re}(\sigma_m)$ and at $\left\{\omega\simeq 2.4,q=0\right\}$ for $\textrm{Im}(\sigma_m)$.
In Figure~\ref{axial-magnetic} we show momenta-dependent $\sigma_a$. Note that $\sigma_a$ is non-vanishing only when $\kappa\bar{\mu}\bar{\mu}_5\neq0$.

\begin{figure}[htbp]
\centering
\includegraphics[width=0.48\textwidth,height=0.40\textwidth]{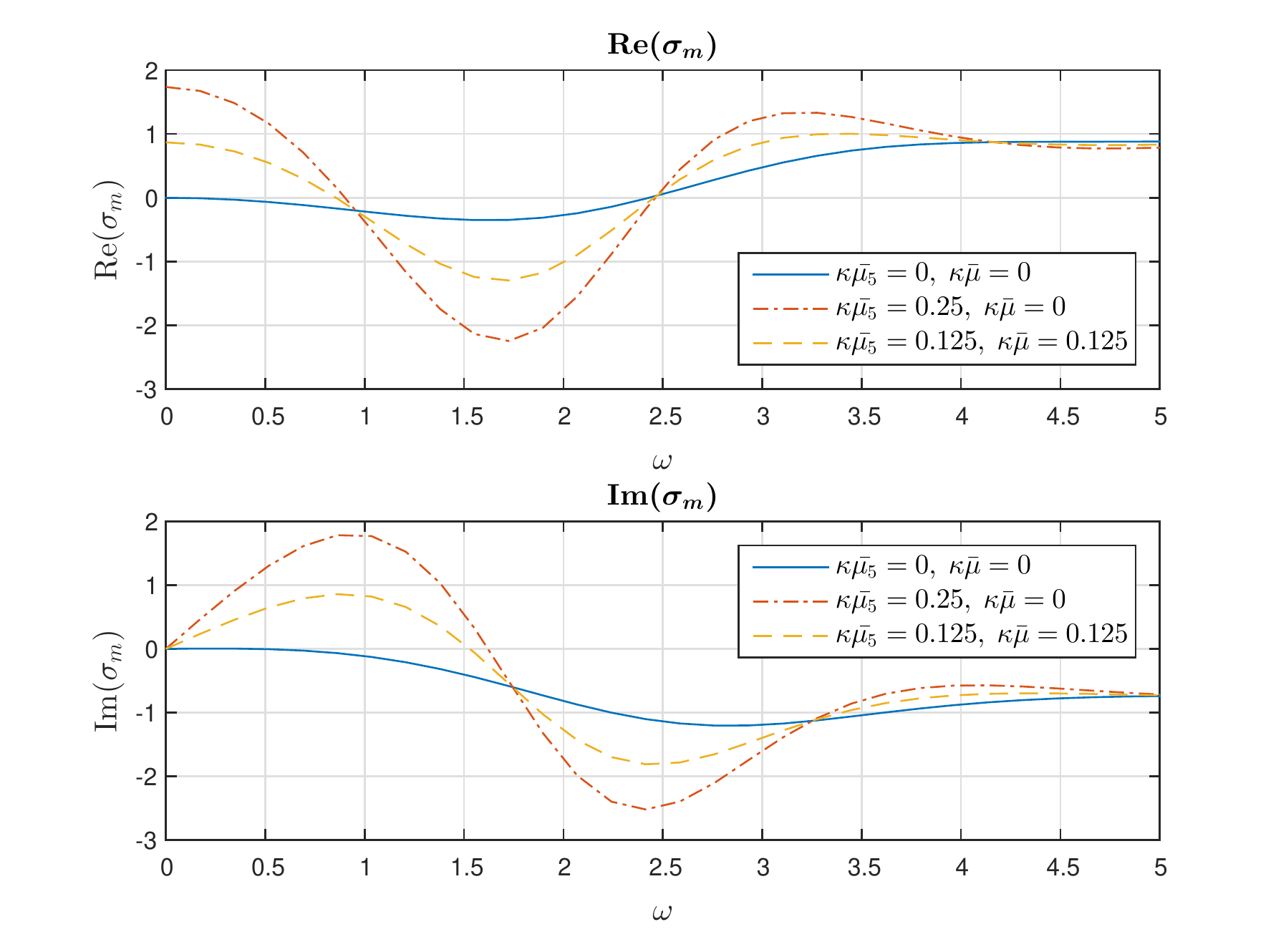}
\includegraphics[width=0.48\textwidth]{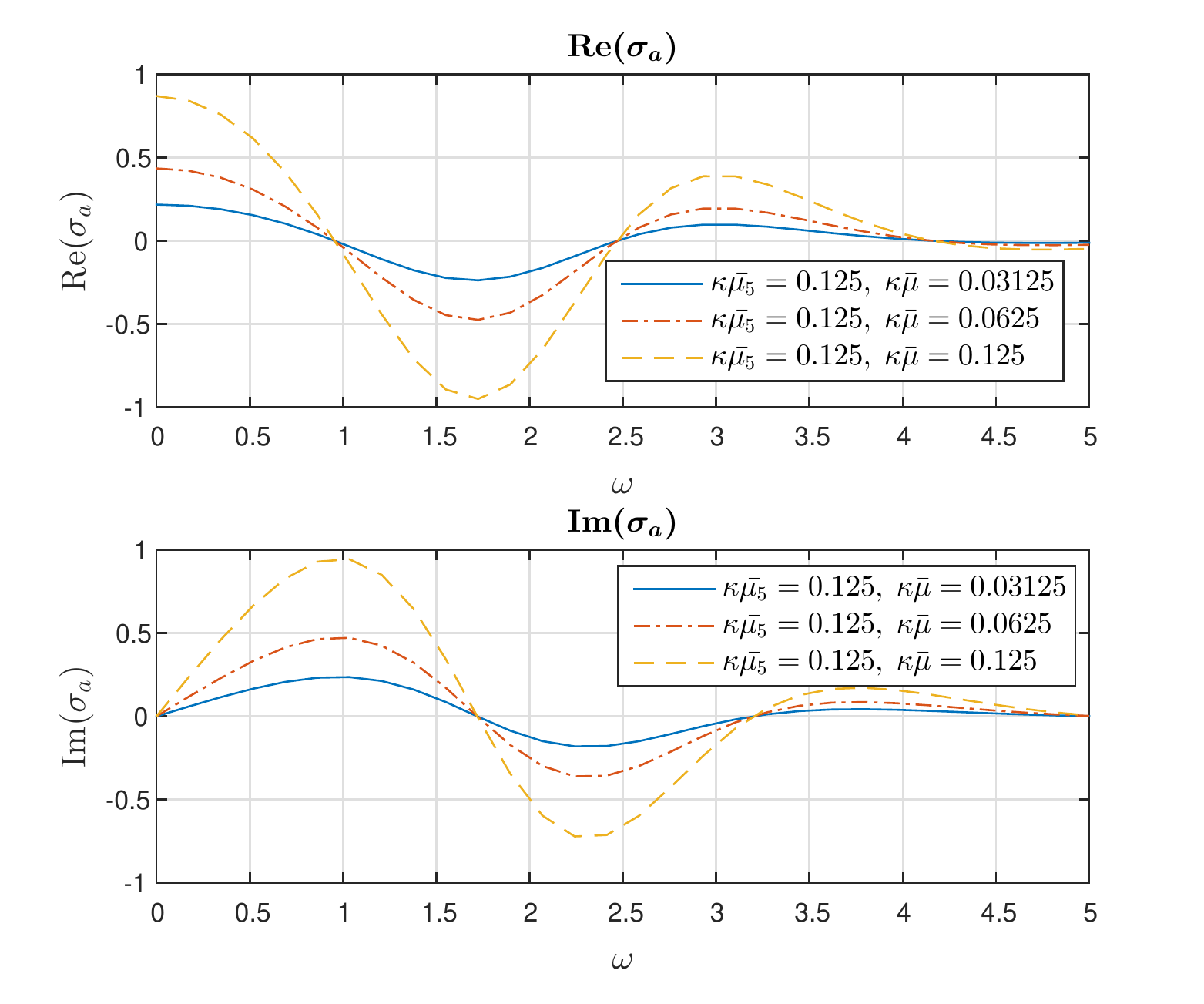}
\caption{$\omega$-dependence of $\sigma_m$ and $\sigma_a$ when $q=0$.}
\label{magnetic-axialw}
\end{figure}

\begin{figure}[htbp]
\centering
\includegraphics[width=0.48\textwidth]{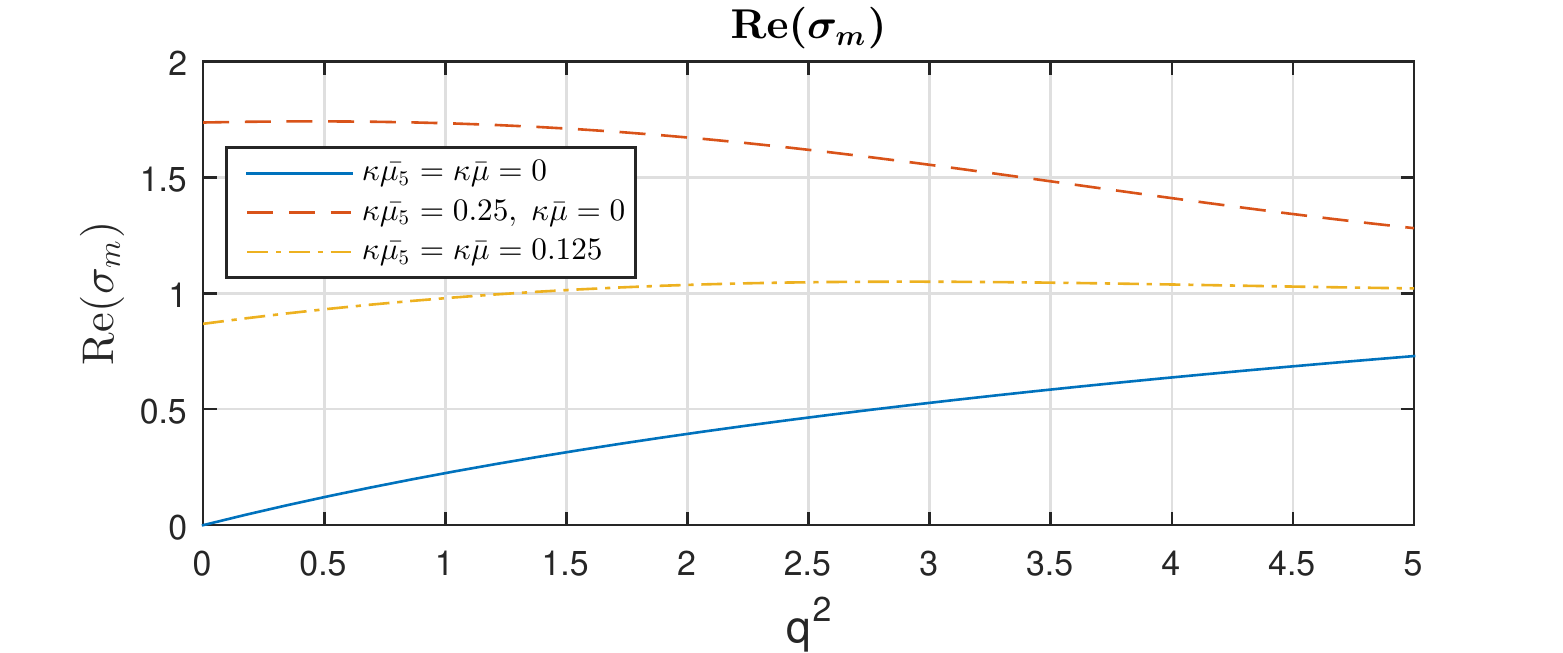}
\includegraphics[width=0.48\textwidth]{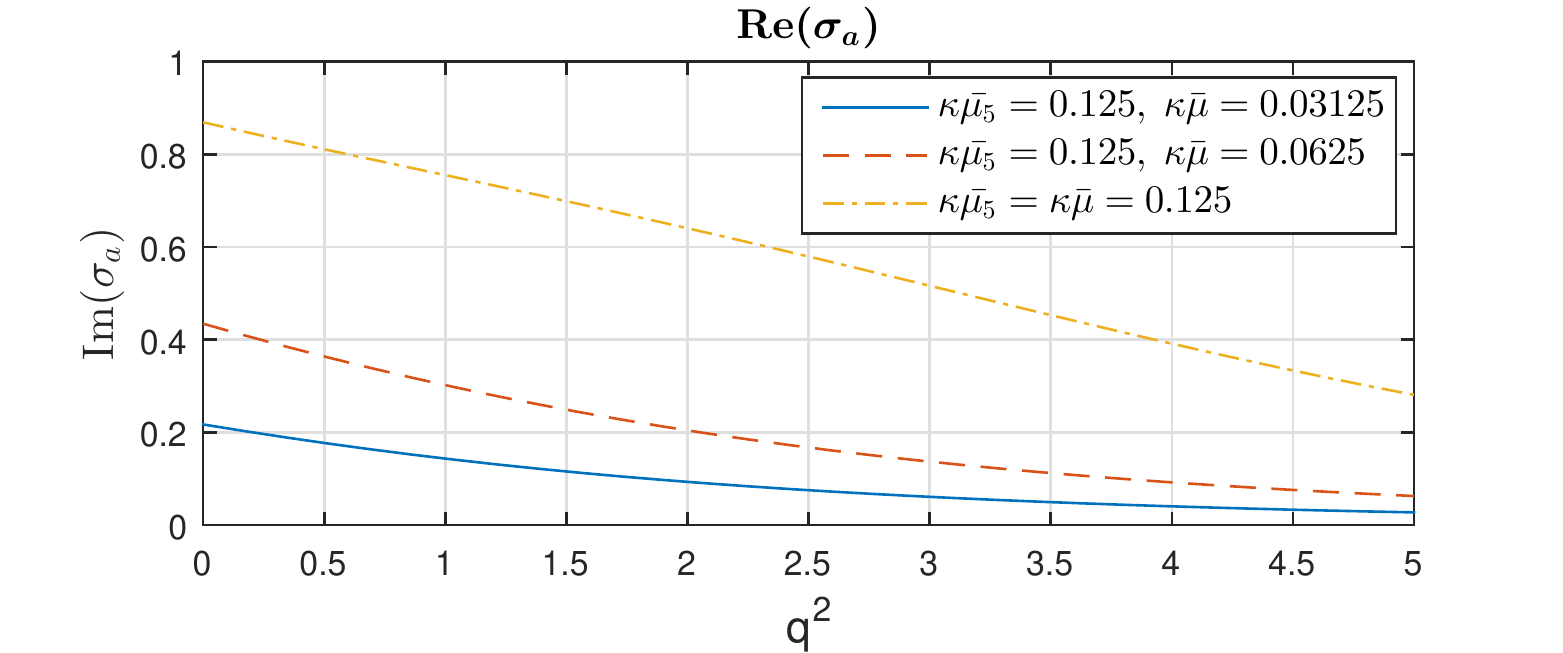}
\caption{$q^2$-dependence of $\sigma_m$ and $\sigma_a$ when $\omega=0$.}
\label{magnetic-axialq}
\end{figure}

\begin{figure}[htbp]
\centering
\includegraphics[width=0.48\textwidth]{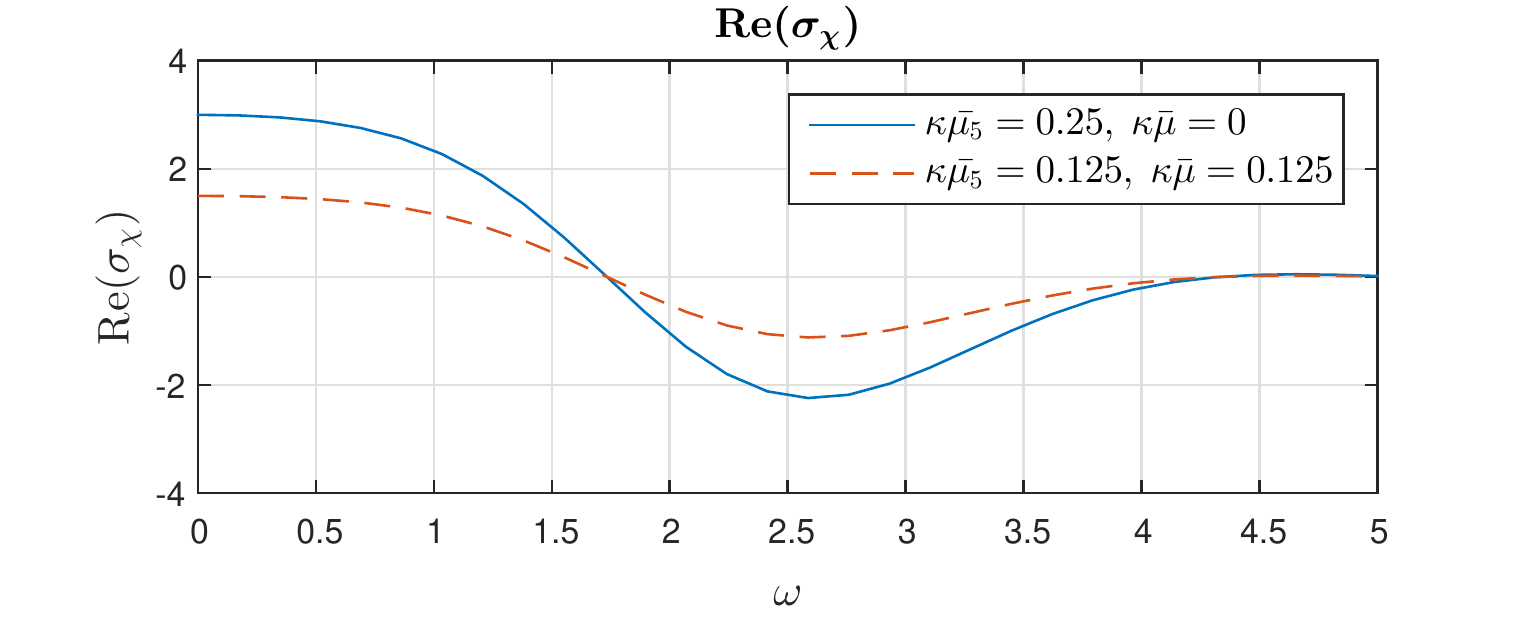}
\includegraphics[width=0.48\textwidth]{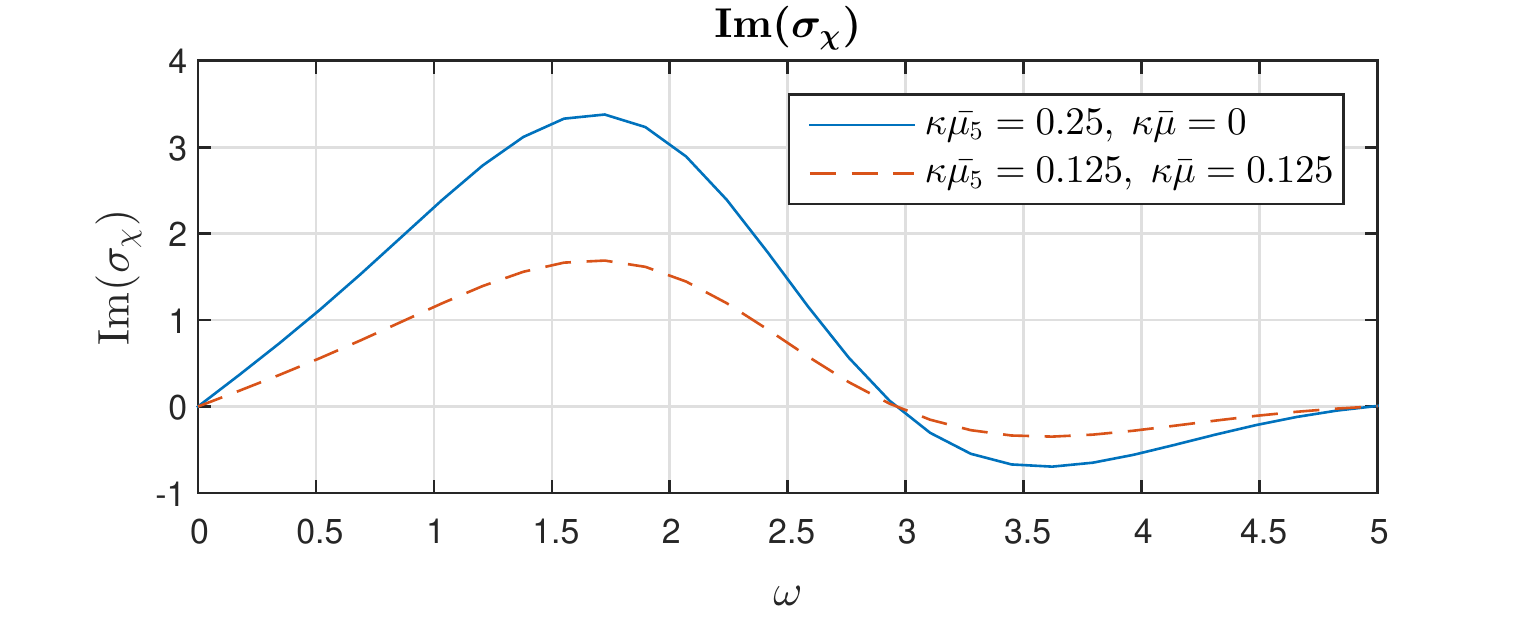}
\includegraphics[width=0.48\textwidth]{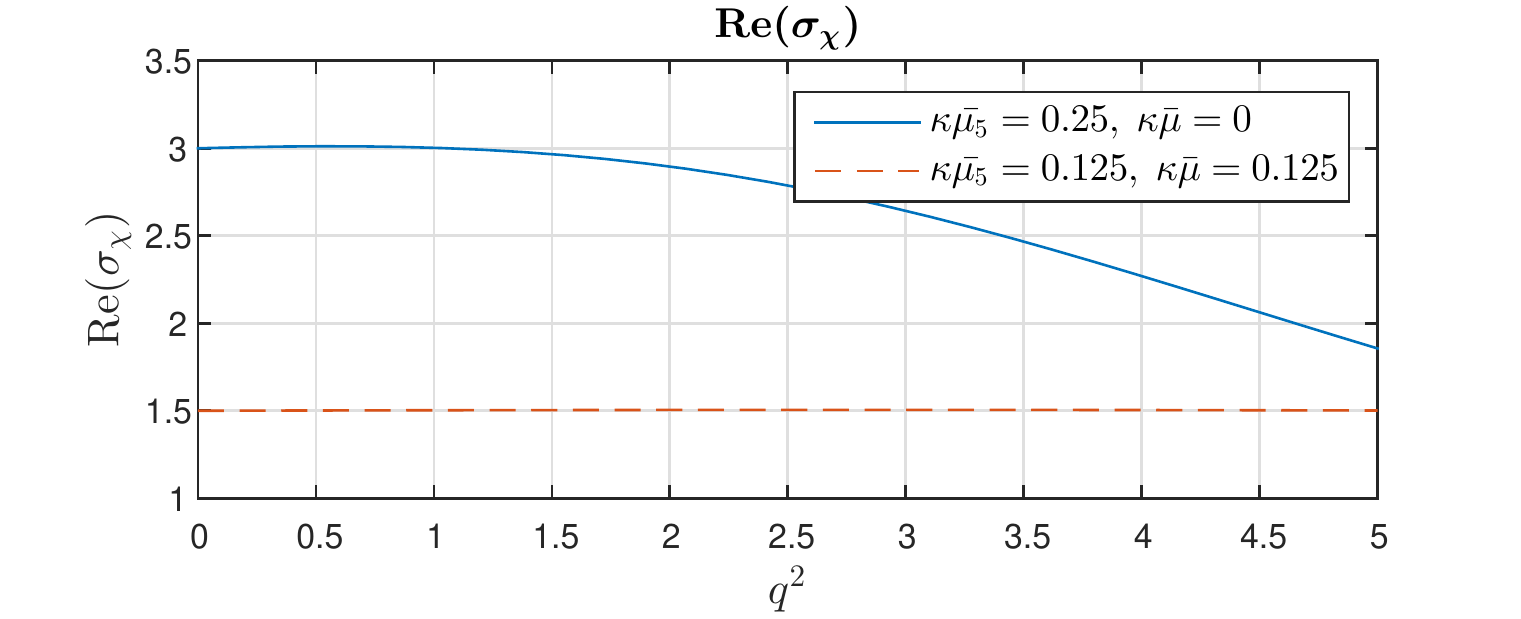}
\caption{$\omega$-dependence ($q$-dependence) of $\sigma_\chi$ when $q=0$ ($\omega=0$).}
\label{cmewq}
\end{figure}

In Figures \ref{magnetic-axialw},\ref{magnetic-axialq} we show 2D slices of Figures~\ref{magnetic1},\ref{magnetic2},\ref{magnetic3},\ref{axial-magnetic} when either $\omega=0$ or $q=0$. It is demonstrated that asymptotically $\sigma_m$ approaches a nonzero value, while $\sigma_a$ goes to zero after damped oscillation. The asymptotic regime is achieved around $\omega \simeq 5$ for both $\sigma_m$ and $\sigma_a$. So, $\sigma_m$ encodes some UV physics while $\sigma_a$ decouples asymptotically.
%While the $q$-dependence is non-oscillating, it is still comparable with the $\omega$-dependence and should not be ignored.
In contrast with the viscosity function~\cite{1406.7222,1409.3095} and the diffusion function~\cite{1511.08789}, Figure~\ref{magnetic-axialq} illustrates that the dependence of $\sigma_m,\sigma_a$ on $q$ is more pronounced. This feature is  shared by the chiral magnetic conductivity $\sigma_\chi$ (see Figure~\ref{cmewq}). As an axial analogue of $\sigma_m$, $\sigma_a$ shows qualitatively similar dependence on $\omega$ as can be seen from Figure~\ref{magnetic-axialw}. However, $q$ dependence of $\sigma_a$ differs from that of  $\sigma_m$ as shown in Figure~\ref{magnetic-axialq}. When $\kappa \bar{\mu}$ and/or $\kappa \bar{\mu}_{_5}$ get increased, $\omega$-dependence of $\sigma_m$ becomes more enhanced, which signifies a stronger response to time-dependent external fields.
%
%\begin{figure}[htbp]
%\centering
%\includegraphics[width=0.48\textwidth]{eps//sigma_mag_omega_mu5_0125_mu_0125}
%\includegraphics[width=0.48\textwidth]{eps//sigma_a_omega_mu5_0125_mu_0125}
%\caption{$\omega$-dependence of $\sigma_m$ and $\sigma_a$ when $q=0$.}
%\label{magnetic-axialw}
%\end{figure}
%

We now turn to the chiral magnetic conductivity $\sigma_\chi$. For weakly coupled theories, momenta-dependence of $\sigma_\chi$ was studied in~\cite{0907.5007,1312.1204}: in high temperature regimes $\textrm{Re}(\sigma_\chi)$ drops from its DC limit $\sigma_\chi^0$ at $\omega=0$ to $\sigma_\chi^0/3$ just away from $\omega=0$. For strongly coupled theories with dual gravity description, frequency-dependence of $\sigma_\chi$ was initially considered in~\cite{0908.4189} for the case $q=0$.  Two different holographic models were considered in \cite{0908.4189}: RN-$AdS_5$ geometry and a finite-temperature Sakai-Sugimoto model~\cite{hep-th/0412141}. Numerical plots of~\cite{0908.4189} look rather similar, suggesting a certain universality of $\sigma_\chi$. Within the RN-$AdS_5$ model, ref.~\cite{1312.1204} also explored the momenta-dependence of anomalous TCFS,
particularly original  results on $\sigma_\kappa(\omega,q)$ were presented there.

\begin{figure}[htbp]
\centering
\includegraphics[scale=0.5]{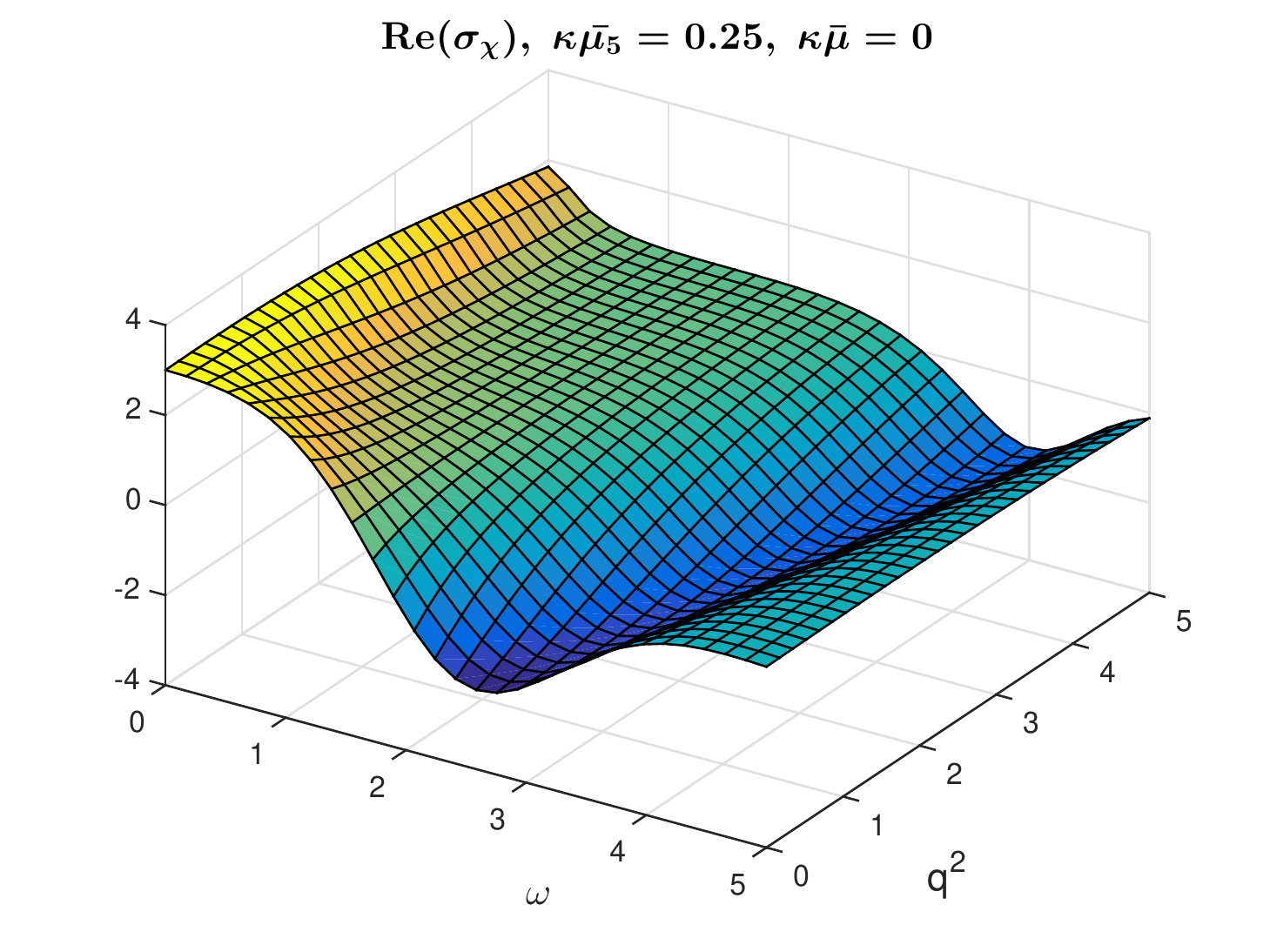}
\includegraphics[scale=0.5]{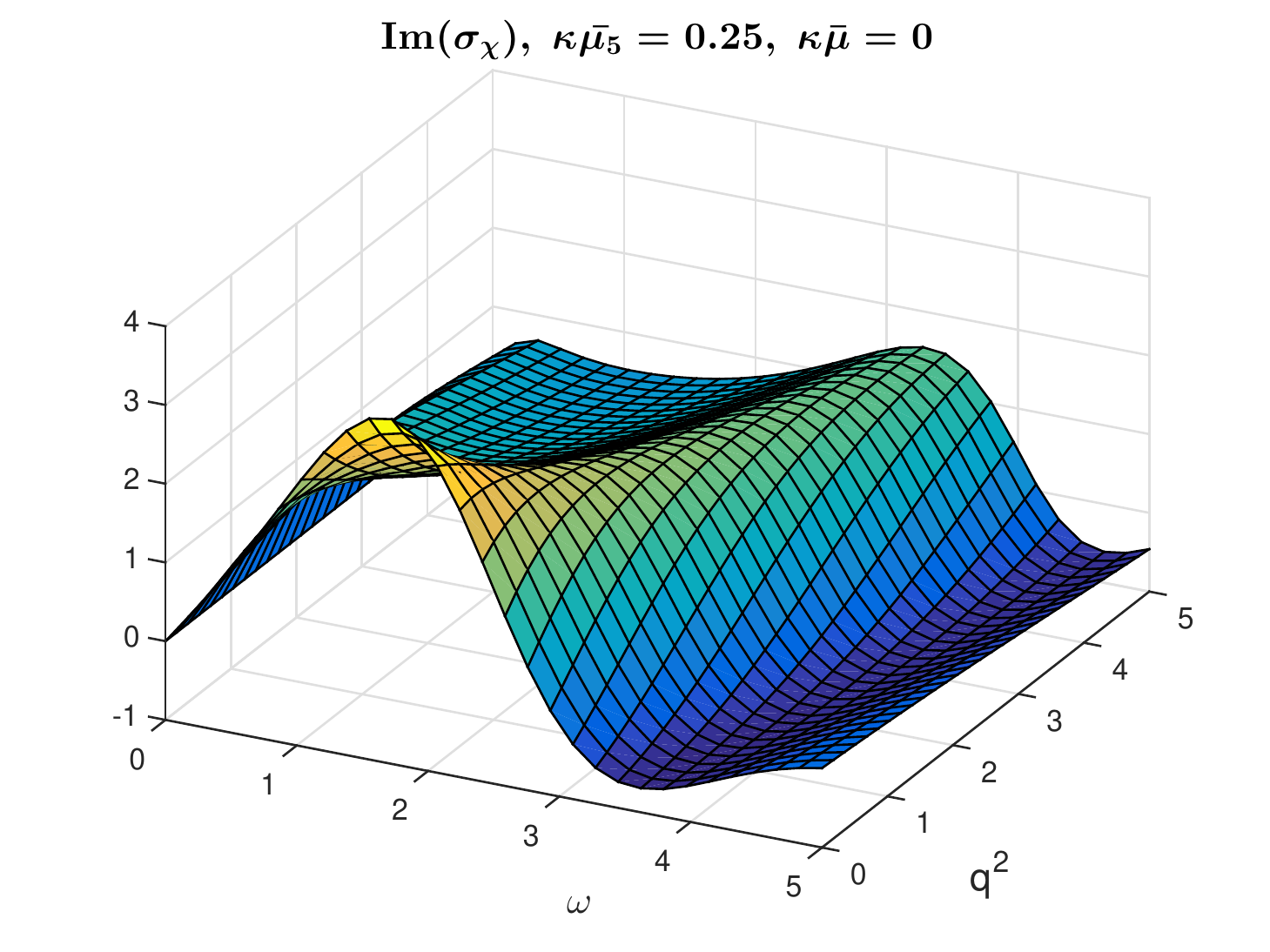}
\caption{Chiral magnetic conductivity $\sigma_\chi$ as function of $\omega$ and $q^2$ when $\kappa\bar{\mu}=0, \kappa\bar{\mu}_{_5}=1/4$.}
\label{cme1}
\end{figure}

\begin{figure}[htbp]
\centering
\includegraphics[scale=0.5]{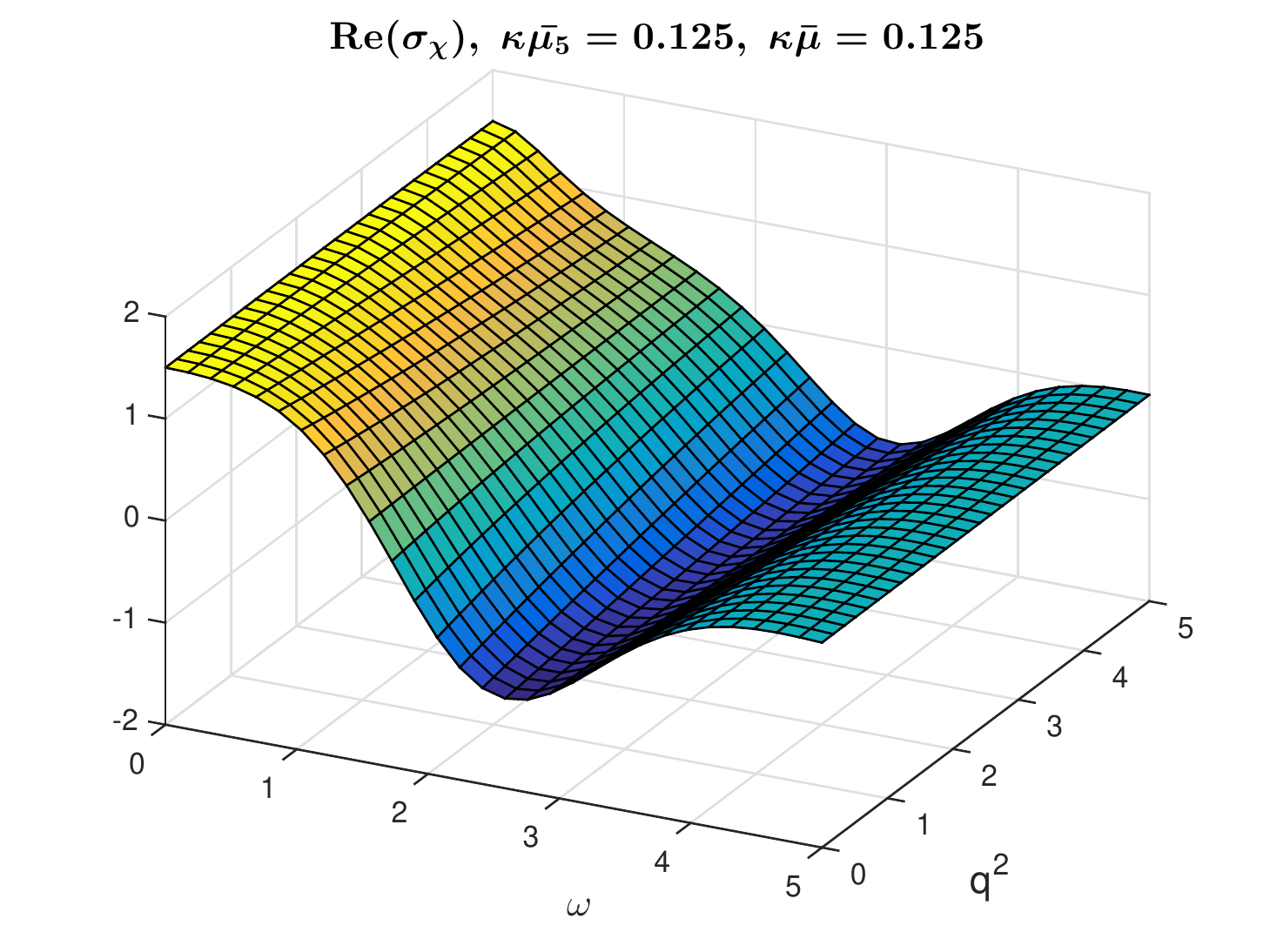}
\includegraphics[scale=0.5]{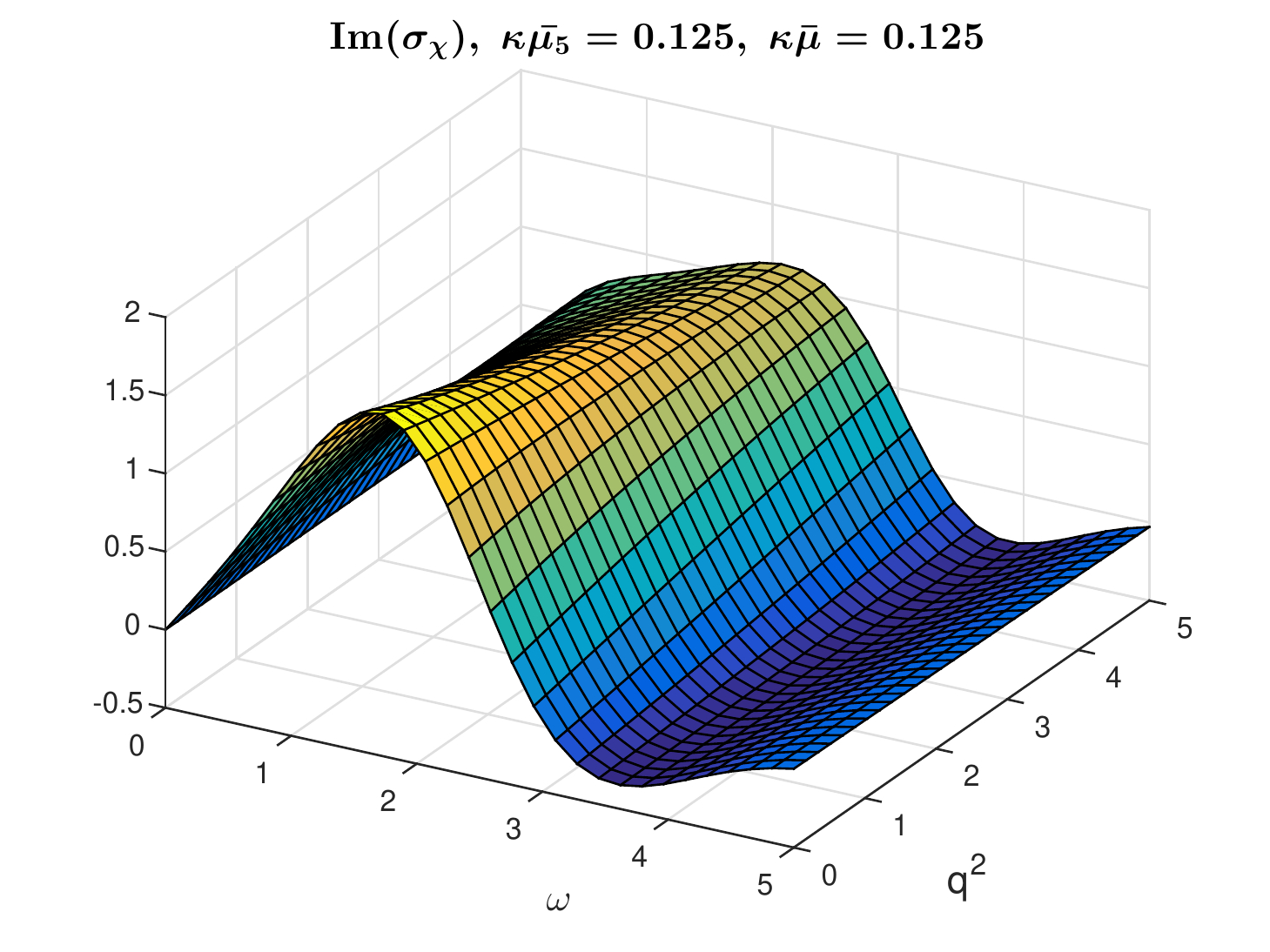}
\caption{Chiral magnetic conductivity $\sigma_\chi$ as function of $\omega$ and $q^2$ when $\kappa\bar{\mu}=\kappa\bar{\mu}_{_5}=1/8$.}
\label{cme2}
\end{figure}

3D plots of $\sigma_\chi$ for representative choices of $\kappa \bar{\mu},\kappa \bar{\mu}_{_5}$ are displayed in Figures~\ref{cme1},\ref{cme2}.
This is in contrast with \cite{1312.1204} which presented  $\textrm{Re}(\sigma_\kappa)$ for some unspecified values of $\bar{\mu},\bar{\mu}_{_5}$.
$\textrm{Re}(\sigma_\chi)$ of~\cite{1312.1204} exhibits similar behaviour to Figures~\ref{cme1},\ref{cme2}. Analogously to $\sigma_a$,
$\sigma_\chi$ approaches zero asymptotically at $\omega\simeq5$, after some damped oscillations. This be seen more clearly in 2D slices of $\sigma_\chi$ (Figures~\ref{cmewq}). Overall, $\omega$-dependence of $\sigma_\chi$ is quite in agreement with early results from holographic models
\cite{0908.4189,1102.4577,1312.1204} . The minimum (maximum) of $\textrm{Re}(\sigma_\chi)$ ($\textrm{Im}(\sigma_\chi)$) is reached at $\omega\simeq 2.6,q=0$ ($\omega\simeq 1.7,q=0$). We do not observe a drop in $\sigma_\chi$ when $\omega$ is away from $0$, which was attributed to the probe limit approximation~\cite{1312.1204}.
%
%\begin{figure}[htbp]
%\centering
%\includegraphics[width=0.48\textwidth]{eps//sigma_chi_q_mu5_025_mu_0}
%\includegraphics[width=0.48\textwidth]{eps//sigma_chi_q_mu5_0125_mu_0125}
%\caption{$q^2$-dependence of chiral magnetic conductivity $\sigma_\chi$ when $\omega=0$.}
%\label{cmeq}
%\end{figure}

\begin{figure}[htbp]
\centering
\includegraphics[width=0.48\textwidth]{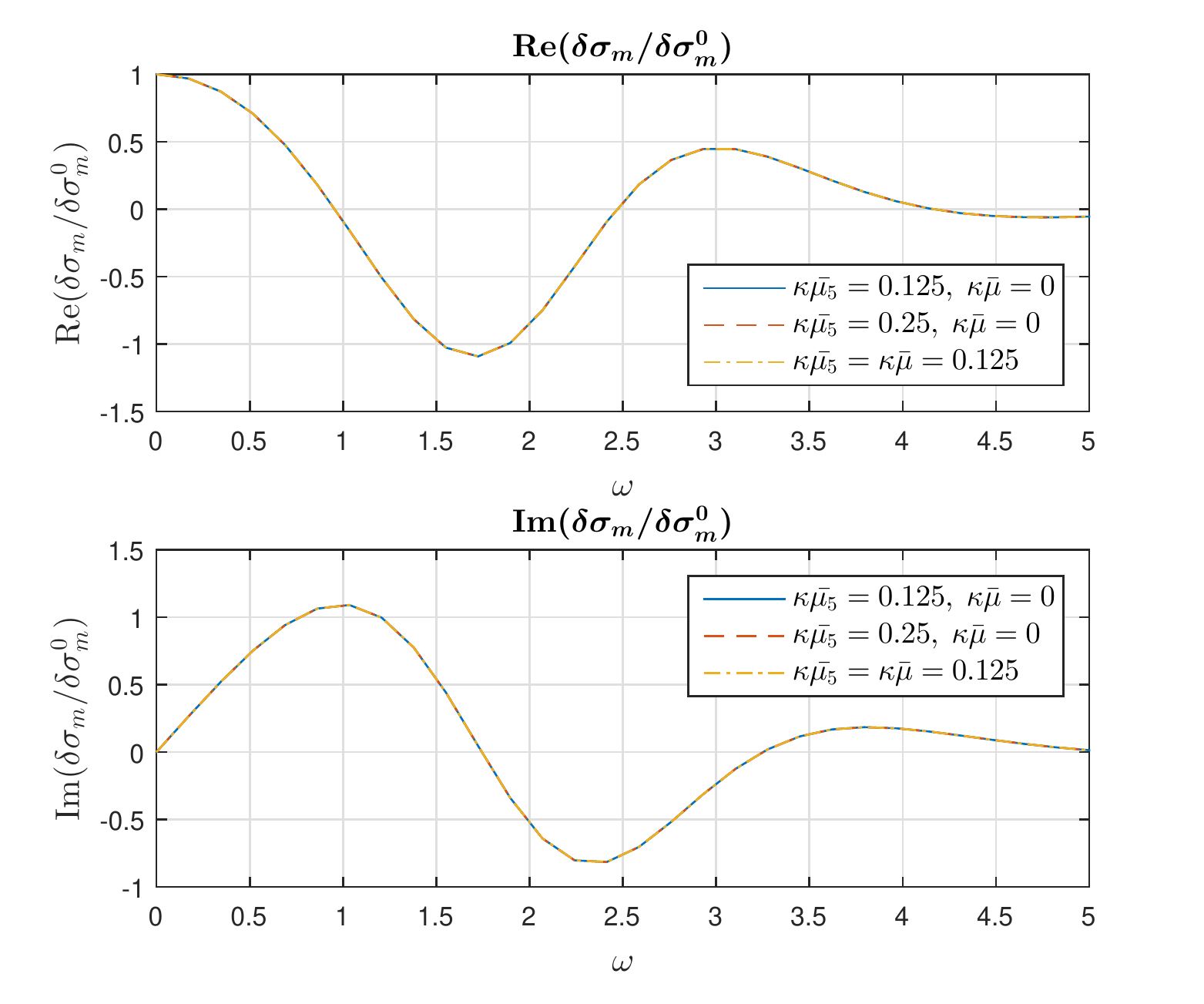}
\includegraphics[width=0.48\textwidth]{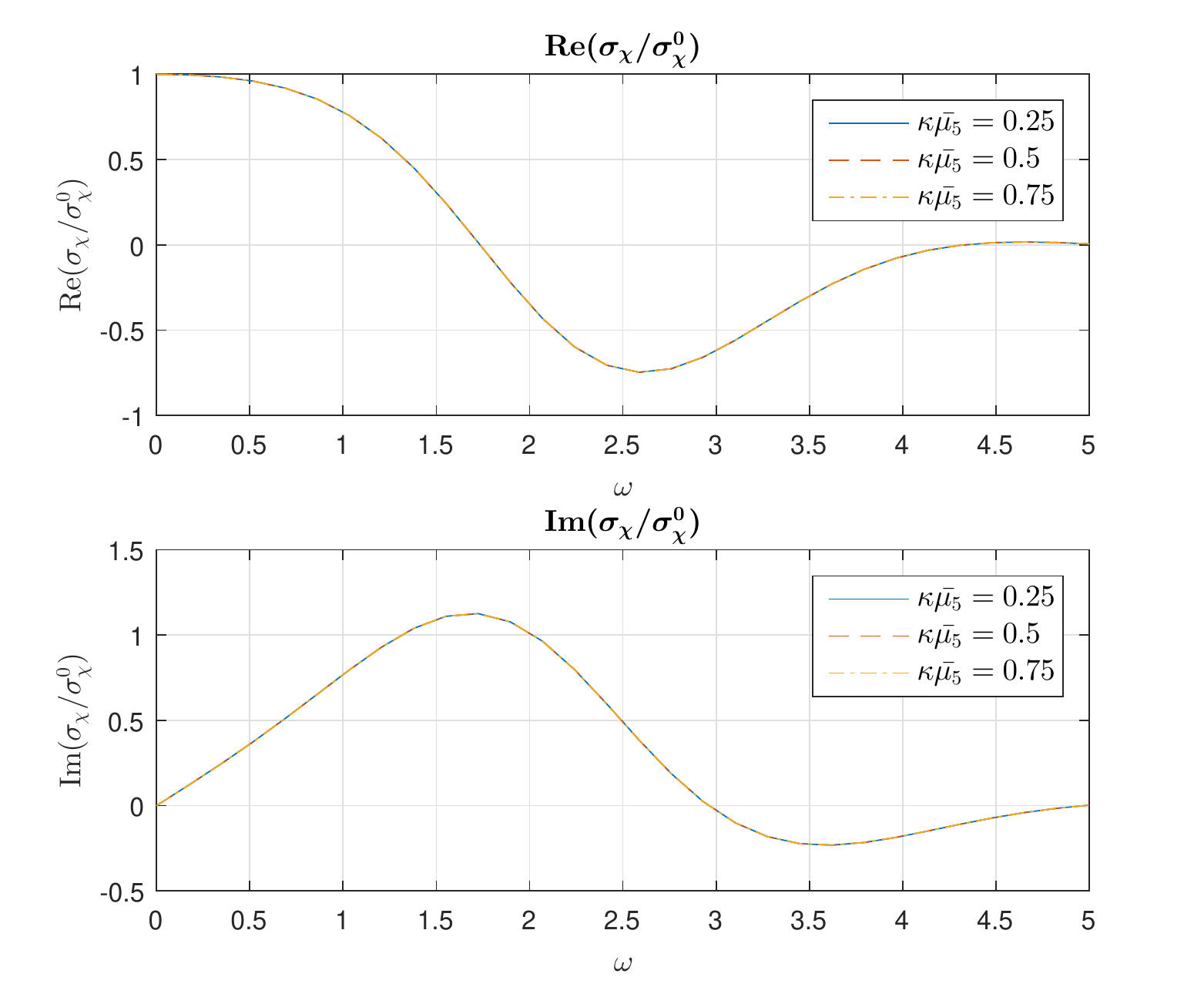}
\caption{$\omega$-dependence of $\delta\sigma_m/\delta\sigma_m^0$ (Left) and $\sigma_\chi/\sigma_\chi^0$ (Right) for different $\kappa \bar{\mu}, \kappa \bar{\mu}_{_5}$; $q=0$. For both left/right plots, the three curves overlap.}
\label{magnetic/cme}
\end{figure}

In Figure~\ref{magnetic/cme} we track the effect of vector/axial chemical potentials on the TCFs. We focus on $q=0$ slices and plot $\omega$-dependence of normalised quantities $\delta\sigma_m/\delta\sigma_m^0$, $\sigma_\chi/\sigma_\chi^0$ where $\delta\sigma_m^0$ and $\sigma_\chi^0$ are the corresponding DC limits. Here $\delta\sigma_m=\sigma_m-\sigma_m(\kappa \bar{\mu}=\kappa \bar{\mu}_{_5}=0)$.
% $\sigma_a$ and $\sigma_\chi$ have similar shapes.
As seen from Figure~\ref{magnetic/cme}, both $\delta\sigma_m/\delta\sigma_m^0$ and $\sigma_\chi/\sigma_\chi^0$ have no dependence on $\kappa\bar{\mu},\kappa \bar{\mu}_{_5}$ (all curves collapse into one). This implies a universal dependence on vector/axial chemical potentials (at $q=0$). Particularly, for $\sigma_\chi$ it is linear in $\kappa \bar{\mu}_5$. As for $\sigma_m$, its anomalous correction is linear in $(\bar{\mu}^2+\bar{\mu}_{_5}^2)$. Both features can actually be realised from the corresponding ODEs (\ref{eom 5}, \ref{eom 6}, \ref{eom 7}, \ref{eom 9}, \ref{eom 10}). When $q=0$, $\bar{V}_1=0$ and $V_1$ are not sensitive to the anomaly. So, $V_5$ (and thus $\sigma_\chi$) does linearly depend on $\kappa \bar{\mu}_{_5}$ and is not affected by $\kappa \bar{\mu}$. From (\ref{eom 9},\ref{eom 10}), $\bar{\mu} V_5=\bar{\mu}_{_5} \bar{V}_5$, which via (\ref{eom 7}) implies that anomalous correction to $V_3$ (and thus $\sigma_m$) is linear in $(\bar{\mu}^2+\bar{\mu}_{_5}^2)$.

\section{Study II: nonlinear transport induced by constant external fields}

\label{section5}

In this section we turn on constant backgrounds for external fields, corresponding to the linearisation scheme (\ref{study2}),
\begin{equation}
\begin{split}
\rho(x)&=\bar{\rho}+\epsilon\, \delta\rho(x),~~~~~~~~~~~~\rho_{_5}(x)=\bar{\rho}_{_5}+ \epsilon \, \delta\rho_{_5}(x),\\
\mathcal{V}_\mu(x)&=\bar{\mathcal{V}}_\mu(x)+\epsilon\, \delta\mathcal{V}_\mu(x), ~~~\mathcal{A}_\mu(x)=\bar{\mathcal{A}}_\mu(x)+\epsilon\, \delta\mathcal{A}_\mu(x),
\end{split}
\end{equation}
where $\bar{\rho}$ and $\bar{\rho}_{_5}$ are  treated as constants. $\bar{\mathcal{V}}_\mu$ and $\bar{\mathcal{A}}_\mu$ depend linearly on $x_\alpha$ so that their field strengths $\bar{\mathcal{F}}^V_{\mu\nu}$ and $\bar{\mathcal{F}}^a_{\mu\nu}$ are constant backgrounds $\vec{{\bf E}},\vec{\bf B},\vec{{\bf E}}^a,\vec{\bf B}^a$. The corrections $\mathbb{V}_\mu,\mathbb{A}_\mu$ of~(\ref{corrections}) are expanded to linear order in $\epsilon$.
\begin{equation} \label{corrections study2}
\mathbb{V}_\mu=\mathbb{V}_\mu^{(0)}+\epsilon \mathbb{V}_\mu^{(1)},~~~~~ \mathbb{A}_\mu= \mathbb{A}_\mu^{(0)}+\epsilon \mathbb{A}_\mu^{(1)}.
\end{equation}

\subsection{Solutions for $\mathbb{V}_\mu^{(0)}$ and $\mathbb{A}_\mu^{(0)}$} \label{subsection51}

We first ignore any derivative corrections.
To order $\mathcal{O}\left(\epsilon^0\right)$, the external fields are constant. Under the frame convention~(\ref{Landau frame}), the corrections $\mathbb{V}_\mu^{(0)}$ and $\mathbb{A}_\mu^{(0)}$ are time-independent and homogeneous, depending on the radial coordinate $r$ only. So, at order $\mathcal{O}\left(\epsilon^0\right)$ the dynamical equations~(\ref{eom Vt}-\ref{eom Ai}) are
\begin{equation}\label{eom Vt0 simp}
0=r^3\partial_r^2 \mathbb{V}_t^{(0)}+3r^2 \partial_r \mathbb{V}_t^{(0)}+ 12\kappa \left(\partial_r \mathbb{A}_k^{(0)} {\bf B}_k+ \partial_r\mathbb{V}_k^{(0)} {\bf B}_k^a \right),
\end{equation}
\begin{equation}\label{eom Vi0 simp}
\begin{split}
0&=(r^5-r)\partial_r^2 \mathbb{V}_i^{(0)}+(3r^4+1)\partial_r \mathbb{V}_i^{(0)}-r^2{\bf E}_i+12\kappa r^2{\bf B}_i \left(\partial_r \mathbb{A}_t^{(0)}+\frac{1}{r^3} \bar{\rho}_{_5} \right)\\
&+12\kappa r^2\epsilon^{ijk} \partial_r\mathbb{A}_j^{(0)} {\bf E}_k-12\kappa r^2 \epsilon^{ijk} \partial_r\mathbb{V}_k^{(0)} {\bf E}_j^a+12\kappa r^2 {\bf B}_i^a \left(\partial_r \mathbb{V}_t^{(0)}+\frac{1}{r^3}\bar{\rho}\right),
\end{split}
\end{equation}
\begin{equation}\label{eom At0 simp}
0=r^3\partial_r^2 \mathbb{A}_t^{(0)}+3r^2 \partial_r \mathbb{A}_t^{(0)}+ 12\kappa \left(\partial_r \mathbb{V}_k^{(0)} {\bf B}_k+ \partial_r\mathbb{A}_k^{(0)} {\bf B}_k^a \right),
\end{equation}
\begin{equation}\label{eom Ai0 simp}
\begin{split}
0&=(r^5-r)\partial_r^2 \mathbb{A}_i^{(0)}+(3r^4+1)\partial_r \mathbb{A}_i^{(0)}-r^2{\bf E}_i^a+12\kappa r^2{\bf B}_i \left(\partial_r \mathbb{V}_t^{(0)}+\frac{1}{r^3} \bar{\rho} \right)\\
&+12\kappa r^2\epsilon^{ijk} \partial_r\mathbb{V}_j^{(0)} {\bf E}_k+12\kappa r^2 \epsilon^{ijk} \partial_r\mathbb{A}_j^{(0)} {\bf E}_k^a+12\kappa r^2 {\bf B}_i^a \left(\partial_r \mathbb{A}_t^{(0)}+\frac{1}{r^3}\bar{\rho}_{_5}\right).
\end{split}
\end{equation}
Under the frame choice~(\ref{Landau frame}), the analysis in~(\ref{asmp cov1}-\ref{asmp cov4}) indicates
\begin{equation}
\mathbb{V}_i^{(0)},\mathbb{A}_i^{(0)}\sim \mathcal{O}\left(\frac{1}{r}\right),~~~ \mathbb{V}_t^{(0)},\mathbb{A}_t^{(0)}\sim \mathcal{O}\left(\frac{\log r}{r^3}\right),~~~ \textrm{as}~~~r\to \infty.
\end{equation}
So, (\ref{eom Vt0 simp}-\ref{eom Ai0 simp}) can be rewritten in  integral forms,
\begin{equation} \label{Vt0 formal}
\mathbb{V}_t^{(0)}(r)=12\kappa \int_r^\infty \frac{dx}{x^3} \left[\mathbb{A}_k^{(0)}(x) {\bf B}_k +\mathbb{V}_k^{(0)}(x) {\bf B}_k^a\right]\underrightarrow{r\to \infty}~ \mathcal{O}\left(\frac{1}{r^3}\right),
\end{equation}
\begin{equation} \label{At0 formal}
\mathbb{A}_t^{(0)}(r)=12\kappa \int_r^\infty \frac{dx}{x^3} \left[\mathbb{V}_k^{(0)}(x) {\bf B}_k +\mathbb{A}_k^{(0)}(x) {\bf B}_k^a\right]\underrightarrow{r\to \infty}~ \mathcal{O}\left(\frac{1}{r^3}\right),
\end{equation}
\begin{equation} \label{Vi0 formal}
\begin{split}
\mathbb{V}_i^{(0)}(r)&=-\frac{1}{4}\left(\pi-2 \arctan(r)+\log\frac{(1+r)^2}{1+r^2} \right) {\bf E}_i\\
&+\int_r^\infty \frac{12\kappa x}{x^4-1}\left\{\left[\mathbb{A}_t^{(0)}(x)- \frac{\bar{\rho}_{_5}}{2x^2}+\mu_{_5}\right]{\bf B}_i +\left[\mathbb{V}_t^{(0)}(x)- \frac{\bar{\rho}}{2x^2}+\mu\right]{\bf B}_i^a\right\} dx\\
&+12\kappa \epsilon^{ijk} \int_r^\infty \frac{xdx}{x^4-1} \left\{\left[ \mathbb{A}_j^{(0)}(x) - \mathbb{A}_j^{(0)}(1) \right]{\bf E}_k- \left[\mathbb{V}_k^{(0)}(x) - \mathbb{V}_k^{(0)}(1) \right]{\bf E}_j^a\right\}\\
&\underrightarrow{r\to\infty}~-\left(\frac{1}{r}-\frac{1}{2r^2}\right) {\bf E}_i+ \frac{6\kappa \mu_{_5} {\bf B}_i}{r^2}+ \frac{6\kappa \mu {\bf B}_i^a}{r^2} -\frac{6\kappa}{r^2}\epsilon^{ijk}\mathbb{A}_j^{(0)}(1) {\bf E}_k +\frac{6\kappa}{r^2}\epsilon^{ijk}\\
&~~~~~~~~~~\times\mathbb{V}_k^{(0)}(1) {\bf E}_j^a+ \mathcal{O} \left(\frac{1}{r^3}\right),
\end{split}
\end{equation}
\begin{equation} \label{Ai0 formal}
\begin{split}
\mathbb{A}_i^{(0)}(r)&=-\frac{1}{4}\left(\pi-2 \arctan(r)+\log\frac{(1+r)^2}{1+r^2} \right) {\bf E}_i^a\\
&+  \int_r^\infty \frac{12\kappa x}{x^4-1}\left\{\left[\mathbb{V}_t^{(0)}(x)- \frac{\bar{\rho}}{2x^2}+\mu\right]{\bf B}_i +\left[\mathbb{A}_t^{(0)}(x)- \frac{\bar{\rho}_{_5}}{2x^2}+\mu_{_5}\right]{\bf B}_i^a \right\}dx\\
&+12\kappa \epsilon^{ijk} \int_r^\infty \frac{xdx}{x^4-1} \left\{ \left[ \mathbb{V}_j^{(0)}(x) - \mathbb{V}_j^{(0)}(1) \right]{\bf E}_k+ \left[\mathbb{A}_j^{(0)}(x) - \mathbb{A}_j^{(0)}(1) \right]{\bf E}_k^a\right\}\\
&\underrightarrow{r\to\infty}~-\left(\frac{1}{r}-\frac{1}{2r^2}\right) {\bf E}_i^a+ \frac{6\kappa \mu {\bf B}_i}{r^2}+ \frac{6\kappa \mu_{_5} {\bf B}_i^a}{r^2} -\frac{6\kappa}{r^2}\epsilon^{ijk}\mathbb{V}_j^{(0)}(1) {\bf E}_k -\frac{6\kappa}{r^2} \epsilon^{ijk}\\
&~~~~~~~~~~\times\mathbb{A}_j^{(0)}(1) {\bf E}_k^a+ \mathcal{O} \left(\frac{1}{r^3}\right),
\end{split}
\end{equation}
where using the definition~(\ref{def potentials}), the chemical potentials $\mu,\mu_{_5}$ are
\begin{equation}
\mu=\frac{1}{2}\bar{\rho}-\mathbb{V}_t^{(0)}(1),~~~~~~~~~~~~~
\mu_{_5}=\frac{1}{2}\bar{\rho}_{_5}-\mathbb{A}_t^{(0)}(1).
\end{equation}
From~(\ref{bdry currents}), the boundary currents are
\begin{equation} \label{jmu EB exact1}
\begin{split}
J^t_{(0)}=\bar{\rho}+6\kappa \epsilon^{tklm}\bar{\mathcal{A}}_k \bar{\mathcal{F}}^{V}_{lm},~~~~
J^i_{(0)}&={\bf E}_i+12\kappa \mu_{_5} {\bf B}_i+ 12\kappa \mu {\bf B}_i^a -12\kappa \epsilon^{ijk}\mathbb{A}_j^{(0)}(1) {\bf E}_k \\
&- 12\kappa \epsilon^{ijk} \mathbb{V}_j^{(0)}(1) {\bf E}_k^a+ 6\kappa \epsilon^{i\nu\rho\lambda} \bar{\mathcal{A}_\nu} \bar{\mathcal{F}}^V_{\rho\lambda},
\end{split}
\end{equation}
\begin{equation} \label{jmu5 EB exact1}
\begin{split}
J^t_{5(0)}=\bar{\rho}_{_5}+2\kappa \epsilon^{tklm}\bar{\mathcal{A}}_k \bar{\mathcal{F}}^{a}_{lm},~~~
J^i_{5(0)}&={\bf E}^a_i+12\kappa \mu_{_5} {\bf B}^a_i+ 12\kappa \mu {\bf B}_i -12\kappa\epsilon^{ijk}\mathbb{V}_j^{(0)}(1) {\bf E}_k \\
&- 12\kappa\epsilon^{ijk} \mathbb{A}_j^{(0)}(1) {\bf E}_k^a+ 2\kappa \epsilon^{i\nu\rho\lambda} \bar{\mathcal{A}_\nu} \bar{\mathcal{F}}^a_{\rho\lambda},
\end{split}
\end{equation}
which reduce to~(\ref{jmu EB exact},\ref{jmu5 EB exact}) when $\bar{\mathcal{A}}_\mu=0$.

Despite the fact that the equations for $\mathbb{V}_i^{(0)},~ \mathbb{A}_i^{(0)}$ are linear, the solutions involve  complex inverse propagators,
which are non-linear functions of the background fields.
To proceed, we resort to  a weak field approximation by introducing yet another formal expansion parameter $\alpha$,
\begin{equation}
\bar{\mathcal{F}}_{\mu\nu}^{V}\to \alpha \bar{\mathcal{F}}_{\mu\nu}^{V},~~~~~~~~~~~~ \bar{\mathcal{F}}_{\mu\nu}^{a} \to \alpha \bar{\mathcal{F}}_{\mu\nu}^{a}.
\end{equation}
Accordingly, $\mathbb{V}_\mu^{(0)}$ and $\mathbb{A}_\mu^{(0)}$ are formally expanded in powers of $\alpha$,
\begin{equation}
\mathbb{V}_\mu^{(0)}= \sum_{n=1}^\infty \alpha^n \mathbb{V}_\mu^{(0)(n)},~~~~~~~~~ \mathbb{A}_\mu^{(0)}= \sum_{n=1}^\infty \alpha^n \mathbb{A}_\mu^{(0)(n)}.
\end{equation}
We analytically solved~(\ref{eom Vt0 simp}-\ref{eom Ai0 simp}) up to order $\mathcal{O}(\epsilon^0 \alpha^2)$. The results are summarised below.
\begin{equation} \label{VtAt01}
\mathbb{V}_t^{(0)(1)}=\mathbb{A}_t^{(0)(1)}=0.
\end{equation}
\begin{equation} \label{Vi01}
\mathbb{V}_i^{(0)(1)}=-\frac{1}{4}\left[\log \frac{(1+r)^2}{1+r^2}-2 \arctan(r)+\pi \right]{\bf E}_i+ 3\kappa \log \frac{1+r^2}{r^2}\left(\bar{\rho}_{_5} {\bf B}_i+ \bar{\rho} {\bf B}_i^a\right).
\end{equation}
\begin{equation} \label{Ai01}
\mathbb{A}_i^{(0)(1)}=-\frac{1}{4}\left[\log \frac{(1+r)^2}{1+r^2}-2 \arctan(r)+\pi \right]{\bf E}_i^a+ 3\kappa \log \frac{1+r^2}{r^2}\left(\bar{\rho} {\bf B}_i+ \bar{\rho}_{_5} {\bf B}_i^a\right).
\end{equation}
\begin{equation} \label{Vt02}
\begin{split}
\mathbb{V}_t^{(0)(2)}&=-\int_r^\infty \frac{dx}{x^3} \int_x^\infty \frac{y dy} {(y^2+1)(y+1)} \left(12\kappa \vec{\bf B} \cdot \vec{\bf E}^a+ 12\kappa \vec{\bf B}^a \cdot \vec{\bf E}\right)\\
&+\int_r^\infty \frac{dx}{x^3} \int_x^\infty \frac{dy} {y(y^2+1)} \left[72\kappa^2\bar{\rho} {\bf B}^2 + 144\kappa^2 \bar{\rho}_{_5} \vec{\bf B} \cdot \vec{\bf B}^a + 72 \kappa^2 \bar{\rho} ({\bf B}^a)^2\right].
\end{split}
\end{equation}
\begin{equation} \label{Vi02}
\mathbb{V}_i^{(0)(2)}=-\int_r^\infty \frac{xdx}{x^4-1} \int_1^x \frac{72 \kappa^2 \epsilon^{ijk} dy}{y(y^2+1)}  \left(\bar{\rho} {\bf B}_j {\bf E}_k + \bar{\rho}_{_5} {\bf B}_j^a {\bf E}_k + \bar{\rho}_{_5} {\bf B}_j {\bf E}_k^a+ \bar{\rho} {\bf B}_j^a {\bf E}_k^a \right).
\end{equation}
\begin{equation} \label{At02}
\begin{split}
\mathbb{A}_t^{(0)(2)}&=-\int_r^\infty \frac{dx}{x^3} \int_x^\infty \frac{y dy} {(y^2+1)(y+1)} \left(12\kappa \vec{\bf B} \cdot \vec{\bf E}+ 12\kappa \vec{\bf B}^a \cdot \vec{\bf E}^a\right)\\
&+\int_r^\infty \frac{dx}{x^3} \int_x^\infty \frac{dy} {y(y^2+1)} \left[72\kappa^2 \bar{\rho}_{_5} {\bf B}^2 + 144 \kappa^2 \bar{\rho} \vec{\bf B} \cdot \vec{\bf B}^a + 72 \kappa^2 \bar{\rho}_{_5} ({\bf B}^a)^2\right].
\end{split}
\end{equation}
\begin{equation} \label{Ai02}
\mathbb{A}_i^{(0)(2)}=-\int_r^\infty \frac{xdx}{x^4-1} \int_1^x \frac{72\kappa^2 \epsilon^{ijk} dy}{y(y^2+1)} \left(\bar{\rho}_{_5} {\bf B}_j {\bf E}_k + \bar{\rho} {\bf B}_j^a {\bf E}_k + \bar{\rho} {\bf B}_j {\bf E}_k^a+ \bar{\rho}_{_5} {\bf B}_j^a {\bf E}_k^a \right).
\end{equation}
These solutions generate the perturbative expansion of (\ref{jmu EB exact1},\ref{jmu5 EB exact1})
\begin{equation}
\begin{split}
J^i_{(0)}&={\bf E}_i+12\kappa \mu_{_5}{\bf B}_i+ 12\kappa \mu{\bf B}_i^a+ 6\kappa \epsilon^{i\nu\rho\lambda}\bar{\mathcal{A}}_\nu\bar{\mathcal{F}}_{\rho\lambda}^V \\
&-72\log 2~\kappa^2 \left(\mu \vec{\bf B}\times \vec{\bf E}+ \mu_{_5} \vec{\bf B}^a\times \vec{\bf E}+ \mu_{_5}\vec{\bf B} \times \vec{\bf E}^a +\mu \vec{\bf B}^a \times \vec{\bf E}^a\right)_i\\
&+18\pi^2\kappa^3 \left\{\left[\mu_{_5}\vec{\bf B}\times \vec{\bf E} +\mu \vec{\bf B}^a \times \vec{\bf E} +\mu \vec{\bf B} \times \vec{\bf E}^a + \mu_{_5} \vec{\bf B}^a\times \vec{\bf E}^a\right] \times \vec{\bf E} \right. \\
&~~~~~~~~~~~~\left.+\left[\mu \vec{\bf B} \times \vec{\bf E} + \mu_{_5} \vec{\bf B}^a \times \vec{\bf E} +\mu_{_5} \vec{\bf B} \times \vec{\bf E}^a +\mu \vec{\bf B}^a \times \vec{\bf E}^a \right] \times \vec{\bf E}^a \right\}_i+\cdots,
\end{split}
\end{equation}
\begin{equation}
\begin{split}
J^i_{5(0)}&={\bf E}^a_i+12\kappa \mu_{_5} {\bf B}^a_i+ 12\kappa \mu {\bf B}_i + 2 \kappa \epsilon^{i\nu\rho\lambda}\bar{\mathcal{A}}_\nu \bar{\mathcal{F}}_{\rho\lambda}^a \\
&-72\log 2~\kappa^2 \left(\mu_{_5} \vec{\bf B}\times \vec{\bf E}+ \mu \vec{\bf B}^a\times \vec{\bf E}+ \mu \vec{\bf B} \times \vec{\bf E}^a \times +\mu_{_5} \vec{\bf B}^a \times \vec{\bf E}^a\right)_i\\
&+18\pi^2\kappa^3 \left\{\left[\mu\vec{\bf B}\times \vec{\bf E} +\mu_{_5} \vec{\bf B}^a \times \vec{\bf E} +\mu_{_5} \vec{\bf B} \times \vec{\bf E}^a + \mu \vec{\bf B}^a \times \vec{\bf E}^a \right]\times\vec{\bf E} \right. \\
&~~~~~~~~~~~~\left.+\left[\mu_{_5} \vec{\bf B} \times \vec{\bf E} + \mu \vec{\bf B}^a \times \vec{\bf E} +\mu \vec{\bf B} \times \vec{\bf E}^a +\mu_{_5}\vec{\bf B}^a \times \vec{\bf E}^a \right] \times \vec{\bf E}^a \right\}_i+\cdots,
\end{split}
\end{equation}
where
\begin{equation}
\begin{split}
\mu&=\frac{1}{2}\bar{\rho}+\frac{3}{2}\left(\pi-2\log 2\right)\kappa \left(\vec{\bf B} \cdot \vec{\bf E}^a + \vec{\bf B}^a \cdot \vec{\bf E}\right) +18\left(1-2\log 2\right) \kappa^2\\
&\times \left(\bar{\rho} {\bf B}^2 + 2 \bar{\rho}_{_5} \vec{\bf B} \cdot {\bf B}^a+ \bar{\rho} \left({\bf B}^a\right)^2\right)+\cdots,
\end{split}
\end{equation}
\begin{equation}
\begin{split}
\mu_{_5}&=\frac{1}{2}\bar{\rho}_{_5}+\frac{3}{2}\left(\pi-2\log 2\right)\kappa \left(\vec{\bf B} \cdot \vec{\bf E} + \vec{\bf B}^a \cdot \vec{\bf E}^a\right) +18\left(1-2\log 2\right)\kappa^2\\
&\times \left(\bar{\rho}_{_5} {\bf B}^2 + 2 \bar{\rho} \vec{\bf B} \cdot {\bf B}^a+ \bar{\rho}_{_5} \left({\bf B}^a\right)^2\right)+\cdots.
\end{split}
\end{equation}
When $\bar{\mathcal{A}}_\mu=0$, the above constitutive relations reduce to ~(\ref{jmu EB},\ref{jmu5 EB},\ref{chem/den2}).

\subsection{Solutions for $\mathbb{V}_\mu^{(1)}$ and $\mathbb{A}_\mu^{(1)}$} \label{subsection52}

The corrections $\mathbb{V}_\mu^{(1)}$ and $\mathbb{A}_\mu^{(1)}$ in (\ref{corrections study2}) are also expandable in powers of $\alpha$,
\begin{equation}
\mathbb{V}_\mu^{(1)}=\sum_{n=0}^\infty \alpha^n \mathbb{V}_\mu^{(1)(n)},~~~~~~
\mathbb{A}_\mu^{(1)}=\sum_{n=0}^\infty \alpha^n \mathbb{A}_\mu^{(1)(n)}.
\end{equation}
At the lowest order $\mathcal{O}\left(\epsilon^1\alpha^0\right)$, the dynamical equations (\ref{eom Vt}-\ref{eom Ai}) are exactly (\ref{eom Vt scheme1}-\ref{eom Ai scheme1}) with $\mathcal{V}_\mu,\mathcal{A}_\mu$ in (\ref{eom Vt scheme1}-\ref{eom Ai scheme1}) replaced by $\delta \mathcal{V}_\mu, \delta \mathcal{A}_\mu$. Therefore, solutions for $\mathbb{V}_\mu^{(1)(0)}$ and $\mathbb{A}_\mu^{(1)(0)}$ are
\begin{equation}
\mathbb{V}_t^{(1)(0)}=S_2\partial_t^{-1} \partial_k \left(\partial_t \delta \mathcal{V}_k -\partial_k \delta \mathcal{V}_t\right)+ S_3 \delta\rho,
\end{equation}
\begin{equation}
\begin{split}
\mathbb{V}_i^{(1)(0)}&=V_2\left(\partial_i \delta \mathcal{V}_t-\partial_t \delta \mathcal{V}_i \right)+V_3\partial_k\left(\partial_i \delta \mathcal{V}_k -\partial_k \delta\mathcal{V}_i \right)+V_4 \partial_i \delta\rho\\
&+V_5\epsilon^{ijk}\partial_j \delta \mathcal{V}_k+\bar{V}_3\partial_k \left(\partial_i \delta \mathcal{A}_k- \partial_k \delta \mathcal{A}_i\right)+ \bar{V}_5 \epsilon^{ijk} \partial_j \delta \mathcal{A}_k,
\end{split}
\end{equation}
\begin{equation}
\mathbb{A}_t^{(1)(0)}=S_2\partial_t^{-1} \partial_k \left(\partial_t \delta \mathcal{A}_k -\partial_k \delta \mathcal{A}_t\right)+ S_3 \delta\rho_{_5},
\end{equation}
\begin{equation}
\begin{split}
\mathbb{A}_i^{(1)(0)}&=V_2\left(\partial_i \delta \mathcal{A}_t-\partial_t \delta \mathcal{A}_i \right)+V_3\partial_k\left(\partial_i \delta \mathcal{A}_k -\partial_k \delta\mathcal{A}_i \right)+V_4 \partial_i \delta\rho_{_5}\\
&+V_5\epsilon^{ijk}\partial_j \delta \mathcal{A}_k+\bar{V}_3\partial_k \left(\partial_i \delta \mathcal{V}_k- \partial_k \delta \mathcal{V}_i\right)+ \bar{V}_5 \epsilon^{ijk} \partial_j \delta \mathcal{V}_k,
\end{split}
\end{equation}
where $S_3,V_2,V_4$ were studied in~\cite{1511.08789} while $S_2,V_3,V_5,\bar{V}_3,\bar{V}_5$ were adressed in section~\ref{subsection42}. As a result, at order $\mathcal{O}\left(\epsilon^1 \alpha^0\right)$, the boundary currents $J^\mu,J^\mu_5$ are
\begin{equation} \label{jmu/jmu5 re}
\begin{split}
&J^t=\delta\rho,~~~~~~
\vec{J}=-\mathcal{D}\vec{\nabla}\delta\rho +\sigma_e \delta \vec{E}+\sigma_m \vec{\nabla}\times \delta \vec{B}+\sigma_{\chi} \delta \vec{B}+\sigma_a \vec{\nabla}\times \delta \vec{B}^a+\sigma_\kappa \delta\vec{B}^a,\\
&J^t_5=\delta \rho_{_5},~~~~
\vec{J}_5=-\mathcal{D}\vec{\nabla}\delta \rho_{_5}+\sigma_e \delta \vec{E}^a+ \sigma_m \vec{\nabla} \times \delta\vec{B}^a+\sigma_\chi \delta\vec{B}^a+\sigma_a \vec{\nabla} \times \delta\vec{B}+\sigma_\kappa \delta\vec{B}.
\end{split}
\end{equation}
As $\alpha\to 0$, the above results coincide with those of section~\ref{section4}.

To the order $\mathcal{O}\left(\epsilon^1\alpha^1\right)$, the dynamical equations~(\ref{eom Vt}-\ref{eom Ai}) are
\begin{equation} \label{eom Vt1}
\begin{split}
0&=r^3\partial_r^2 \mathbb{V}_t^{(1)(1)}+3r^2 \partial_r \mathbb{V}_t^{(1)(1)}+ r\partial_r \partial_k \mathbb{V}_k^{(1)(1)}+12\kappa \epsilon^{ijk}\left(\partial_r \mathbb{A}_i^{(0)(1)} \partial_j \delta \mathcal{V}_k  \right.\\
&\left.+\partial_r \mathbb{A}_i^{(1)(0)} \partial_j \bar{\mathcal{V}}_k+ \partial_r \mathbb{A}_i^{(0)(1)} \partial_j \mathbb{V}_k^{(1)(0)}+\partial_r \mathbb{V}_i^{(0)(1)} \partial_j \delta\mathcal{A}_k+\partial_r \mathbb{V}_i^{(1)(0)} \partial_j \bar{\mathcal{A}}_k\right.\\
&\left.+\partial_r \mathbb{V}_i^{(0)(1)} \partial_j  \mathbb{A}_k^{(1)(0)}\right),
\end{split}
\end{equation}
\begin{equation} \label{eom Vi1}
\begin{split}
0&=(r^5-r)\partial_r^2 \mathbb{V}_i^{(1)(1)}+(3r^4+1)\partial_r \mathbb{V}_i^{(1)(1)}+ 2r^3 \partial_r \partial_t\mathbb{V}_i^{(1)(1)}-r^3\partial_r\partial_i \mathbb{V}_t^{(1)(1)}\\
&+r^2 \partial_r\left(\partial_t \mathbb{V}_i^{(1)(1)} -\partial_i \mathbb{V}_t^{(1)(1)} \right)+ r\left(\partial^2 \mathbb{V}_i^{(1)(1)}- \partial_i \partial_k \mathbb{V}_k^{(1)(1)} \right) \\
& + 12 \kappa r^2 \epsilon^{ijk}\left\{\frac{1}{r^3}\delta \rho_{_5} \partial_j \bar{\mathcal{V}}_k+\frac{1}{r^3} \bar{\rho}_{_5} \partial_j \mathbb{V}_k^{(1)(1)} +\partial_r \mathbb{A}_t^{(1)(0)} \partial_j\bar{\mathcal{V}}_k \right.\\
&\left.- \partial_r \mathbb{A}_j^{(0)(1)} \left[\left(\partial_t\delta\mathcal{V}_k-\partial_k \delta \mathcal{V}_t \right) + \left(\partial_t \mathbb{V}_k^{(1)(0)}-\partial_k \mathbb{V}_t^{(1)(0)} \right)+ \frac{1}{2r^2}\partial_k \delta\rho\right]\right.\\
&\left.-\partial_r \mathbb{A}_j^{(1)(0)}\left(\partial_t\bar{\mathcal{V}}_k- \partial_k \bar{\mathcal{V}}_t\right)+ \partial_r \mathbb{V}_k^{(0)(1)} \left[\left(\partial_t \delta \mathcal{A}_j- \partial_j \delta \mathcal{A}_t\right)+\frac{1}{2r^2}\partial_j \delta \rho_{_5}\right.\right.\\
&\left.\left.+ \left(\partial_t \mathbb{A}_j^{(1)(0)} -\partial_j \mathbb{A}_t^{(1)(0)} \right) \right]+ \partial_r \mathbb{V}_k^{(1)(0)}\left(\partial_t \bar{\mathcal{A}}_j- \partial_j \bar{\mathcal{A}}_t \right)+\frac{1}{r^3}  \delta \rho \partial_j \bar{\mathcal{A}}_k\right.\\
&\left.+\frac{1}{r^3}\bar{\rho} \partial_j \mathbb{A}_k^{(1)(1)} + \partial_r \mathbb{V}_t^{(1)(0)} \partial_j \bar{\mathcal{A}}_k\right\},
\end{split}
\end{equation}
\begin{equation} \label{eom At1}
\begin{split}
0&=r^3\partial_r^2 \mathbb{A}_t^{(1)(1)}+3r^2 \partial_r \mathbb{A}_t^{(1)(1)}+ r\partial_r \partial_k \mathbb{A}_k^{(1)(1)}+12\kappa \epsilon^{ijk}\left(\partial_r \mathbb{V}_i^{(0)(1)} \partial_j \delta\mathcal{V}_k  \right.\\
&\left.+\partial_r \mathbb{V}_i^{(1)(0)} \partial_j \bar{\mathcal{V}}_k + \partial_r \mathbb{V}_i^{(0)(1)} \partial_j \mathbb{V}_k^{(1)(0)}+\partial_r \mathbb{A}_i^{(0)(1)} \partial_j \delta\mathcal{A}_k +\partial_r \mathbb{A}_i^{(1)(0)} \partial_j \bar{\mathcal{A}}_k \right.\\
&\left.+\partial_r \mathbb{A}_i^{(0)(1)} \partial_j \mathbb{A}_k^{(1)(0)}\right),
\end{split}
\end{equation}
\begin{equation} \label{eom Ai1}
\begin{split}
0&=(r^5-r)\partial_r^2 \mathbb{A}_i^{(1)(1)}+(3r^4+1)\partial_r \mathbb{A}_i^{(1)(1)}+ 2r^3 \partial_r \partial_t\mathbb{A}_i^{(1)}-r^3\partial_r\partial_i \mathbb{A}_t^{(1)(1)} \\
&+r^2 \partial_r\left(\partial_t \mathbb{A}_i^{(1)(1)} -\partial_i \mathbb{A}_t^{(1)(1)} \right) +r \left(\partial^2 \mathbb{A}_i^{(1)(1)}- \partial_i \partial_k \mathbb{A}_k^{(1)(1)}\right) \\
&+ 12 \kappa r^2 \epsilon^{ijk}\left\{\frac{1}{r^3}\delta \rho \partial_j \bar{\mathcal{V}}_k+\frac{1}{r^3} \bar{\rho} \partial_j \mathbb{V}_k^{(1)(1)}+\partial_r \mathbb{V}_t^{(1)(0)} \partial_j\bar{\mathcal{V}}_k\right.\\
&\left.- \partial_r \mathbb{V}_j^{(0)(1)} \left[\left(\partial_t \delta \mathcal{V}_k -\partial_k \delta \mathcal{V}_t \right) + \left(\partial_t \mathbb{V}_k^{(1)(0)}- \partial_k \mathbb{V}_t^{(1)(0)} \right)+\frac{1}{2r^2}\partial_k \delta\rho\right] \right.\\
&\left.-\partial_r \mathbb{V}_j^{(1)(0)}\left(\partial_t\bar{\mathcal{V}}_k- \partial_k \bar{\mathcal{V}}_t\right)- \partial_r \mathbb{A}_j^{(0)(1)} \left[\left(\partial_t \delta \mathcal{A}_k- \partial_k \delta \mathcal{A}_t\right)+\frac{1}{2r^2}\partial_k \delta \rho_{_5}\right.\right.\\
&\left.\left.+ \left(\partial_t \mathbb{A}_k^{(1)(0)} -\partial_k \mathbb{A}_t^{(1)(0)} \right)\right]- \partial_r \mathbb{A}_j^{(1)(0)}\left(\partial_t \bar{\mathcal{A}}_k- \partial_k \bar{\mathcal{A}}_t \right)+\frac{1}{r^3} \delta \rho_{_5} \partial_j \bar{\mathcal{A}}_k \right.\\
&\left. +\frac{1}{r^3}\bar{\rho}_{_5} \partial_j \mathbb{A}_k^{(1)(1)} + \partial_r \mathbb{A}_t^{(1)(0)} \partial_j \bar{\mathcal{A}}_k\right\}.
\end{split}
\end{equation}
The source terms in (\ref{eom Vt1}-\ref{eom Ai1}) introduce all the basic structures in solutions for $\mathbb{V}_{\mu}^{(1)(1)}$ and $\mathbb{A}_{\mu}^{(1)(1)}$, which then get propagated into the constitutive relations for the currents $J^\mu$ and $J^\mu_5$. The full solutions can be constructed in parallel with section~\ref{section4}.
%With $\mathbb{V}_{\mu}^{(0)(1)}$, $\mathbb{A}_{\mu}^{(0)(1)}$, $\mathbb{V}_{\mu}^{(1)(0)}$, $\mathbb{A}_{\mu}^{(1)(0)}$ at hand,
We have decided not to solve (\ref{eom Vt1}-\ref{eom Ai1}) in this publication and to leave a comprehensive study of $\mathbb{V}_{\mu}^{(1)(1)}$ and $\mathbb{A}_{\mu}^{(1)(1)}$ and corresponding transport coefficients for future work. Here,
we merely list all the basic structures that would emerge in the constitutive relations, as dictated by the source terms in (\ref{eom Vt1}-\ref{eom Ai1}).
In $\vec{J}$ we anticipate to have the following terms each multiplied by its own TCF,
\begin{equation} \label{structure of jmu}
\begin{split}
&\delta \rho_{_5} \vec{\bf B},~~\delta \rho\,\vec{\bf B}^a,~~\left(\vec{\nabla}\cdot \delta \vec{E}^a\right)\vec{\bf B},~~\left(\vec{\nabla}\cdot \delta \vec{E}\right) \vec{\bf B}^a,~~\vec{\bf E}^a\times \delta \vec{E},~~\vec{\bf E}^a\times \vec{\nabla} \delta\rho,\\
&\vec{\bf E}^a\times\left(\vec{\nabla}\times \delta \vec{B}\right),~~\vec{\bf E}^a\times \delta \vec{B},~~\vec{\bf E}^a\times \left(\vec{\nabla}\times \delta \vec{B}^a\right),~~ \vec{\bf E}^a \times \delta \vec{B}^a,\\
&\vec{\bf E}^a\times \vec{\nabla}\left( \vec{\nabla} \cdot \delta \vec{E}\right),~\left(\bar{\rho} \vec{\bf B} + \bar{\rho}_{_5} \vec{\bf B}^a\right)\times \delta \vec{E}, ~~\left(\bar{\rho} \vec{\bf B} + \bar{\rho}_{_5} \vec{\bf B}^a\right) \times\vec{\nabla} \delta\rho, \\
&\left(\bar{\rho} \vec{\bf B} + \bar{\rho}_{_5} \vec{\bf B}^a\right) \times \left(\vec{\nabla}\times \delta \vec{B}\right),~~
\left(\bar{\rho} \vec{\bf B} + \bar{\rho}_{_5} \vec{\bf B}^a\right)\times \delta \vec{B},~~
\left(\bar{\rho} \vec{\bf B} + \bar{\rho}_{_5} \vec{\bf B}^a\right)\times \left(\vec{\nabla}\times \delta \vec{B}^a\right),\\
&\left(\bar{\rho} \vec{\bf B} + \bar{\rho}_{_5} \vec{\bf B}^a\right)\times \delta \vec{B}^a,~\left(\bar{\rho} \vec{\bf B} + \bar{\rho}_{_5} \vec{\bf B}^a\right)\times \vec{\nabla}\left( \vec{\nabla} \cdot \delta \vec{E}\right),~~ \vec{\bf E}\times \delta \vec{E}^a,~~ \vec{\bf E}\times \vec{\nabla} \delta\rho_{_5}, \\
&\vec{\bf E} \times \left(\vec{\nabla} \times\delta \vec{B}^a\right),~~\vec{\bf E}\times\delta \vec{B}^a,~\vec{\bf E}\times \left(\vec{\nabla}\times \delta \vec{B}\right),~~ \vec{\bf E} \times \delta \vec{B},~~\vec{\bf E} \times \vec{\nabla} \left(\vec{\nabla}\cdot \delta \vec{E}^a\right),\\
&\left(\bar{\rho}_{_5} \vec{\bf B} + \bar{\rho} \vec{\bf B}^a\right) \times\delta \vec{E}^a,~\left(\bar{\rho}_{_5} \vec{\bf B} + \bar{\rho} \vec{\bf B}^a\right)\times \vec{\nabla} \delta\rho_{_5},~~
\left(\bar{\rho}_{_5} \vec{\bf B} + \bar{\rho} \vec{\bf B}^a\right)\times \left(\vec{\nabla} \times \delta \vec{B}^a\right),\\
&\left(\bar{\rho}_{_5} \vec{\bf B} + \bar{\rho} \vec{\bf B}^a\right)\times \delta \vec{B}^a,~\left(\bar{\rho}_{_5} \vec{\bf B} + \bar{\rho} \vec{\bf B}^a\right)\times \left(\vec{\nabla}\times \delta \vec{B}\right),~~
\left(\bar{\rho}_{_5} \vec{\bf B} + \bar{\rho} \vec{\bf B}^a\right)\times \delta \vec{B},\\
&\left(\bar{\rho}_{_5} \vec{\bf B} + \bar{\rho} \vec{\bf B}^a\right)\times \vec{\nabla} \left(\vec{\nabla}\cdot \delta \vec{E}^a\right).
\end{split}
\end{equation}
In the axial current $\vec{J}_5$,
\begin{equation}\label{structure of jmu5}
\begin{split}
&\delta \rho\,\vec{\bf B},~~\delta \rho_{_5}\vec{\bf B}^a,~~\left(\vec{\nabla}\cdot \delta \vec{E}\right)\vec{\bf B},~~\left(\vec{\nabla}\cdot \delta \vec{E}^a\right) \vec{\bf B}^a,~~\vec{\bf E}\times \delta \vec{E},~~\vec{\bf E}\times \vec{\nabla} \delta\rho,\\
&\vec{\bf E}\times\left(\vec{\nabla}\times \delta \vec{B}\right),~~\vec{\bf E}\times \delta \vec{B},~~\vec{\bf E}\times \left(\vec{\nabla}\times \delta \vec{B}^a\right),~~ \vec{\bf E} \times \delta \vec{B}^a,~~\vec{\bf E}\times \vec{\nabla}\left( \vec{\nabla} \cdot \delta \vec{E}\right),\\
&\left(\bar{\rho}_{_5} \vec{\bf B} + \bar{\rho} \vec{\bf B}^a\right)\times \delta \vec{E}, ~~
\left(\bar{\rho}_{_5} \vec{\bf B} + \bar{\rho} \vec{\bf B}^a\right) \times\vec{\nabla} \delta\rho,~~
\left(\bar{\rho}_{_5} \vec{\bf B} + \bar{\rho} \vec{\bf B}^a\right)\times \left(\vec{\nabla}\times \delta \vec{B}\right),\\
&\left(\bar{\rho}_{_5} \vec{\bf B} + \bar{\rho} \vec{\bf B}^a\right)\times \delta \vec{B},~~
\left(\bar{\rho}_{_5} \vec{\bf B} + \bar{\rho} \vec{\bf B}^a\right)\times \left(\vec{\nabla}\times \delta \vec{B}^a\right),~~
\left(\bar{\rho}_{_5} \vec{\bf B} + \bar{\rho} \vec{\bf B}^a\right)\times \delta \vec{B}^a,\\
&\left(\bar{\rho}_{_5} \vec{\bf B} + \bar{\rho} \vec{\bf B}^a\right)\times \vec{\nabla}\left( \vec{\nabla} \cdot \delta \vec{E}\right),~ \vec{\bf E}^a\times \delta \vec{E}^a,~ \vec{\bf E}^a\times \vec{\nabla} \delta\rho_{_5},~ \vec{\bf E}^a \times \left(\vec{\nabla} \times\delta \vec{B}^a\right),\\
&\vec{\bf E}^a\times\delta \vec{B}^a, ~~\vec{\bf E}^a\times \left(\vec{\nabla}\times \delta \vec{B}\right),~~ \vec{\bf E}^a \times \delta \vec{B},~~\vec{\bf E}^a \times \vec{\nabla} \left(\vec{\nabla}\cdot \delta \vec{E}^a\right),\\
&\left(\bar{\rho} \vec{\bf B} + \bar{\rho}_{_5} \vec{\bf B}^a\right) \times\delta \vec{E}^a,~
\left(\bar{\rho} \vec{\bf B} + \bar{\rho}_{_5} \vec{\bf B}^a\right)\times \vec{\nabla} \delta\rho_{_5},~~
\left(\bar{\rho} \vec{\bf B} + \bar{\rho}_{_5} \vec{\bf B}^a\right)\times \left(\vec{\nabla} \times \delta \vec{B}^a\right),\\
&\left(\bar{\rho} \vec{\bf B} + \bar{\rho}_{_5} \vec{\bf B}^a\right)\times \delta \vec{B}^a,~
\left(\bar{\rho} \vec{\bf B} + \bar{\rho}_{_5} \vec{\bf B}^a\right)\times \left(\vec{\nabla}\times \delta \vec{B}\right),~~
\left(\bar{\rho} \vec{\bf B} + \bar{\rho}_{_5} \vec{\bf B}^a\right)\times \delta \vec{B},\\
&\left(\bar{\rho} \vec{\bf B} + \bar{\rho}_{_5} \vec{\bf B}^a\right)\times \vec{\nabla} \left(\vec{\nabla}\cdot \delta \vec{E}^a\right).
\end{split}
\end{equation}

%By basic structures, we mean each term in~(\ref{structure of jmu},\ref{structure of jmu5}) will generate infinitely many ones through a scalar functional like $\mathcal{D}[\partial_t,\vec{\partial}^2]$ acting on this term.
The terms $\delta \rho_{_5}\vec{\bf B},\delta\rho \vec{\bf B}$ in~(\ref{structure of jmu},\ref{structure of jmu5}) would  lead to the chiral magnetic wave~\cite{1012.6026},
%Actually, to study chiral magnetic wave, one considers
which reflects density fluctuations $\delta \rho,\delta\rho_{_5}$ at constant external magnetic field.
According to~\cite{1012.6026}, the speed of the chiral magnetic wave can depend on ${\bf B}$ nonlinearly, the property which is inherited
from a nonlinear ${\bf B}$-dependence of $\mu,\mu_{_5}$.
In~\cite{1012.6026}, this nonlinearity of $\mu,\mu_{_5}$  was realised in a top-down holographic QCD model based on a DBI action for the bulk gauge fields. In contrast, working with the canonical Maxwell action, our study demonstrates that similar non-linear phenomena
can emerge solely from the Chen-Simons term.

The terms $\vec \nabla\delta \rho_5\times \vec{\bf E}$ and $\vec \nabla\delta\rho\times \vec{\bf E}$ in~(\ref{structure of jmu},\ref{structure of jmu5})
were studied in \cite{1603.03620} within the chiral kinetic theory. It would be interesting to compare the results once the corresponding transport coefficients are computed in the holographic model.

%dependence on ${\bf B}$ of chiral magnetic wave can be introduced by including nonlinear corrections such as ${\bf B}^2 \vec{\bf B}$-term to $\vec{J},\vec{J}_5$.
%Obviously, the nonlinearity is important when magnetic field is strong.
%
%When there are external fields of left-handed chirality only, the authors of~\cite{1105.6360} wrote down all terms up to second order in gradients for the current. After making the identification $E=E^a,B=B^a,\rho=\rho_{_5}$, we do reproduce all the relevant terms in~\cite{1105.6360} but get one more term corresponding to relaxation time of charge density diffusion. We realized that this relaxation term was thought of as a third order structure in gradient once making use of hydro equation in~\cite{1105.6360}.
%
%It is worth comparing with~\cite{1105.6360} where the authors wrote down all possible terms for a $U(1)$ current (with left-handed external fields only) up to second order in the hydrodynamic gradient expansion. To make this comparison, we should identify the vector e/m fields with the axial ones and sets $\rho=\rho_{_5}$. Doing so, we recovered all the six terms for the anomalous $U(1)$ current of~\cite{1105.6360}. However, we got one more term which encodes the relaxation time of charge diffusion. We realized that the authors of~\cite{1105.6360} treated this relaxation term as a third order structure via hydrodynamic equation.

\section{Conclusions} \label{section6}

In this paper, we have revised anomaly induced  transport in a holographic model containing two $U(1)$ fields interacting via Chern-Simons terms.
For a finite temperature system, we have computed  off-shell constitutive relations  for the vector and axial currents responding to
 external vector and axial electromagnetic fields.

All-order gradients can be resummed in a weak field (linear response) approximation.
%linear response  are evaluated using the fluid/gravity correspondence. The calculations involve an all-order gradient resummation. As a result, at linear level in external fields and charge densities, the most general off-shell
Thus obtained  constitutive relations (\ref{jmu},\ref{jmu5}) for the vector/axial currents are parameterised by six independent momenta-dependent TCFs: the diffusion $\mathcal{D}(\omega,q^2)$, the electric/magnetic conductivities $\sigma_{e/m}(\omega,q^2)$, chiral magnetic/separation conductivities $\sigma_{\chi/\kappa}(\omega,q^2)$, and an axial analogue of the magnetic conductivity $\sigma_a(\omega,q^2)$.

%These transport coefficient functions are explicitly computed within a specific holographic model.
Within the linear approximation, the TCFs $\mathcal{D}(\omega,q^2)$ and $\sigma_e(\omega,q^2)$ are left unaffected by the anomaly and
were computed previously in \cite{1511.08789}. While $\sigma_m(\omega,q^2)$ gets an anomaly induced correction, the remaining TCFs, $\sigma_{\chi/\kappa/a}$,  are induced by the anomaly.
%Computation of anomaly-related transports is our main result.
In the hydrodynamic regime, we have analytically reproduced all the known results in the literature and succeeded to extended the gradient expansion to third order, see (\ref{D expansion}-\ref{sigma_kappa expansion}). Beyond the hydrodynamic regime, these transport coefficient functions were numerically calculated up to large values of momenta so that the asymptotic regime is reached. The results are displayed by the  plots in section \ref{subsection42}. The electric/magnetic conductivities $\sigma_{e/m}$ are the only TCFs that survive at asymptotically large $\omega$ (the asymptotics is reached around $\omega \simeq 5$).
%while all the rest decouple.

Nonlinear transport has been studied in a specific setting of the external fields having constant backgrounds. When the only non-vanishing external field is a constant magnetic field, the CME has  been shown to be exact, relating the induced vector current and the  magnetic field, see (\ref{jmu EB exact}). This exact relation is nonlinear in the magnetic field and the entire nonlinearity is absorbed into the axial chemical potential $\mu_{_5}$, see (\ref{chem/den2}). Electric fields lead to new  nonlinear effects, see (\ref{jmu EB}). Small time-dependent/non-homogenous perturbations introduce many more interesting anomaly-induced  structures in the constitutive relations.
% In this step we first reproduce the results of section \ref{section4}, see (\ref{jmu/jmu5 re}); then, we primarily outlined all the basic structures for vector/axial currents at the order $\mathcal{O}(\epsilon^1 \alpha^1)$, see (\ref{structure of jmu},\ref{structure of jmu5}).
We have merely listed these structures, leaving determination of  associated new transport coefficients for a future study.

%It was claimed that chiral magnetic effect is exact even when the magnetic field is promoted to
Additional non-linear anomaly-induced effects are explored in our forthcoming publication \cite{BLS}. Particularly,
CME in presence of a space-varying magnetic field $\vec{B}(\vec{x})$
%(so that external electric field can be set to zero)
is found to be modified by derivative corrections.  An interplay between constant magnetic and time-dependent electric fields is another focus of \cite{BLS}.

%There are some open issues worthy to be addressed in the future.

At the lowest order in derivative expansion, the anomalous transport coefficients  $\sigma_{\chi/\kappa}^0$ are known to be dissipationless \cite{0906.5044,1105.6360}. Particularly, they do not contribute to entropy production in a hydrodynamic system. It would be interesting to classify  all the higher derivatives/nonlinear terms in the constitutive relations in accord with  their dissipative nature \cite{1502.00636}. We do expect that the higher order gradient terms
would introduce dissipation and this would potentially affect various phenomena such as the chiral drag force \cite{1505.07379,1511.08794}.

Our study has been carried out in the probe limit, in which the currents get decoupled from the dynamics of the energy-momentum tensor.
Beyond this limit, new phenomena emerge such as the normal Hall  and  chiral vortical effects
%certain interesting terms will appear in both the currents and the energy-momentum tensor at nonlinear level
\cite{1105.6360,1304.5529}.
Interplay between the vorticity and strong magnetic field \cite{1607.01513} is an interesting direction worth further study in a holographic setup beyond the probe limit.

\appendix

\section{ODEs and the constraints for the decomposition coefficients in (\ref{decomp V},\ref{decomp A})} \label{appendix1}
In this appendix, we collect the ODEs satisfied by the decomposition coefficients in~(\ref{decomp V},\ref{decomp A}), and derive the constraint relations relating these  coefficients. Substituting~(\ref{decomp V},\ref{decomp A}) into (\ref{eom Vt scheme1}-\ref{eom Ai scheme1}) and making Fourier transform $\partial_\mu\to (-i\omega,i\vec{q})$, we arrive at the ODEs, which are grouped into several partially decoupled sub-sectors.\\
$\left\{S_1,V_2\right\}$
\begin{equation} \label{eom 1}
0=r^2\partial^2_rS_1+3r\partial_r S_1-q^2\partial_r V_2,
\end{equation}
\begin{equation} \label{eom 2}
\begin{split}
0=\left(r^5-r\right)\partial_r^2 V_2+\left(3r^4+1\right)\partial_r V_2-2i\omega r^3 \partial_r V_2-i\omega r^2 V_2-r^3\partial_rS_1-r^2S_1-r^2.
\end{split}
\end{equation}
$\left\{S_2,\bar{S}_2,V_1,\bar{V}_1,V_3,\bar{V}_3,V_5,\bar{V}_5\right\}$
\begin{equation} \label{eom 3}
0=r^2\partial^2_rS_2+3r\partial_r S_2+\partial_r V_1-q^2\partial_r V_3,
\end{equation}
\begin{equation} \label{eom 4}
0=r^2\partial^2_r\bar{S}_2+3r\partial_r \bar{S}_2+\partial_r \bar{V}_1-q^2\partial_r \bar{V}_3,
\end{equation}
\begin{equation} \label{eom 5}
\begin{split}
0&=\left(r^5-r\right)\partial_r^2V_1+\left(3r^4+1\right)\partial_rV_1-2i\omega r^3 \partial_r V_1-i\omega r^2 V_1-q^2rV_1-i\omega r^2-q^2r\\
&+\frac{12\kappa q^2}{r}\left(\bar{\rho}_{_5}V_5 +\bar{\rho} \bar{V}_5\right),
\end{split}
\end{equation}
\begin{equation} \label{eom 6}
0=\left(r^5-r\right)\partial_r^2\bar{V}_1+\left(3r^4+1-2i\omega r^3\right) \partial_r\bar{V}_1  -(i\omega r^2+q^2r) \bar{V}_1+\frac{12\kappa q^2}{r} \left(\bar{\rho}V_5 +\bar{\rho}_{_5} \bar{V}_5\right),
\end{equation}
\begin{equation} \label{eom 7}
\begin{split}
0&=\left(r^5-r\right)\partial_r^2V_3+\left(3r^4+1\right)\partial_rV_3-2i\omega r^3 \partial_r V_3-i\omega r^2 V_3-r^3\partial_rS_2-r^2S_2 -rV_1\\ &-r+\frac{12\kappa}{r}\left(\bar{\rho}_{_5}V_5 +\bar{\rho} \bar{V}_5\right),
\end{split}
\end{equation}
\begin{equation} \label{eom 8}
\begin{split}
0&=\left(r^5-r\right)\partial_r^2\bar{V}_3+\left(3r^4+1\right)\partial_r\bar{V}_3-2i\omega r^3 \partial_r \bar{V}_3-i\omega r^2 \bar{V}_3-r^3\partial_r\bar{S}_2-r^2\bar{S}_2 -r\bar{V}_1\\
&+\frac{12\kappa}{r} \left(\bar{\rho}V_5 +\bar{\rho}_{_5} \bar{V}_5\right),
\end{split}
\end{equation}
\begin{equation} \label{eom 9}
\begin{split}
0&=\left(r^5-r\right)\partial_r^2V_5+\left(3r^4+1\right)\partial_rV_5-2i\omega r^3 \partial_r V_5 -i\omega r^2 V_5-q^2rV_5 +\frac{12\kappa}{r}\left(\bar{\rho}_5\right.\\
&\left.+ \bar{\rho}_5V_1+\bar{\rho}\bar{V}_1\right),
\end{split}
\end{equation}
\begin{equation} \label{eom 10}
\begin{split}
0&=\left(r^5-r\right)\partial_r^2\bar{V}_5+\left(3r^4+1\right)\partial_r\bar{V}_5-2i\omega r^3 \partial_r \bar{V}_5-i\omega r^2 \bar{V}_5-q^2r\bar{V}_5+\frac{12\kappa}{r} \left(\bar{\rho}\right.\\
&\left.+\bar{\rho}V_1+\bar{\rho}_5 \bar{V}_1\right).
\end{split}
\end{equation}
$\left\{S_3,V_4\right\}$
\begin{equation} \label{eom 11}
0=r^2\partial_r^2S_3+3r\partial_rS_3-q^2\partial_rV_4,
\end{equation}
\begin{equation} \label{eom 12}
0=\left(r^5-r\right)\partial_r^2V_4+\left(3r^4+1\right)\partial_rV_4-2i\omega r^3 \partial_rV_4-i\omega r^2V_4-r^3\partial_rS_3-r^2S_3-\frac{1}{2}.
\end{equation}
$\left\{S_5,\bar{S}_5,V_6,\bar{V}_6,V_8,\bar{V}_8,V_{10},\bar{V}_{10}\right\}$
\begin{equation} \label{eom 13}
0=r^2\partial_r^2S_5+3r\partial_rS_5+\partial_rV_6-q^2\partial_rV_8,
\end{equation}
\begin{equation} \label{eom 14}
0=r^2\partial_r^2\bar{S}_5+3r\partial_r\bar{S}_5+\partial_r\bar{V}_6-q^2\partial_r \bar{V}_8,
\end{equation}
\begin{equation} \label{eom 15}
\begin{split}
0=\left(r^5-r\right)\partial_r^2V_6+\left(3r^4+1-2i\omega r^3\right)\partial_rV_6 -(i\omega r^2+q^2 r )V_6+\frac{12\kappa q^2}{r}\left(\bar{\rho}_5 V_{10} +\bar{\rho} \bar{V}_{10} \right),
\end{split}
\end{equation}
\begin{equation} \label{eom 16}
\begin{split}
0&=\left(r^5-r\right)\partial_r^2\bar{V}_6+\left(3r^4+1\right)\partial_r\bar{V}_6-2i\omega r^3 \partial_r\bar{V}_6-i\omega r^2 \bar{V}_6-q^2 r\bar{V}_6-i\omega r^2-q^2r\\
&+\frac{12\kappa q^2}{r}\left(\bar{\rho} V_{10} +\bar{\rho}_5 \bar{V}_{10} \right),
\end{split}
\end{equation}
\begin{equation} \label{eom 17}
\begin{split}
0&=\left(r^5-r\right)\partial_r^2V_8+\left(3r^4+1\right)\partial_rV_8-2i\omega r^3 \partial_rV_8-i\omega r^2 V_8-r^3\partial_rS_5-r^2S_5-rV_6\\
&+\frac{12\kappa}{r}\left(\bar{\rho}_5 V_{10} +\bar{\rho} \bar{V}_{10} \right),
\end{split}
\end{equation}
\begin{equation} \label{eom 18}
\begin{split}
0&=\left(r^5-r\right)\partial_r^2\bar{V}_8+\left(3r^4+1\right)\partial_r\bar{V}_8-2i\omega r^3 \partial_r\bar{V}_8-i\omega r^2 \bar{V}_8-r^3\partial_r\bar{S}_5 -r^2\bar{S}_5 -r\bar{V}_6\\
&-r+\frac{12\kappa}{r}\left(\bar{\rho} V_{10} +\bar{\rho}_5 \bar{V}_{10} \right),
\end{split}
\end{equation}
\begin{equation} \label{eom 19}
\begin{split}
0=\left(r^5-r\right)\partial_r^2V_{10}+\left(3r^4+1-2i\omega r^3\right)\partial_rV_{10} -(i\omega r^2+q^2 r) V_{10}+\frac{12\kappa}{r}\left(\bar{\rho}_5 V_6+\bar{\rho} +\bar{\rho} \bar{V}_6 \right),
\end{split}
\end{equation}
\begin{equation} \label{eom 20}
0=\left(r^5-r\right)\partial_r^2\bar{V}_{10}+\left(3r^4+1-2 i\omega r^3\right) \partial_r\bar{V}_{10}-(i\omega r^2+q^2 r) \bar{V}_{10}+\frac{12\kappa}{r} \left(\bar{\rho} V_6+\bar{\rho}_5 +\bar{\rho}_5 \bar{V}_6 \right).
\end{equation}
$\left\{S_4,V_7\right\}$
\begin{equation} \label{eom 21}
0=r^2\partial_r^2S_4+3r\partial_rS_4-q^2\partial_rV_7,
\end{equation}
\begin{equation} \label{eom 22}
0=\left(r^5-r\right)\partial_r^2V_7+\left(3r^4+1\right)\partial_rV_7-2i\omega r^3 \partial_rV_7-i\omega r^2V_7-r^3\partial_rS_4-r^2S_4.
\end{equation}
For the remaining coefficients: $\left\{S_6,V_9\right\}$, $\left\{\bar{S}_1,\bar{V}_2\right\}$, and $\left\{\bar{S}_3,\bar{V}_4\right\}$ obey the same ODEs as $\left\{S_4,V_7\right\}$; $\left\{\bar{S}_4,\bar{V}_7\right\}$ satisfy the same equations as $\left\{S_1,V_2\right\}$; $\left\{\bar{S}_6,\bar{V}_9\right\}$ and $\left\{S_3,V_4\right\}$ obey the same ODEs as well.

Certain constraint relations among the coefficients in~(\ref{decomp V},\ref{decomp A}) can be established. First, (\ref{Lcons1}) combined with
the boundary conditions~(\ref{asymp cond},\ref{regular cond}) and  homogeneity of the ODEs (\ref{eom 21},\ref{eom 22}) for $\left\{S_4,V_7\right\}$, $\left\{S_6,V_9\right\}$, $\left\{\bar{S}_1,\bar{V}_2\right\}$ and $\left\{\bar{S}_3,\bar{V}_4\right\}$ result in the following identities
\begin{equation}\label{constraint1}
S_4=S_6=\bar{S}_1=\bar{S}_3=0,~~~~~~~~~~V_7=V_9=\bar{V}_2=\bar{V}_4=0.
\end{equation}
Furthermore,
\begin{equation}\label{constraint2}
S_1=\bar{S}_4,~~~~~~~~~~~~V_2=\bar{V}_7;~~~~~~~~~~~S_3=\bar{S}_6,~~~~~~~~~~V_4=\bar{V}_9.
\end{equation}
This comes from the fact that these functions satisfy identical ODEs (\ref{eom 1},\ref{eom 2}) with identical  boundary conditions. Additionally, as in \cite{1511.08789} consider the combinations
\begin{equation}
X_1=i\omega S_1+q^2 S_2,~~~~~\bar{X}_1=\bar{S}_2,~~~~~Y_1=i\omega V_2+q^2V_3-V_1,~~~~~\bar{Y}_1=q^2\bar{V}_3-\bar{V}_1,
\end{equation}
\begin{equation}
X_2=S_5,~~~~~\bar{X}_2=i\omega \bar{S}_4+q^2\bar{S}_5,~~~~~ Y_2=q^2V_8-V_6,~~~~~\bar{Y}_2=i\omega \bar{V}_7+q^2 \bar{V}_8-\bar{V}_6,
\end{equation}
which satisfy homogeneous equations. Therefore, under (\ref{Lcons2},\ref{Lcons3}) and boundary conditions (\ref{asymp cond},\ref{regular cond}) we have
\begin{equation}\label{constraint3}
i\omega S_1+q^2 S_2=0,~~~~~\bar{S}_2=0,~~~~~i\omega V_2+q^2V_3-V_1=0,~~~~~q^2\bar{V}_3- \bar{V}_1=0,
\end{equation}
\begin{equation}\label{constraint4}
S_5=0,~~~~~i\omega \bar{S}_4+q^2\bar{S}_5=0,~~~~~q^2V_8-V_6=0,~~~~~i\omega \bar{V}_7+q^2 \bar{V}_8-\bar{V}_6=0.
\end{equation}
The relations~(\ref{constraint1}, \ref{constraint2}, \ref{constraint3}, \ref{constraint4}) could be further translated into constraints among $v_{_i},\bar{v}_{_i}$,
\begin{equation}\label{LD cons}
\begin{split}
&v_{_7}=v_{_9}=\bar{v}_{_2}=\bar{v}_{_4}=0,~~~~~v_{_2}=\bar{v}_{_7},~~~~~v_{_4}=\bar{v}_{_9},~~~~~
i\omega v_{_2}+q^2 v_{_3}-v_{_1}=0,\\
&q^2\bar{v}_{_3}-\bar{v}_{_1}=0,~~~~~q^2v_{_8}-v_{_6}=0,~~~~~i\omega \bar{v}_{_7}+q^2 \bar{v}_{_8}-\bar{v}_{_6}=0.
\end{split}
\end{equation}

The relations~(\ref{constraint1}, \ref{constraint2}, \ref{constraint3}, \ref{constraint4}) reduce the number of equations that needs to be solved.
The remaining independent sub-sectors are
\begin{equation}
\left\{S_1,V_2\right\},~~~~~\left\{S_3,V_4\right\},~~~~~\left\{V_1,\bar{V}_1,V_5,\bar{V}_5\right\},
~~~~~~\left\{V_6,\bar{V}_6,V_{10},\bar{V}_{10}\right\}.
\end{equation}
Note that under the interchange
\begin{equation}
V_1 \leftrightarrow \bar{V}_6,~~~~~\bar{V}_1\leftrightarrow V_6,~~~~~V_5 \leftrightarrow \bar{V}_{10},~~~~~\bar{V}_5 \leftrightarrow V_{10},
\end{equation}
The ODEs satisfied by the sub-sectors $\left\{V_1,\bar{V}_1,V_5,\bar{V}_5\right\}$ and
$\left\{V_6,\bar{V}_6,V_{10},\bar{V}_{10}\right\}$ get exchanged in the following way,
$$
\textrm{\ref{eom 5}}\leftrightarrow \textrm{\ref{eom 16}},~~~~~
\textrm{\ref{eom 6}}\leftrightarrow \textrm{\ref{eom 15}},~~~~~
\textrm{\ref{eom 9}}\leftrightarrow \textrm{\ref{eom 20}},~~~~~
\textrm{\ref{eom 10}}\leftrightarrow \textrm{\ref{eom 19}}.
$$
Given that $\left\{V_1,\bar{V}_1,V_5,\bar{V}_5\right\}$ and
$\left\{V_6,\bar{V}_6,V_{10},\bar{V}_{10}\right\}$ obey the same boundary conditions (\ref{asymp cond},\ref{regular cond}),
the following "symmetric" relations hold
\begin{equation}\label{symmt1}
\begin{split}
&V_1=\bar{V}_6,~~~~~\bar{V}_1=V_6,~~V_5=\bar{V}_{10},~~~~~\bar{V}_5=V_{10},\\
\Rightarrow&v_{_{_1}}=\bar{v}_{_6},~~~~~\bar{v}_{_1}=v_{_6},~~~~~v_{_{_5}}=\bar{v}_{_{10}},~~~~~ \bar{v}_{_{5}}=v_{_{10}}.
\end{split}
\end{equation}
Based on~(\ref{constraint3},\ref{constraint4}), (\ref{symmt1}) implies
\begin{equation}\label{symmt2}
V_3=\bar{V}_8\Longrightarrow v_{_3}=\bar{v}_{_8}.
\end{equation}
Eventually, we only need to solve $\left\{S_1,V_2\right\}$, $\left\{S_3,V_4\right\}$, $\left\{V_1,\bar{V}_1,V_5,\bar{V}_5\right\}$ and obtain all the other
functions through the relations revealed above.

\section{Perturbative solutions} \label{appendix2}
Perturbative solutions for $\left\{S_2,V_1,\bar{V}_1,V_3,\bar{V}_3, V_5,\bar{V}_5\right\}$ when $\omega,q \ll 1$
are summarised in this appendix. First, let introduce a formal expansion parameter $\lambda$
\begin{equation}
\omega \to \lambda \omega,~~~~~~~~~~~~~\vec{q}\to \lambda \vec{q}.
\end{equation}
Expanding these functions in powers of $\lambda$,
\begin{equation}
\begin{split}
&S_2=\sum_{n=0}^{\infty}\lambda^n S_2^{(n)},~~~~~V_1=\sum_{n=0}^{\infty}\lambda^n V_1^{(n)},~~~~~\bar{V}_1=\sum_{n=0}^{\infty}\lambda^n \bar{V}_1^{(n)},~~~~~
V_3=\sum_{n=0}^{\infty}\lambda^n V_3^{(n)}\\
&\bar{V}_3=\sum_{n=0}^{\infty}\lambda^n \bar{V}_3^{(n)},~~~~~
V_5=\sum_{n=0}^{\infty}\lambda^n V_5^{(n)},~~~~~\bar{V}_5=\sum_{n=0}^{\infty}\lambda^n \bar{V}_5^{(n)},
\end{split}
\end{equation}
we solve (\ref{eom 3},\ref{eom 5}-\ref{eom 10}) perturbatively, order by order in $\lambda$. Final results are quoted below.
\begin{equation}
V_1^{(0)}=\bar{V}_1^{(0)}=\bar{V}_1^{(1)}=S_2^{(0)}=0,
\end{equation}
\begin{equation}
V_5^{(0)}=3\kappa \bar{\rho}_{_5}\log {\frac{1+r^2}{r^2}}\,\underrightarrow{r\to \infty}\, \frac{3\kappa \bar{\rho}_{_5}}{r^2}+\mathcal{O}\left(\frac{1}{r^3}\right),
\end{equation}
\begin{equation}
\bar{V}_5^{(0)}=3\kappa \bar{\rho}\log {\frac{1+r^2}{r^2}} \,\underrightarrow{r\to \infty}\, \frac{3\kappa \bar{\rho}}{r^2}+\mathcal{O}\left(\frac{1}{r^3}\right),
\end{equation}
\begin{equation}
V_1^{(1)}=-\frac{1}{4}i\omega \left[\pi-2 \arctan (r)+\log \frac{(1+r)^2}{1+r^2}\right] \,\underrightarrow{r\to \infty}\,-\frac{i\omega}{r}+\frac{i\omega}{2r^2}+\mathcal{O} \left(\frac{1}{r^3}\right),
\end{equation}
\begin{equation}
\begin{split}
V_5^{(1)}&=-\int_r^{\infty}\frac{xdx}{x^4-1}\int_1^x \left[2i\omega y \partial_y V_5^{(0)}+i\omega V_5^{(0)}-\frac{12\kappa \bar{\rho}_{_5}}{y^3} V_1^{(1)}\right]dy\\
&\underrightarrow{r\to \infty}\, \frac{3i\omega \kappa \bar{\rho}_{_5}\log 2}{r^2}+\mathcal{O} \left(\frac{1}{r^3}\right),
\end{split}
\end{equation}
\begin{equation}
\begin{split}
\bar{V}_5^{(1)}&=-\int_r^{\infty}\frac{xdx}{x^4-1}\int_1^x \left[2i\omega y \partial_y \bar{V}_5^{(0)}+i\omega \bar{V}_5^{(0)}-\frac{12\kappa \bar{\rho}}{y^3}V_1^{(1)} \right] dy \\
&\underrightarrow{r\to \infty}\, \frac{3i\omega \kappa \bar{\rho}\log 2}{r^2} +\mathcal{O} \left(\frac{1}{r^3}\right),
\end{split}
\end{equation}
\begin{equation}
\begin{split}
V_1^{(2)}&=-\int_r^{\infty} \frac{xdx}{x^4-1}\int_1^xdy \left[2i\omega y \partial_y V_1^{(1)}+i\omega V_1^{(1)}+\frac{q^2}{y}+\frac{q^2}{y}V_1^{(0)} -\frac{12\kappa q^2}{y^3}\right.\\
&~~~~~~~~~~~~~~~~~~~~~~~~~~~~\left.\times\left(\bar{\rho}_{_5}V_5^{(0)} +\bar{\rho} \bar{V}_5^{(0)} \right) \right]\\
&\underrightarrow{r\to\infty}\, -\frac{1}{4r^2}\left\{\omega^2\left(1+\log 2\right)+q^2 \left[1-36\kappa^2\left(\bar{\rho}^2+\bar{\rho}_{_5}^2\right)\left(2\log 2-1\right) \right] \right\}\\
&~~~~~~+\frac{1}{2}\left(\omega^2-q^2\right)\frac{\log r}{r^2}+\mathcal{O} \left(\frac{\log r}{r^3}\right),
\end{split}
\end{equation}
\begin{equation}
\bar{V}_1^{(2)}=\int_r^\infty \frac{xdx}{x^4-1}\int_1^x \frac{12\kappa q^2}{y^3} \left(\bar{\rho}V_5^{(0)}+\bar{\rho}_{_5}V_5^{(0)}\right)\,\underrightarrow{r\to \infty}\, \frac{18q^2\kappa^2 \bar{\rho}\bar{\rho}_{_5}}{r^2}\left(2\log 2-1 \right) +\mathcal{O}\left(\frac{1}{r^3}\right),
\end{equation}
\begin{align}
V_5^{(2)}&=-\int_r^\infty \frac{xdx}{x^4-1} \int_1^x dy\left[2i\omega y \partial_y V_5^{(1)} + i\omega V_5^{(1)}+ \frac{q^2}{y}V_5^{(0)}-\frac{12\kappa}{y^3} \left(\bar{\rho}_{_5} V_1^{(2)}+\bar{\rho} \bar{V}_1^{(2)}\right)\right] \nonumber\\
&\underrightarrow{r\to \infty} \,-\frac{\kappa \bar{\rho}_{_5}}{8r^2} \left\{6\omega^2 \log^22+ q^2\left[\pi^2-432 \kappa^2 \left(\log 2-1\right)^2 \left(\bar{\rho}_{_5}^2 +3 \bar{\rho}^2\right)\right]\right\} +\mathcal{O}\left(\frac{1}{r^3}\right),
\end{align}
\begin{align}
\bar{V}_5^{(2)}&=-\int_r^\infty \frac{xdx}{x^4-1} \int_1^x dy\left[2i\omega y \partial_y \bar{V}_5^{(1)} + i\omega \bar{V}_5^{(1)}+ \frac{q^2}{y}\bar{V}_5^{(0)} -\frac{12\kappa} {y^3} \left(\bar{\rho} V_1^{(2)}+\bar{\rho}_{_5} \bar{V}_1^{(2)}\right)\right] \nonumber \\
&\underrightarrow{r\to \infty}\,- \frac{\kappa \bar{\rho}}{8r^2}\left\{6\omega^2\log^22+ q^2 \left[\pi^2-432 \kappa^2 \left(\log 2-1\right)^2\left(\bar{\rho}^2 +3 \bar{\rho}_{_5}^2 \right)\right]\right\}+\mathcal{O}\left(\frac{1}{r^3}\right),
\end{align}
\begin{equation}
\begin{split}
V_3^{(0)}&=-\int_r^\infty \frac{xdx}{x^4-1}\int_1^x \left[\frac{1}{y}-\frac{12\kappa} {y^3} \left(\bar{\rho}_{_5} V_5^{(0)}+\bar{\rho} \bar{V}_{_5}^{(0)}\right)\right]\\
&\underrightarrow{r\to \infty}\, -\frac{1}{4r^2}\left[1-36\kappa^2 \left(\bar{\rho}^2+\bar{\rho}_{_5}^2 \right)\left(2\log 2-1\right)\right]-\frac{\log r}{2r^2}+\mathcal{O}\left(\frac{\log r} {r^3}\right),
\end{split}
\end{equation}
\begin{equation}
\begin{split}
S_2^{(1)}=&\int_r^\infty \frac{dx}{x^3} \int_1^x y \partial_y V_1^{(1)} dy+ \frac{i\omega} {16r^2} \left(\pi+ 6\log 2\right)\underrightarrow{r\to \infty} \, \frac{i\omega}{4r^2}(1+2\log r) + \mathcal{O} \left(\frac{1}{r^3}\right),
\end{split}
\end{equation}
\begin{equation}
\begin{split}
V_3^{(1)}=&-\int_r^\infty \frac{xdx}{x^4-1} \int_1^x dy \left[2i\omega y \partial_y V_3^{(0)} +i\omega V_3^{(0)}+y \partial_y S_2^{(1)} +S_2^{(1)} +\frac{1}{y} V_1^{(1)}\right.\\
&\left.~~~~~~~~~~~~~~~~~~~~~~~~~~~~-\frac{12\kappa}{y^3}\left(\bar{\rho}_{_5} V_5^{(1)}+ \bar{\rho}\bar{V}_5^{(1)}\right)\right],\\
&\underrightarrow{r\to \infty}\, \frac{i\omega}{32r^2}\left[(2\pi-\pi^2+4\log 2)+ \mathcal{O}\left(\bar{\rho}^2,\bar{\rho}_{_5}^2\right)\right]+ \mathcal{O} \left(\frac{1}{r^3}\right),
\end{split}
\end{equation}
\begin{equation}
\bar{V}_3^{(0)}=\int_r^\infty \frac{xdx}{x^4-1}\int_1^x \frac{12\kappa}{y^3} \left(\bar{\rho}V_5^{(0)}+\bar{\rho}_{_5}\bar{V}_5^{(0)}\right)\underrightarrow{r\to \infty} \frac{18\kappa^2 \bar{\rho}\bar{\rho}_{_5}}{r^2} \left(2\log 2-1\right) +\mathcal{O}\left(\frac{1}{r^3}\right),
\end{equation}
\begin{equation}
\begin{split}
\bar{V}_3^{(1)}=&-\int_r^\infty \frac{xdx}{x^4-1} \int_1^x dy \left[2i\omega y \partial_y \bar{V}_3^{(0)}+i\omega \bar{V}_3^{(0)} -\frac{12\kappa}{y^3} \left(\bar{\rho} V_5^{(1)} + \bar{\rho}_{_5} \bar{V}_5^{(1)} \right) \right].
\end{split}
\end{equation}
These perturbative solutions generate the hydrodynamic expansion of $v_{_3},\bar{v}_{_3},v_5,\bar{v}_{_5}$ as summarized in section~\ref{sss421}.

\section{Frame-independence of the TCFs in~(\ref{jmu},\ref{jmu5})} \label{appendix5}

When the anomaly coefficient $\kappa=0$, frame-independence of $\mathcal{D},\sigma_e,\sigma_m$ was proved in~\cite{1511.08789}. In this appendix, we show that in the presence of triangle anomaly all the TCFs entering ~(\ref{jmu},\ref{jmu5}) are uniquely fixed, even when the frame convention~(\ref{Landau frame}) is relaxed. The proof below goes in parallel to that of~\cite{1511.08789}, but the algebra is more involved.

Under the scheme~(\ref{linear shceme1}), $J^\mu, J^\mu_5$ are conserved as seen from (\ref{continuity}). Relaxing the frame choice (\ref{Landau frame}), we instead require that $J^\mu,J^\mu_5$ of~(\ref{jmu/jmu5}) are invariant under the gauge transformations
\begin{equation}
\mathcal{V}_\mu\to \mathcal{V}_\mu+\partial_\mu \phi,~~~~~~~~~~~~~
\mathcal{A}_\mu\to \mathcal{A}_\mu+\partial_\mu \varphi.
\end{equation}
The gauge invariance of $J^\mu,J^\mu_5$ gives rise to the following relations,
\begin{equation}
i\omega s_{_1}+q^2 s_{_2}=0,~~~~~~i\omega s_{_4}+q^2 s_{_5}=0,
\end{equation}
\begin{equation}
i\omega \bar{s}_{_1}+q^2 \bar{s}_{_2}=0,~~~~~~i\omega \bar{s}_{_4}+q^2 \bar{s}_{_5}=0,
\end{equation}
\begin{equation}
v_{_1}-i\omega v_{_2}-q^2 v_{_3}=0,~~~~~~v_{_6}- i\omega v_{_7}-q^2 v_{_8}=0,
\end{equation}
\begin{equation}
\bar{v}_{_1}-i\omega \bar{v}_{_2}-q^2 \bar{v}_{_3}=0,~~~~~~\bar{v}_{_6}- i\omega \bar{v}_{_7}-q^2 \bar{v}_{_8}=0.
\end{equation}
Substituting  $\delta \rho,\delta \rho_{_5}$ by $J^t$ and $J^t_5$, (\ref{jmu/jmu5}) becomes
\begin{equation}
\begin{split}
J^i=&-D_1 \partial_iJ^t-D_2 \partial_i J^t_5+\sigma_1 \left(E_i-\frac{\partial_i \partial_k }{\partial^2}E_k\right)+ \sigma_2 \frac{\partial_i \partial_k } {\partial^2}E_k +\sigma_3 \left(E_i^a-\frac{\partial_i \partial_k } {\partial^2} E_k^a\right)\\
&+ \sigma_4 \frac{\partial_i \partial_k }{\partial^2}E_k^a+\sigma_5 B_i+ \sigma_6 B_i^a,
\end{split}
\end{equation}
\begin{equation}
\begin{split}
J^i_5=&-\bar{D}_1 \partial_iJ^t_5-\bar{D}_2 \partial_i J^t+\bar{\sigma}_1 \left(E_i^a-\frac{\partial_i \partial_k }{\partial^2}E_k^a\right)+ \bar{\sigma}_2 \frac{\partial_i \partial_k }{\partial^2}E_k^a +\bar{\sigma}_3 \left(E_i-\frac{\partial_i \partial_k } {\partial^2}E_k\right)\\
&+ \bar{\sigma}_4 \frac{\partial_i \partial_k }{\partial^2}E_k+\bar{\sigma}_5 B_i^a+ \bar{\sigma}_6 B_i,
\end{split}
\end{equation}
where
\begin{equation}
D_1=-\frac{2v_{_4}(1-2\bar{s}_{_6})+2v_{_9}2\bar{s}_{_3}}{(1-2s_{_3})(1-2\bar{s}_{_6})-2 \bar{s}_{_3} 2s_{_6}},~~~~~~
D_2=-\frac{2v_{_9}(1-2s_{_3})+2v_{_4}2s_{_6}}{(1-2s_{_3})(1-2\bar{s}_{_6})-2 \bar{s}_{_3} 2s_{_6}},
\end{equation}
\begin{equation}
\sigma_1=\frac{2v_{_1}+(\omega^2+q^2)/2}{i\omega},~~~~~
\sigma_2=2v_{_2}-\frac{1}{2}i\omega-D_1 \left(2s_{_1}+\frac{1}{2}q^2\right)- D_2 2\bar{s}_{_1},~~
\sigma_3=\frac{2v_{_6}}{i\omega},
\end{equation}
\begin{equation}
\sigma_4=2v_{_7}-D_12s_{_4}-D_2 \left(2\bar{s}_{_4}+\frac{1}{2}q^2\right),~~~~~~
\sigma_5=2v_{_5},~~~~~~
\sigma_6=2v_{_{10}},
\end{equation}
\begin{equation}
\bar{D}_1=-\frac{2\bar{v}_{_9}(1-2s_{_3})+2\bar{v}_{_4}2s_{_6}} {(1-2s_{_3})(1-2\bar{s}_{_6})-2\bar{s}_{_3}2s_{_6}},~~~~
\bar{D}_2=-\frac{2\bar{v}_{_4}(1-2\bar{s}_{_6})+2\bar{v}_{_9}2\bar{s}_{_3}} {(1-2s_{_3})(1-2\bar{s}_{_6})-2\bar{s}_{_3}2s_{_6}},
\end{equation}
\begin{equation}
\bar{\sigma}_1=\frac{2\bar{v}_{_6}+(\omega^2+q^2)/2}{i\omega},~~~~
\bar{\sigma}_2=2\bar{v}_{_7}-\frac{1}{2}i\omega-\bar{D}_1\left(2\bar{s}_{_4}+\frac{1}{2} q^2\right)-\bar{D}_22s_{_4},~~~~
\bar{\sigma}_3=\frac{2\bar{v}_{_1}}{i\omega},
\end{equation}
\begin{equation}
\bar{\sigma}_4=2\bar{v}_2- \bar{D}_12\bar{s}_{_1}- \bar{D}_2 \left(2s_{_1}+ \frac{1}{2} q^2\right),~~~~~~
\bar{\sigma}_5=2\bar{v}_{_{10}},~~~~~~
\bar{\sigma}_6=2\bar{v}_{_5}.
\end{equation}

Note that the ODEs (\ref{eom 5}, \ref{eom 6}, \ref{eom 9}, \ref{eom 10}) for $\left\{V_1,\bar{V}_1,V_5,\bar{V}_5\right\}$ and the ODEs (\ref{eom 15}, \ref{eom 16}, \ref{eom 19}, \ref{eom 20}) for $\left\{V_6,\bar{V}_6,V_{10},\bar{V}_{10}\right\}$ are decoupled from all the $S_i,\bar{S}_i$. The boundary conditions (\ref{asymp cond},\ref{regular cond}) are sufficient to completely determine these two sub-sectors. So, the following TCFs get uniquely fixed without imposing the frame convention~(\ref{Landau frame}):
\begin{equation}
\sigma_1,~\sigma_3,~~\sigma_5,~~\sigma_6,~~\bar{\sigma}_1, ~~\bar{\sigma}_3,~~
\bar{\sigma}_5,~~\bar{\sigma}_6.
\end{equation}
The symmetric relations~(\ref{symmt1}) still hold. From~(\ref{symmt1}), one identifies
\begin{equation}
\sigma_5=\bar{\sigma}_5\equiv \sigma_\chi,~~~~~~~\sigma_6=\bar{\sigma}_6\equiv \sigma_\kappa,~~~~~~~~
\sigma_1=\bar{\sigma}_1\equiv \sigma^{\textrm{T}},~~~~~~~~\sigma_3=\bar{\sigma}_3\equiv \sigma^{\textrm{T}}_a.
\end{equation}
To proceed, redefine $S_3$ and $\bar{S}_6$,
\begin{equation}
S_3^*=S_3-\frac{1}{2r^2},~~~~~~~~~\bar{S}_6^*=\bar{S}_6-\frac{1}{2r^2}.
\end{equation}
Then the sub-sectors $\left\{S_3^*,V_4\right\}$, $\left\{S_6,V_9\right\}$, $\left\{\bar{S}_3,\bar{V}_4\right\}$, $\left\{\bar{S}_6^*, \bar{V}_9\right\}$ obey the same homogeneous ODEs. Furthermore, these sub-sectors satisfy the same boundary conditions at $r=\infty$,
\begin{equation}\label{bdy1}
\begin{split}
&S_3^*\to 0,~~~~V_4\to 0;~~~~~~~~~S_6\to 0,~~~~V_9\to 0,~~~~~~~~\textrm{as}~~~r\to \infty,\\
&\bar{S}_3\to 0,~~~~\bar{V}_4\to 0;~~~~~~~~~\bar{S}_6^*\to 0,~~~~\bar{V}_9\to 0,~~~~~~~~~\textrm{as}~~~r\to \infty.
\end{split}
\end{equation}
Regularity at the horizon $r=1$ fixes one more integration constant for $V_4$, $V_9$, $\bar{V}_4$, $\bar{V}_9$, respectively. As a result, relaxing the frame convention~(\ref{Landau frame}), solutions for $\left\{S_3^*,V_4\right\}$, $\left\{S_6,V_9\right\}$, $\left\{\bar{S}_3,\bar{V}_4\right\}$, $\left\{\bar{S}_6^*, \bar{V}_9\right\}$ are parameterized by a choice of $s_{_3}$, $s_{_6}$, $\bar{s}_{_3}$, $\bar{s}_{_6}$, respectively. Homogeneity of the ODEs obeyed by $\left\{S_3^*,V_4\right\}$, $\left\{S_6,V_9\right\}$, $\left\{\bar{S}_3,\bar{V}_4\right\}$, $\left\{\bar{S}_6^*, \bar{V}_9\right\}$, along with the boundary conditions~(\ref{bdy1}), admits a family of solutions which have $r$-independent scaling symmetry for these sub-sectors. In other words, the ratios $V_4/S_3^*$, $V_9/S_6$, $\bar{V}_4/\bar{S}_3$ and $\bar{V}_9/\bar{S}_6^*$ are uniquely fixed to the same value. At the boundary $r=\infty$, these ratios translate into the statement that
\begin{equation}\label{relation1}
\frac{2v_{_4}}{-1+2s_{_3}}=\frac{v_{_9}}{s_{_6}}=\frac{\bar{v}_4}{\bar{s}_{_3}}= \frac{2\bar{v}_{_9}}{-1+2\bar{s}_{_6}}
\end{equation}
are uniquely fixed and independent of the choice of $s_{_3}$, $s_{_6}$, $\bar{s}_{_3}$, $\bar{s}_{_6}$. The relations~(\ref{relation1}) imply that
\begin{equation}
D_2=\bar{D}_2=0,~~~~~~~~
D_1=\frac{2v_{_4}}{-1+2s_{_3}}=\frac{2\bar{v}_{_9}}{-1+2\bar{s}_{_6}}=\bar{D}_1.
\end{equation}
Thus, $D_1, \bar{D}_1$ are identified as the diffusion $\mathcal{D}$ of~(\ref{jmu},\ref{jmu5}). This proves that the diffusion TCF $\mathcal{D}$ is uniquely fixed and  frame-independent quantity.

The remaining transport coefficients are
\begin{equation}
\begin{split}
&\sigma_2=2v_{_2}-\frac{1}{2}i\omega-\mathcal{D}\left(2s_{_1}+\frac{1}{2}q^2\right),~~
\sigma_4=2v_{_7}-\mathcal{D}2s_{_4},\\
&\bar{\sigma}_2=2\bar{v}_{_7}-\frac{1}{2}i\omega-\mathcal{D}\left(2\bar{s}_{_4}+\frac{1} {2}q^2 \right),~~\bar{\sigma}_4=2\bar{v}_{_2}-\mathcal{D}2\bar{s}_{_1}.
\end{split}
\end{equation}
Notice that $\left\{S_4,V_7\right\}$ and $\left\{\bar{S}_1,\bar{V}_2\right\}$ satisfy the same ODEs as $\left\{S_3^*,V_4\right\}$. Near $r=\infty$, $\left\{S_4,V_7\right\}$ and $\left\{\bar{S}_1,\bar{V}_2\right\}$ satisfy the boundary condition
\begin{equation}
S_4\to0,~V_7\to 0;~~~\bar{S}_1\to 0,~\bar{V}_2\to 0.
\end{equation}
With the same analysis leading to~(\ref{relation1}), we have
\begin{equation}
\frac{v_{_7}}{s_{_4}}=\frac{\bar{v}_{_2}}{\bar{s}_{_1}}=\frac{2v_{_4}}{-1+2s_{_3}}
\Longrightarrow \sigma_4=\bar{\sigma}_4=0.
\end{equation}

The situation for $\sigma_2,\bar{\sigma}_2$ is more complicated. Since the sub-sectors $\left\{S_1,V_2\right\}$ and $\left\{\bar{S}_4,\bar{V}_7\right\}$ satisfy the same ODEs, we focus on $\left\{S_1,V_2\right\}$. The case  is exactly the same as considered in~\cite{1511.08789}. Therefore,
\begin{equation}
\sigma_2=\bar{\sigma}_2\equiv \sigma^{\textrm{L}}
\end{equation}
are uniquely fixed and  frame-independent.

The currents $J^\mu$ and $J^\mu_5$ are
\begin{equation}
J^i=-\mathcal{D} \partial_iJ^t+\sigma^{\textrm{T}} \left(E_i-\frac{\partial_i \partial_k }{\partial^2}E_k\right)+ \sigma^{\textrm{L}} \frac{\partial_i \partial_k } {\partial^2}E_k+\sigma^{\textrm{T}}_a \left(E_i^a-\frac{\partial_i \partial_k } {\partial^2}E_k^a\right)+\sigma_\chi B_i+ \sigma_\kappa B_i^a,
\end{equation}
\begin{equation}
J^i_5=-\mathcal{D} \partial_iJ^t_5+\sigma^{\textrm{T}} \left(E_i^a-\frac{\partial_i \partial_k }{\partial^2}E_k^a\right)+ \sigma^{\textrm{L}} \frac{\partial_i \partial_k }{\partial^2}E_k^a+\sigma^{\textrm{T}}_a \left(E_i-\frac{\partial_i \partial_k } {\partial^2}E_k\right)+\sigma_\chi B_i^a+ \sigma_\kappa B_i,
\end{equation}
which can be put into the forms of~(\ref{jmu},\ref{jmu5}) under the identification
\begin{equation}
\sigma_e=\sigma^{\textrm{L}},~~~\sigma_m=\frac{i\omega}{q^2}\left(\sigma^{\textrm{T}}- \sigma^{\textrm{L}}\right),~~~\sigma_a=\frac{i\omega}{q^2} \sigma^{\textrm{T}}.
\end{equation}
This completes the proof.

\section*{Acknowledgements}
We would like to thank Dmitri E. Kharzeev, Alex Kovner, Derek Teaney, and Ho-Ung Yee for useful discussions related to this work. YB would like to thank KITPC (Beijing) for financial support and hospitality, Physics Department of the University of Connecticut for hospitality where part of this work was done. This work was supported by the ISRAELI SCIENCE FOUNDATION grant \#87277111, BSF grant \#012124, the People Program (Marie Curie Actions) of the European Union's Seventh Framework under REA grant agreement \#318921; and the Council for Higher Education of Israel under the PBC Program of Fellowships for Outstanding Post-doctoral Researchers from China and India (2015-2016).

\providecommand{\href}[2]{#2}\begingroup\raggedright\endgroup

%\bibliographystyle{utphys}
%\bibliography{reference}
\end{document}